\documentclass[fleqn,usenatbib]{mnras}

\usepackage{fix-cm}
\usepackage{newtxtext,newtxmath}

\usepackage[T1]{fontenc}

\DeclareRobustCommand{\VAN}[3]{#2}
\let\VANthebibliography\thebibliography
\def\thebibliography{\DeclareRobustCommand{\VAN}[3]{##3}\VANthebibliography}

\usepackage{graphicx}	
\usepackage{amsmath}	
\usepackage{booktabs} 
\usepackage{makecell} 

\defcitealias{jiang_statistics_2016}{JvB~I}
\defcitealias{vandenbosch_statistics_2016}{JvB~II}
\defcitealias{jiang_statistics_2017}{JvB~III}
\defcitealias{Dekel2007}{DB07}

\title[Dynamical friction heating in early Universe]{Impact of subhalo dynamical friction heating on the formation of the first structures in the Universe}

\author[Zhenyu Wu et al.]{
Zhenyu Wu,$^{1}$\thanks{E-mail: Zhenyu.Wu@ed.ac.uk}
Sadegh Khochfar,$^{1}$
Muhammad A. Latif,$^{2}$
Ben Morton,$^{1}$ and Britton Smith$^{1}$
\\
$^{1}$Institute for Astronomy, University of Edinburgh, Royal Observatory, Blackford Hill, Edinburgh EH9 3HJ, UK\\
$^{2}$Physics Department, College of Science, United Arab Emirates University, PO Box 15551, Al-Ain, UAE\\
}

\date{Accepted July 30. Received July 30; in original form March 19}

\pubyear{\the\year{}}

\begin{document}
\label{firstpage}
\pagerange{\pageref{firstpage}--\pageref{lastpage}}
\maketitle

\begin{abstract}
We present a model for gas heating, driven by dynamical friction from orbiting subhalos within dark matter halos. Using data from the TNG50 simulation, we derive the subhalo mass function and calculate the dynamical friction heating rate for a wide range of halo masses and redshifts from $z = 15$ to 0. Our results show that, by converting gravitational potential energy into thermal energy, dynamical friction is an important mechanism for galaxy quenching in massive halos at low redshifts, consistent with previous studies. Additionally, we find that in the early Universe at $z \sim 15$, heating rates can be comparable to the molecular hydrogen cooling rates in metal-free minihalos.
This can suppress gas cooling and fragmentation and does increase the critical molecular fraction for Pop III star formation by up to one order of magnitude, thereby making Pop III star formation more difficult. In combination with the Lyman-Werner background, the dynamical friction heating mechanism favors the formation of direct-collapse black hole (DCBH) seeds in atomic cooling halos, even when the average H$_2$ fraction is $\sim 10^{-5}$ during the minihalo progenitor phase. Dynamical friction heating at a fixed host halo mass can vary by two orders of magnitude due to the scatter in the number of subhalos. To capture dynamical friction heating in simulations, it is necessary to resolve subhalos with a subhalo to host halo mass ratio $\psi \gtrsim 0.05$.

\end{abstract}




\section{Introduction}

The thermal history of baryons in dark matter halos fundamentally regulates star formation. 
In the presence of efficient cooling, gas can collapse under its self-gravity to form gas clouds in which star formation will take place. Competing with cooling are various feedback processes that can suppress star formation  via thermal energy deposition, kinetic processes, or radiation \citep[see e.g.][for a review]{Somerville2015, Naab2017}. Below $M_{\mathrm{halo}} \sim 10^{12} M_{\odot}$, photo-ionization and supernova feedback play a dominant role \citep[e.g.][]{Dekel86,Davis14}, while in more massive halos above the threshold, AGN feedback and gravitational heating contribute more \citep{Dekel2006,Khochfar2008}.

In this work, we systematically investigate the heating of gas via dynamical friction (DF) by subhalos as a form of gravitational heating. In the standard $\Lambda$CDM framework, structures form in a  hierarchical sequence. Small subhalos are accreted and merge with other haloes to form larger haloes. As subhalos orbit around their host halos, they gravitationally perturb the ambient gas, inducing a dynamical friction wake \citep{ostriker_dynamical_1999}.
This wake exerts a drag force on the subhalos, leading to the dissipation of their orbital kinetic energy and the deposition of thermal energy into the ambient gas. 

The process of gaseous dynamical friction is a variation of the collisionless dynamical friction originally proposed by Chandrasekhar \citep{Chandrasekhar1943}. Using time-dependent linear perturbation theory, \citet{ostriker_dynamical_1999} derives the DF drag force and density wake structure for a perturber moving along a straight line through a uniform gaseous medium. 
The analytic results of \citet{ostriker_dynamical_1999}  have been compared to numerical simulations using extended perturbers  and shown to yield good agreement  \citep[e.g.][]{SnchezSalcedo1999,Morton2025}.
Further studies have extended Ostriker's work to more realistic scenarios, such as a perturber on an orbital trajectory \citep{SnchezSalcedo2001, kim_dynamical_2007, Desjacques2022, oneill_gaseous_2024}, binary perturber \citep{Kim2008}, nonlinear perturbations \citep{kim_nonlinear_2009, bernal_gravitational_2013}, mass accretion onto the object \citep{Lee2011, suzuguchi_gas_2024}, perturber moving at a relativistic speed \citep{Barausse2007}, DF in a turbulent gaseous medium \citep{lescaudron_dynamical_2023}, and DF with radiative feedback \citep{Park2017, velasco-romero_dynamical_2020, Toyouchi2020}. 
Nevertheless, the analytic solution of \citet{ostriker_dynamical_1999} remains one of the most widely used models and serves as a baseline for comparison with more complicated cases. 
The numerical experiments in \citet{Thun2016} confirm the scaling laws for the dependence of the DF force on the basic physical parameters as derived in the previous analytic estimates. 
\citet{Morton2025} finds a broad consistency between different numerical studies and the predictions from the linear perturbation theory, despite minor differences in the fitting values in the DF formula. 
 
The impact of gaseous DF has been investigated in different astrophysical scenarios such as e.g. massive black hole mergers in a gas-rich environment \citep{chen_dynamical_2021, Mayer2013, Bogdanovi2022} or stellar-mass black hole binary formation \citep{DeLaurentiis2023}. 
In terms of gas heating, there is consensus in the literature that gravitational heating is more significant in massive halos above $10^{11-12} M_{\odot}$ 
\citep{Birnboim2007, Dekel2007, Khochfar2008, Johansson2009, Birnboim2011}. \citet{Dekel2007} (hereafter DB07) claim that DF can convert up to one-third of the gravitational potential energy gained during infall into heat. This is likely an upper limit  because they do not include  mass loss  of subhalos due to the tidal field \citep{Zhao2004}. The magnitude of gaseous dynamical friction  heating in massive haloes is large enough to explain the energy balance in X-ray clusters \citep{ElZant2004, kim_dynamical_2005}. 

Although the case of massive haloes at $z=0$ has been investigated in detail, the redshift evolution and the upper and lower halo mass limits of  DF heating have not received similar attention. 
An interesting regime where gaseous DF heating from subhalos may play a role is the formation of Population III (Pop III) stars in the very early Universe. If DF heating is large enough it can overcome cooling and prevent Pop III star formation. 

Pop III stars, as the first generation of stars, form in primordial gas at redshift z $\sim$ 30 \citep[see][e.g.]{Barkana2001,Yoshida2012,klessen_first_2023}.
Below $z = 15$, Pop II star formation becomes more dominant due to metal enrichment from the first stars, though there may be lingering Pop III star formation pockets until the end of the epoch of reionization \citep{Maio10,Maio11,Smith15,Zier2025}. 
Pop III stars are expected to form in minihalos with $T_{\mathrm{vir}} \lesssim 10^4$K.
At these temperatures, the most efficient way for gas to radiate away its internal energy is through ro-vibrational transitions of molecular Hydrogen H$_2$.

However, gas can only cool efficiently if the H$_2$ fraction in a halo reaches a critical value \citep{tegmark_how_1997, yoshida_simulations_2003}.
In the absence of sufficient cooling from molecular Hydrogen, the minihalo can eventually grow into the atomic cooling regime with $T_{\mathrm{vir}} \gtrsim 10^4$ K without forming Pop III stars. The gas can then collapse isothermally via efficient atomic Hydrogen cooling into a $\sim 10^5 M_{\odot}$ direct collapse black hole (DCBH) seed, via the formation of a short-lived supermassive star (SMS) or a quasi-star \citep[see][for a review]{latif_formation_2016, woods_titans_2019, Inayoshi2020review}.

H$_2$ mainly forms via H$^{-}$ and H$_2^{+}$ as catalysts \citep{Abel1997, Galli1998}. One efficient way of reducing the H$_2$ fraction is through
 a Lyman-Werner (LW) photon  background\footnote{Here and in the following we will always refer to H$^{-}$ photo-detachment photons, $>0.76$ eV, as well when citing LW radiation which nominally only covers photon energies in the range $11.2 - 13.6 $ eV.}.  
Lyman-Werner photons cause H$_2$ photo-dissociation via the Solomon process and H$^{-}$ photo-detachment, which suppresses or delays star formation \citep{machacek_simulations_2001, Johnson2012, kulkarni_fragmentation_2019, latif_uv_2019}. 
This process is mostly effective in lower-density gas where self-shielding of the gas against LW radiation is not important \citep{draine_structure_1996, wolcott-green_photodissociation_2011, wolcott-green_h2_2019}. 
The birthplaces of DCBHs typically require no metal enrichment from nearby stars and the suppression of H$_2$ cooling via LW radiation \citep{shang_supermassive_2010, agarwal_ubiquitous_2012, agarwal_first_2014, latif_uv_2014, regan_direct_2014, sugimura_critical_2014, glover_simulating_2015} coming from a nearby galaxy \citep{agarwal_optimal_2019}. 
Typically the critical LW intensity in units of $10^{-21} \mathrm{erg}\ \mathrm{cm}^{-2} \mathrm{~s}^{-1} \mathrm{~Hz}^{-1} \mathrm{sr}^{-1}$ is in the range of $ J_{\mathrm{crit}} \sim 10 - 5\times 10^4$ and depends on the underlying spectrum of the external source \citep{latif_how_2015, agarwal_new_2016, wolcott-green_beyond_2017}.  

Besides DF, other mechanisms related to structure formation have been discussed in the literature that can hinder Pop III star formation and promote DCBH. One mechanism, dubbed {\it dynamical heating}, suggests that rapid mass accretion can be an effective source of heat that delays star formation
\citep{yoshida_simulations_2003, hosokawa_formation_2013, schleicher_massive_2013, lupi_forming_2021}, even in the presence of H$_2$ \citep{wise_formation_2019, sakurai_radiative_2020}. However, dynamical heating mainly accounts for the increase in $T_{\mathrm{vir}}$ due to accretion shocks at $R_{\mathrm{vir}}$, whereas DF includes the heating generated by subhalos orbiting inside $R_{\mathrm{vir}}$ and can operate even in the absence of active accretion.
Other scenarios include streaming velocities \citep{schauer_influence_2021}, accretion shocks \citep{inayoshi_supermassive_2012, Fernandez2014}, and turbulence \citep{latif_turbulent_2022, ho_turbulence_2025}. 

The formation of heavy black hole seeds such as DCBHs has received increased attention as a promising way to explain the number density of the observed quasars powered by black holes $\ge 10^9 M_{\odot}$ at $z \approx 6- 7$ \citep{habouzit_number_2016, Valiante2017, regan_emergence_2020, lupi_forming_2021, Trinca2026}. 
Recent studies also find that DCBHs can potentially lead to systems resembling the spectra of JWST's Little Red Dots (LRDs) \citep{Naidu2025, Cenci2025, Santarelli2025, Pacucci2026}. 

In this work, we investigate dynamical friction heating from subhaloes and present an analytic model to capture its effect on the formation of the first stars and DCBHs.
Section~\ref{Section:DF_method} describes the modeling details including a prescription for the subhalo mass function. 
Based on the model, Section~\ref{Section:massive_halo} compares the heating and cooling flow in massive halos. 
Section~\ref{Section:minihalo} focuses at the  high redshift Universe and discusses on the impact of dynamical friction heating on Pop III stars and DCBH seed formation in minihalos. Section~\ref{Section:heating_profile} examines radial profile corrections to the DF heating model and tests the central heating--cooling balance within halos.
Finally, Section~\ref{Section:conclusion} summarizes the main conclusions and discusses possible improvements in future work. 

Throughout the paper, we adopt the cosmological parameters from the Planck 2015 results \citep{2015Planck}, namely $\Omega_{\mathrm{m}}= 0.3089$, $\Omega_{\mathrm{b}}= 0.0486$, and $H_0=67.74 \mathrm{~km} \mathrm{~s}^{-1} \mathrm{Mpc}^{-1}$. 

\section{Analytic modelling of dynamical friction heating rates in halos}
\label{Section:DF_method}

Modeling the dynamical friction heating rate from subhalos depends crucially on the abundance of subhalos and their mass spectrum. In this section, we first review the analytic dynamical friction model by \citet{ostriker_dynamical_1999} before providing estimates for the subhalo mass function (SHMF) and its scatter across redshift from the TNG50-1\footnote{\url{https://www.tng-project.org/data/}} simulation data \citep{nelson_illustristng_2021, nelson_first_2019, pillepich_first_2019}. These will form the basis of estimates for the evolution of the total DF heating rate across redshift.

\subsection{Analytic model for dynamical friction heating}
\label{section:ostriker}

Using time-dependent linear perturbation theory, \citet{ostriker_dynamical_1999} derives the DF drag force and density wake structure for a perturber in a uniform gaseous medium:

\begin{equation}
\mathbf{F}_{\mathrm{DF}}=-\frac{4 \pi \rho\left(G m\right)^2 I}{v^3} \mathbf{v} \ ,
\label{eq:Ostriker_DF}
\end{equation}

where $\rho$ is the background gas density, $m$ is the perturber mass, and $\mathbf{v}$ is the perturber velocity moving in a straight line. The efficiency factor $I$ is defined similarly to the Coulomb logarithm $\ln (r_{\max}/r_{\min})$ in the collisionless case \citep{Chandrasekhar1943, 2008Binney}, which depends on the Mach number $\mathcal{M}$.

\begin{equation}
I \equiv \begin{cases}\frac{1}{2} \ln \left(\frac{1+\mathcal{M}}{1-\mathcal{M}}\right)-\mathcal{M}, & \mathcal{M}<1 \\ \frac{1}{2} \ln \left(1-\mathcal{M}^{-2}\right)+\ln \left(v t / r_{\min }\right), & \mathcal{M}>1\end{cases}
\label{eq:I_DF_Ostriker99}
\end{equation}

$I$ corresponds to the integral over the wake, which in the sub-sonic case reduces to a contribution only from the non-symmetric region due to the finite time perturbation. In the supersonic case, it sums the contributions from the Mach cone and the sound-speed sphere. 

Following \citet{kim_dynamical_2005}, we adopt the \citet{ostriker_dynamical_1999} model to calculate the DF heating rate from the substructure of halos. For each host halo, the total heating rate due to the gaseous DF of subhalos is 
\begin{equation}
\dot{E}_{\mathrm{DF}}=\sum_m \left (-\mathbf{F}_{\mathrm{DF}} \cdot \mathbf{v}\right) \approx \int \frac{4 \pi \rho\left(G m\right)^2}{c_s} \left\langle I/\mathcal{M}\right\rangle \frac{dN}{dm} dm
\label{eq:Kim_DFheating}
\end{equation}
where $\rho$ is the gas density, $m$ is the mass of the subhalo, $c_s$ is the sound speed in the gas, $\left\langle I/\mathcal{M}\right\rangle$ is the average correction factor for the Mach number distribution of the subhalos (see Section \ref{section:DF_correction}), and $dN/dm$ is the subhalo mass function. Note, that the right-hand side of Eq. \ref{eq:Kim_DFheating} predicts the SHMF average heating rate in a host halo. In Section \ref{section:SHMF_scatter} we will present the scatter around the average heating rate.  
We treat the total mass $m$ of the subhalo collectively as a gravitational perturber and assume that the ambient gas density around a subhalo is $\rho = 200 f_{\rm g} \rho_{\mathrm{crit}} (z)$ with $f_{\rm g}$ as a parameter that allows the mass fraction of gas to vary from the universal baryon fraction in a halo. The sound speed $c_s$ is calculated from the virial temperature of the host halo $T_{\mathrm{vir}} (M, z)$. For the DF efficiency factor $I$ in the supersonic case, the maximum gaseous medium size, $r_{\max} \sim vt$,  is characterized by the host halo size, while the perturber size is the effective size of the subhalo. Using the halo catalog in the TNG50 simulations, we present the subhalo mass function across redshifts in Section~\ref{section:SHMF}. 

Since Equation~\ref{eq:Kim_DFheating} adopts a characteristic ambient gas density and does not include the radial gas-density profile, we refer to this baseline calculation as the \textit{global DF heating model} for a specific host halo. We use this \textit{global model} throughout Sections~\ref{Section:massive_halo} and \ref{Section:minihalo}, and discuss the effects of the DF heating profile in Section~\ref{Section:heating_profile}.



\subsection{Subhalo Mass Function across redshifts}

\label{section:SHMF}

In  \citet{kim_dynamical_2005}, the total dynamical friction heating rate in a galaxy cluster is calculated by summing the contribution from subhalos with mass $m > 10^{11} M_{\odot}$.
In this work, in order to model the contribution of dynamical friction heating from subhalos at different epochs of the Universe, we first need to investigate the subhalo mass function for a wide range of host halo masses $M < 10^{12} M_{\odot} $ from $z = 15$ to $z=0$. 

The statistics of the subhalo abundance in dark matter halos, as well as their properties like the velocity function, has been studied with both N-body simulations \citep[e.g.][]{Gao2004, Diemand2007, giocoli_population_2008, giocoli_substructure_2010, boylan-kolchin_theres_2010, rodriguez-puebla_halo_2016, vandenbosch_statistics_2016, jiang_statistics_2017, moline_cdm_2022} and semi-analytic models \citep[e.g.][]{penarrubia_effects_2005, van_den_bosch_mass_2005, yang_analytical_2011, jiang_statistics_2016, salvador-sole_accurate_2021, hiroshima_semi-analytical_2022}. Numerical simulations identify sub-structures via halo finders, such as \textsc{BDM} \citep{Klypin1999}, \textsc{SUBFIND} \citep{springel_populating_2001}, \textsc{ROCKSTAR} \citep{behroozi_rockstar_2013}, and \textsc{HBT} \citep{Han2017, ForouharMoreno2025}, while semi-analytic models usually construct the merger trees based on the extended Press-Schechter (EPS) formalism \citep[e.g.][]{Bond1991, Lacey1993, Khochfar2001}  and account for halo mass loss due to processes like tidal stripping and dynamical friction \citep{2008Binney, jiang_statistics_2016, Hiroshima2018}. The so-called unevolved SHMF describes the subhalo mass distribution in a host halo before their tidal evolution at the time of accretion, while the evolved SHMF takes into account the evolution of subhalo masses under the influence of tidal forces. The latter is relevant for the dynamical friction heating calculation.

We mainly follow the series of works by \citet{jiang_statistics_2016}, \citet{vandenbosch_statistics_2016}, and \citet{jiang_statistics_2017} (hereafter \citetalias{jiang_statistics_2016}, \citetalias{vandenbosch_statistics_2016}, and \citetalias{jiang_statistics_2017}, respectively) to study the evolution of SHMF.
They use a universal fitting formula for the average SHMF, which is a power law plus an exponential decay when the subhalo mass approaches the host halo mass:

\begin{equation}
\frac{\mathrm{d} N}{\mathrm{~d} \ln \psi}=A \psi^{-\alpha} \exp \left(-\beta \psi^\omega\right)
\label{eq:SHMF_form}
\end{equation}

Here $\psi = m/M$ is the mass ratio of subhalo to host halo, and $A$, $\alpha$, $\beta$, $\omega$ are free parameters depending on host halo mass $M$, redshift $z$, and halo formation time (evaluated when the host halo formed half of its current mass).  

Most of the earlier work focused on $z = 0$ or low redshifts, and Milky Way-sized or cluster-sized halos. Therefore, instead of using the previous fitting values directly, we investigate the evolution of the average SHMF in a wider range of redshifts up to $z = 15$. We derive the SHMF from the halo catalog in the TNG50 data \citep{nelson_illustristng_2021, nelson_first_2019, pillepich_first_2019}. Since the lowest resolved halo mass in the TNG50 is $\sim 10^{7.5} M_{\odot}/h$ and therefore does not extend to the minihalo mass range considered in Section~\ref{Section:minihalo}, we also investigate the SHMF using the high-resolution \textsc{Pop2Prime} simulations of minihalos presented in \cite{Smith15} (see Appendix~\ref{appendix:SHMF_Pop2Prime} for details).


TNG50 has the highest mass resolution in the IllustrisTNG project, making it suitable for the examination of the SHMF in lower-mass host halos. 
The hydrodynamical simulation TNG50-1 consists of $2160^3$ dark matter particles with a mass of $m_{\rm{DM}} = 3.1 \times 10^5 \rm{M}_{\odot}/h$ and $2160^3$ baryonic fluid elements with $m_{\rm{b}} = 5.7 \times 10^4 \rm{M}_{\odot}/h$ in a comoving box of $(51.7 ~\mathrm{cMpc})^3$. We select the 20 full snapshots from $z = 12$ to $z = 0$ along with the ``mini" snapshot at $z = 15$, and use the FoF host halo catalog and the subhalo catalog detected with the \textsc{SUBFIND} algorithm \citep{springel_populating_2001}. 
At each redshift, we require a host halo to contain at least a total mass of $100 m_{\rm{DM}}$ to be considered as resolved, and a subhalo to contain at least $50 m_{\rm{DM}}$, a similar threshold to that of \citetalias{jiang_statistics_2016}. We only consider halos that have at least one subhalo, excluding the central one.

At each redshift, we divide the host halos into 5  mass bins, which are spaced at equal intervals on the log scale from the minimum resolved halo mass to the maximum at a given snapshot. For each mass bin of the host halos, we collect all of their resolved subhalos, and plot the histogram of their $\psi = m/M$, the ratio between the total mass of a subhalo and its host halo. We divide the $\psi$ distribution by the number of host halos to get the average SHMF $\frac{dN}{d\lg\psi}$.

\begin{figure}
 \includegraphics[width=\columnwidth]{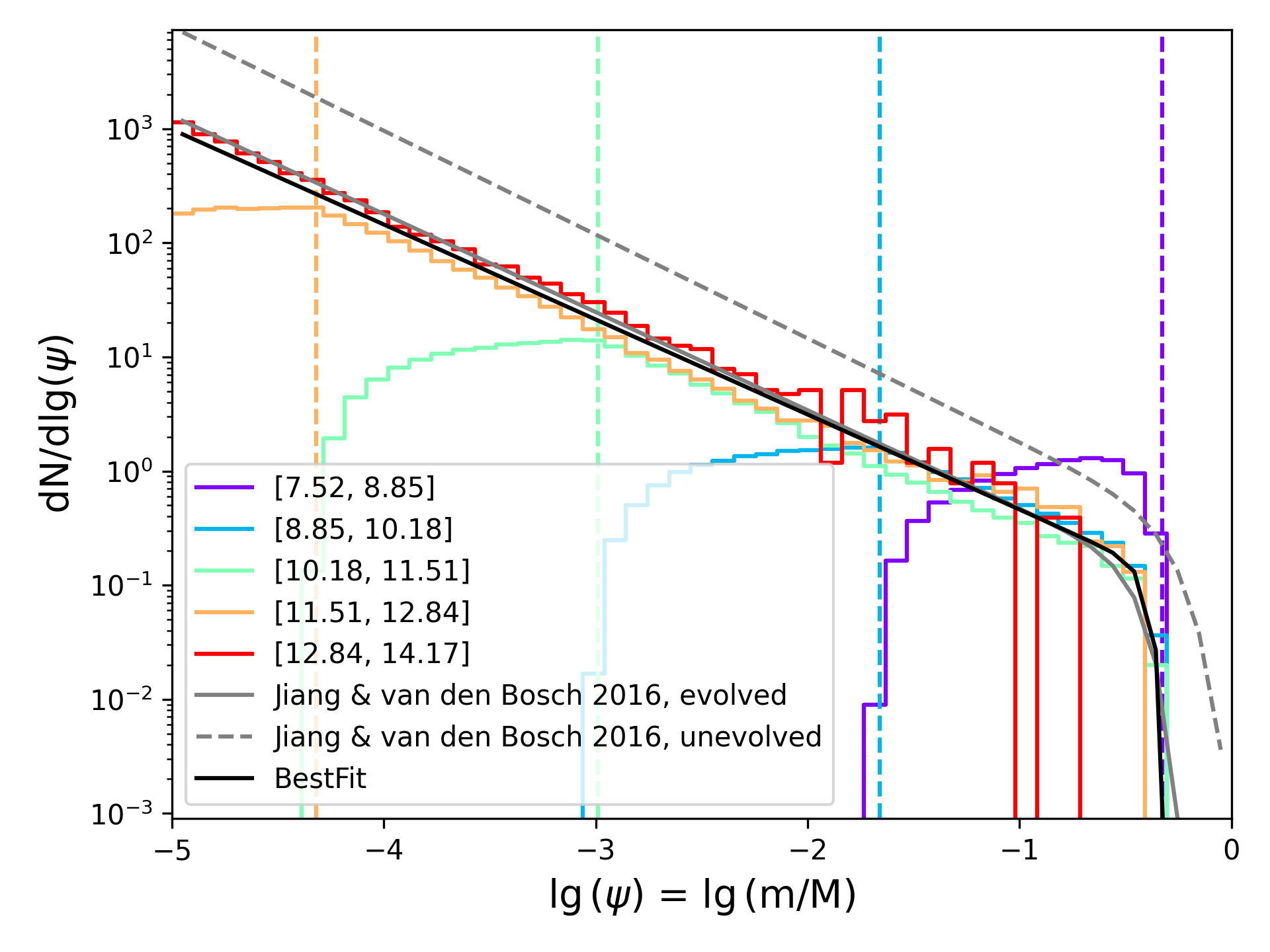}
 \caption{Subhalo mass function in TNG50-1 at $z = 0$. We group the host halos in 5 mass bins separated equally in logarithmic scale, covering the minimum to maximum mass, as shown by the different colors. The solid and dashed grey lines are the evolved and unevolved subhalo mass function, respectively, in \citetalias{jiang_statistics_2016, vandenbosch_statistics_2016}. The vertical dashed lines indicate the critical mass ratio where for the left boundary of the host mass range, its subhalo reaches the resolution limit of 50 dark matter particle mass. Using only the resolved results, i.e. the data to the right of the vertical lines, we show the best-fit subhalo mass function in the solid black line. Our best-fit subhalo mass function naturally becomes almost the same as the evolved subhalo mass function in \citetalias{jiang_statistics_2016, vandenbosch_statistics_2016}. }
 \label{fig:SHMF_snap_99}
\end{figure}

Figure \ref{fig:SHMF_snap_99} shows the average SHMF at $z = 0$. The different colors correspond to the 5 mass bins of the host halos, which are equally spaced in logarithmic mass, with ($\lg (M/ [h^{-1}\rm{M}_{\odot}]) =[7.52;8.85;10.18;11.51;12.84;14.17]$).
Ideally, it is better to group the host halos into more bins to represent the halo mass range more accurately, but that would lead to a smaller number of halos in each bin and a larger statistical error at higher $\psi$ due to the low abundance of subhalos. We therefore use 5 bins, and this results in 62008, 49109, 5877, 447, and 25 host halos, respectively, in each bin by increasing mass order. Except the fluctuations at higher $\psi$, the amplitude of the SHMF shows a weak dependence on host halo mass, i.e. a higher $A$ for the larger mass host halos. This trend is consistent with the findings in \citetalias{vandenbosch_statistics_2016} for $10^{12}-10^{15} \rm{M}_{\odot}/h$ halo mass. 
However, we mainly focus on the redshift evolution of the SHMF, so we ignore the differences in SHMF due to host halo mass or formation redshift, but use a single fitting for each redshift. The vertical lines in the plot correspond to the mass ratios where the subhalo mass reaches the resolution limit $50 m_{\rm{DM}}$ of a given host halo mass bin. The lowest mass bin, i.e., the purple line is excluded for fitting due to lack of resolution. By using the data points to the right of these vertical lines, where the subhalos can be fully resolved, we fit the SHMFs with Equation \ref{eq:SHMF_form}. At $z = 0$, the best-fit result is $A = 0.073, \alpha = 0.81, \beta = 223, \omega =6$, as shown by the black solid line in the figure. We also overplot the baseline evolved SHMF (grey solid) and the unevolved SHMF (grey dashed) from \citetalias{jiang_statistics_2016}, and our results agree quite well with their evolved SHMF. 
The differences in the best-fit $\omega$ and $\beta$ values are reflected by the minor difference at $\lg\psi \sim -0.5$, which may be attributed to the small number of our halo examples there, but do not have a substantial impact on the SHMF.

\begin{figure}
 \includegraphics[width=\columnwidth]{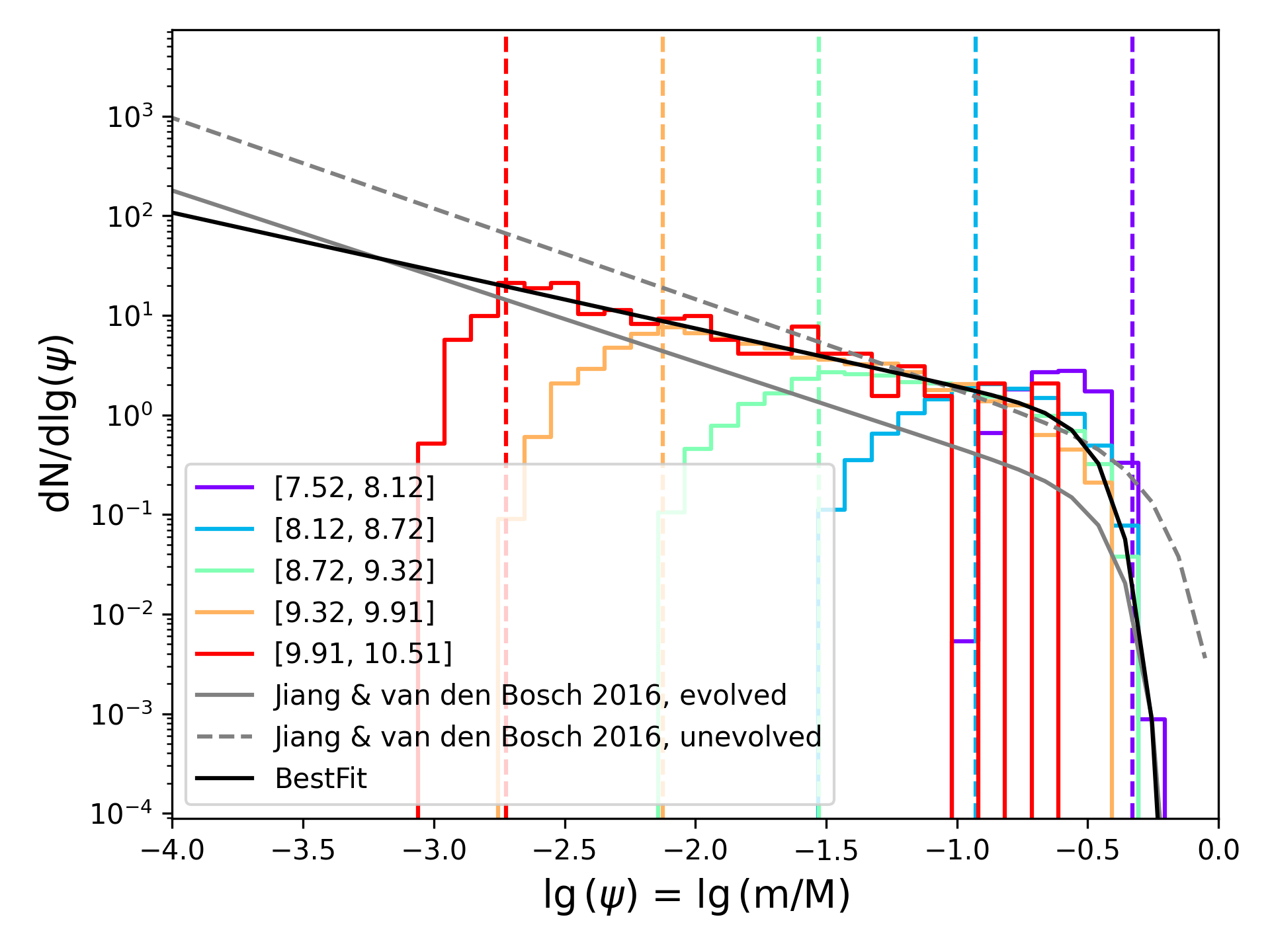}
 \caption{Same as Figure \ref{fig:SHMF_snap_99} but at $z = 12$. The slope of the SHMF is shallower  compared to $z = 0$ because at larger $\psi$, $dN/d\lg \psi$ is closer to the unevolved SHMF in \citetalias{jiang_statistics_2016}.}
 \label{fig:SHMF_snap_2}
\end{figure}

Using the same procedure, we further investigate the SHMF at higher redshifts, and it turns out that the SHMF gradually deviates from the $z = 0$ results as the redshift increases. As an example, we show the SHMF at the highest redshift of the full snapshots, $z = 12$, in Figure \ref{fig:SHMF_snap_2}. For larger $\psi$, the best-fit SHMF (solid black) is closer to the unevolved SHMF in \citetalias{jiang_statistics_2016}, and the slope is shallower than the $z=0$ result. This is mainly because the formation times of halos at $z = 12$ do not allow enough time for tidal mass stripping compared to halos at $z = 0$.

\begin{figure}
    \centering
    \includegraphics[width=\linewidth]{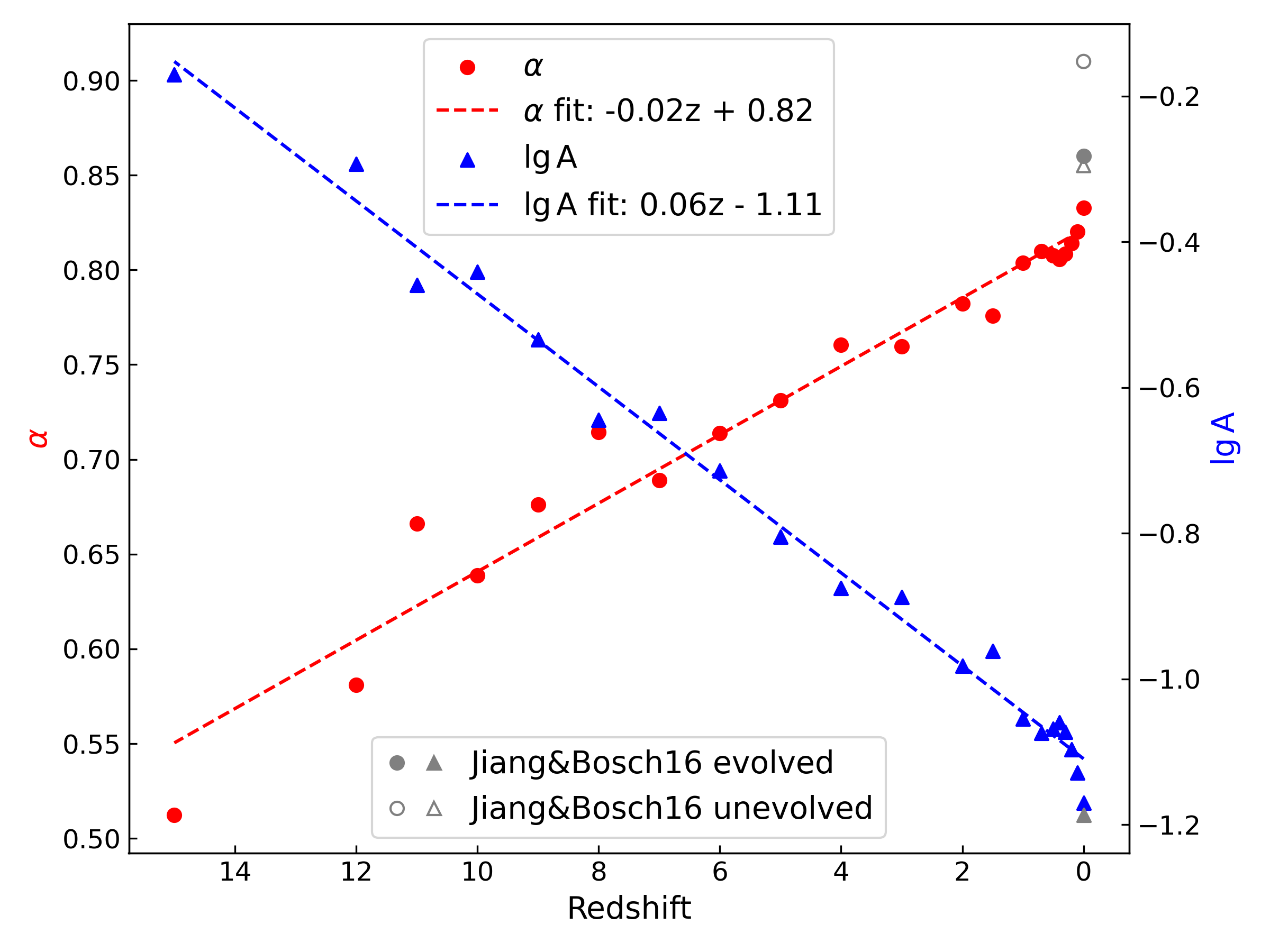}
    \caption{The evolution of the best-fit $\alpha$ (red dots) and $\lg A$ (blue triangles) as a function of redshift. From $z = 15$ to $z = 0$, the subhalos of higher mass ratios transform from being similar to the unevolved SHMF to being similar to the evolved SHMF, leading to a steeper slope, or a larger $\alpha$. In the fitting for SHMF, $\lg A$ is anti-correlated with $\alpha$. At $z = 0$, our results agree with the evolved SHMF in \citetalias{jiang_statistics_2016, vandenbosch_statistics_2016}, which are indicated by the filled grey circles and triangles. The dashed lines show the fit for the trend of the redshift evolution of the best-fit $\alpha$ and $\lg A$. }
    \label{fig:SHMF_redshift_evolution_alpha_lgA}
\end{figure}

\begin{figure}
    \centering
    \includegraphics[width=\linewidth]{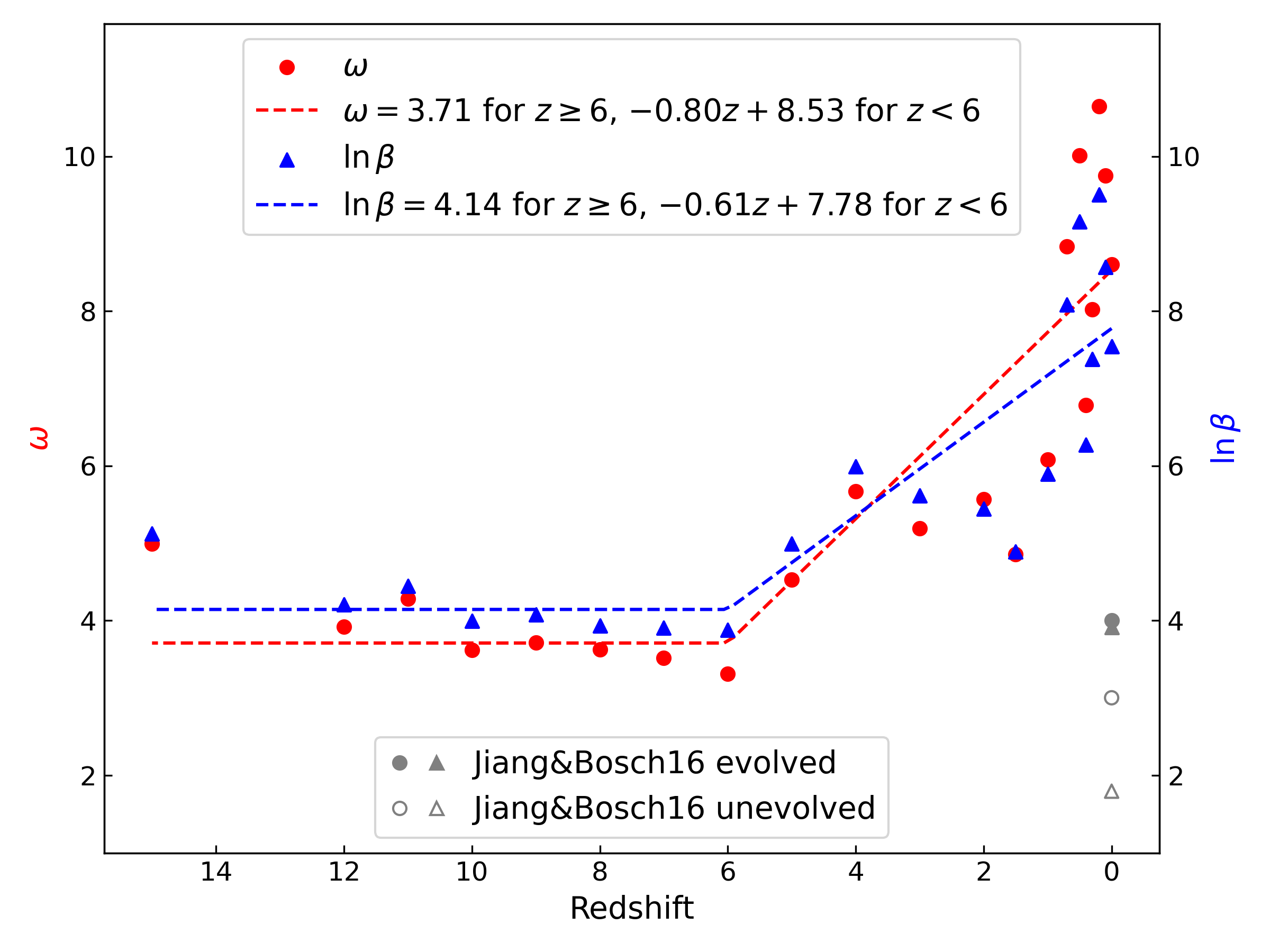}
    \caption{Same as Figure \ref{fig:SHMF_redshift_evolution_alpha_lgA} but for the evolution of best-fit $\omega$ and $\ln \beta$ in SHMF. We use a piecewise function to fit their evolution, which is a constant for $z > 6$ and a linearly increasing line below $z = 6$. }
    \label{fig:SHMF_redshift_evolution_omega_beta}
\end{figure}

In Figure \ref{fig:SHMF_redshift_evolution_alpha_lgA} and Figure \ref{fig:SHMF_redshift_evolution_omega_beta}, we plot the redshift evolution of the best-fit SHMF parameters in Equation \ref{eq:SHMF_form}, respectively. The slope $\alpha$ increases from about 0.5 to 0.81 as the redshift decreases from $z = 15$ to 0, while the amplitude $\lg A$ decreases. This can be fitted with the linear functions:

\begin{eqnarray}
\alpha  = -0.02 z + 0.82, \nonumber \\
\lg A = 0.06z -1.11 
\label{eq:SHMF_alpha_lgA_z} 
\end{eqnarray}

Note that $\alpha$ and $A$ are not independent parameters in the fitting. In fact, if we take the logarithm of Equation \ref{eq:SHMF_form}, we can observe the anti-correlation between $\alpha$ and $\lg A$. When the exponential decay in the SHMF remains fixed, if we increase the slope, the amplitude should naturally reduce for a smaller $\chi^2$. In Figure \ref{fig:SHMF_redshift_evolution_alpha_lgA}, we also plot the $\alpha$ and $\lg A$ corresponding to the evolved SHMF in \citetalias{jiang_statistics_2016} with the filled circle and triangle, which are very close to our best-fit results.

In Figure \ref{fig:SHMF_redshift_evolution_omega_beta}, we show the redshift evolution of $\omega$ and $\ln \beta$, the parameters responsible for the exponential decay at large $\psi$. These two correlated parameters remain roughly constant at high redshifts, but with an upward trend near $z = 0$. Therefore, we use a piecewise linear function to describe their redshift evolution. 
\begin{equation}
\begin{aligned}
\omega &=
\begin{cases}
3.71, & z \geq 6, \\[6pt]
-0.80z + 8.53, & z < 6
\end{cases}
\\
\ln\beta &=
\begin{cases}
4.14, & z \geq 6, \\[6pt]
-0.61z + 7.78, & z < 6
\end{cases}
\end{aligned}
\end{equation}

The fitted cutoff parameters, $\omega$ and $\ln\beta$, are less robust than $ \alpha$ and $A$, because they are mainly constrained by the rare high-mass-ratio subhalos near the exponential cutoff and are partially degenerate through the combination $\exp(-\beta\psi^\omega)$. The apparent transition at $z=6$ should therefore be regarded only as a convenient empirical parametrization, not as a physical critical redshift. Note that our fitted cutoff parameters are not fully consistent with those of \citetalias{jiang_statistics_2016} at $z=0$. However, the resulting SHMF itself remains similar over the resolved mass-ratio range, as shown by the comparison in Figure~\ref{fig:SHMF_snap_99}; noticeable differences occur only near the high-mass-ratio end, $\lg\psi \gtrsim -0.5$, where the number of resolved subhalos is small.


\subsection{The scatter of SHMF}
\label{section:SHMF_scatter}
By fitting the halo catalog in the TNG50 simulations, we obtain the average evolved SHMF across a wide range of redshifts in Section \ref{section:SHMF}. However, the evolved SHMF also has a considerable halo-to-halo variance, which contributes a lot to the variance in the DF heating rate (see Section~\ref{Section:massive_halo} and Section~\ref{Section:minihalo}). 

In semi-analytic models, the number of subhalos of a host halo $ N(\geq \psi)$ mainly exhibits a Poisson fluctuation, regardless of the details of the subhalo tidal evolution model (\citetalias{jiang_statistics_2017}; \citealt{hiroshima_semi-analytical_2022}). For host halos in different mass bins, the scatter in the overall SHMF is also contributed by the scatter of host mass as we can see from Section \ref{section:SHMF}, though in this paper we mainly focus on the redshift evolution.

\begin{figure}
 \includegraphics[width=\columnwidth]{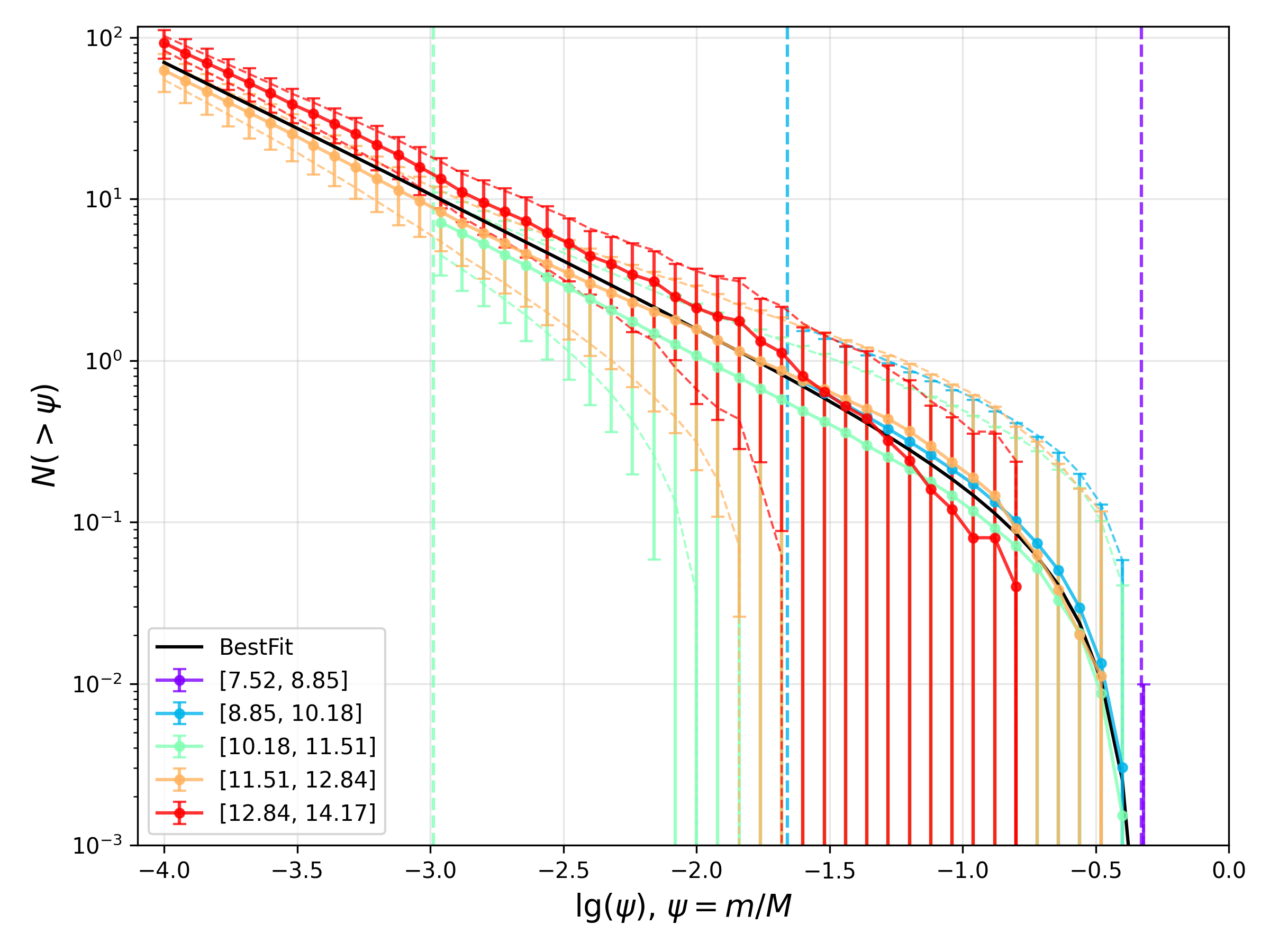}
 \caption{The scatter of the cumulative SHMF at $z = 0$. Same as Figure \ref{fig:SHMF_snap_99}, the rainbow colors represent different host halo mass bins, and the black solid line is overall best-fit SHMF. The vertical lines correspond to mass resolution. We show the scatter of the SHMF in the TNG data with the error bars. As a comparison, we use dashed lines to indicate the range of Poisson fluctuations. The scatter becomes super-Poissonian when $\psi \lesssim 10^{-2}$, especially for the lowest mass ratios.}
 \label{fig:SHMF_cumulative_snap_99}
\end{figure}

\begin{figure}
 \includegraphics[width=\columnwidth]{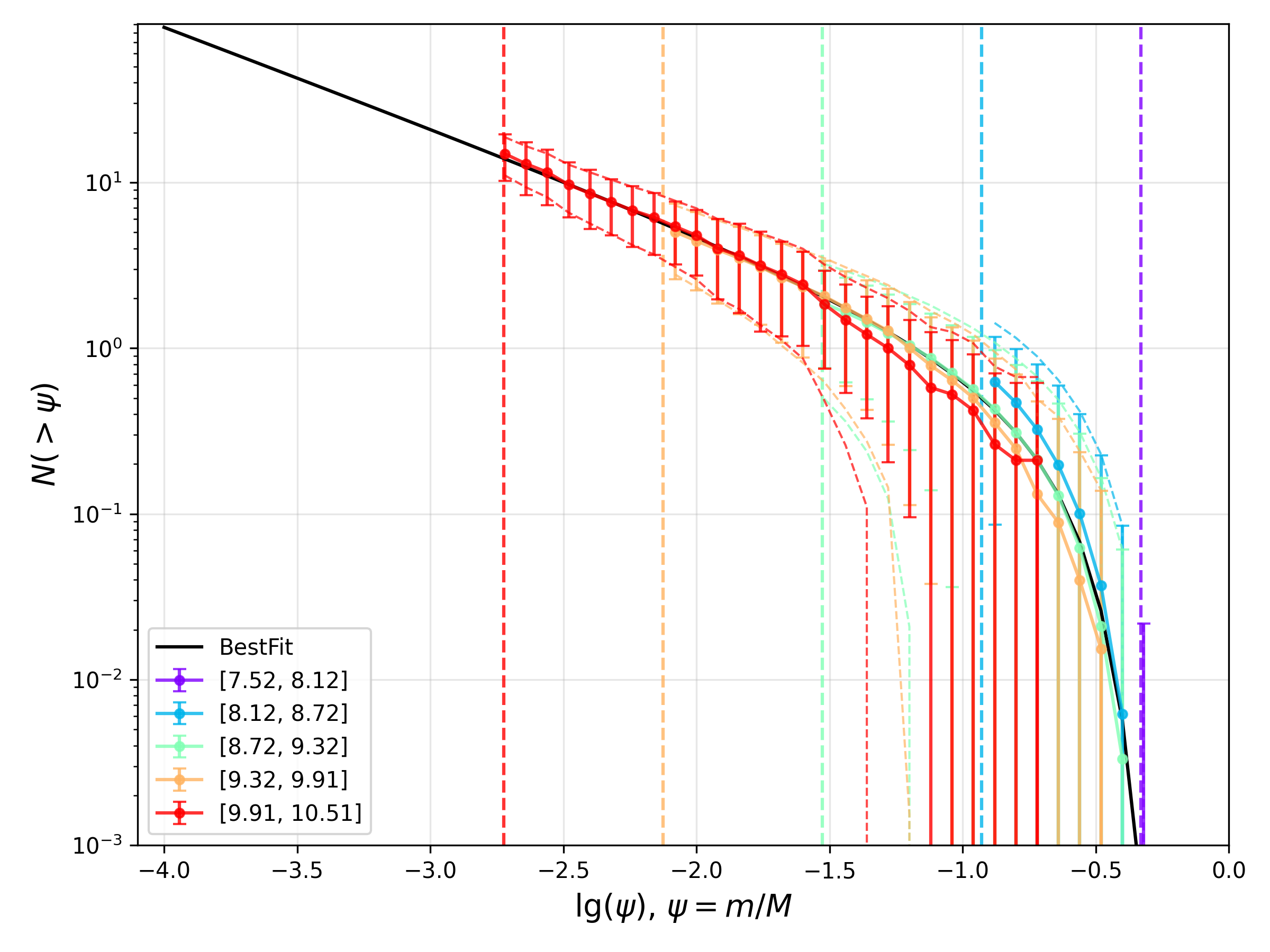}
 \caption{Same as Figure \ref{fig:SHMF_cumulative_snap_99} but at $z = 12$. The sub-Poissonian feature is more obvious at the higher mass ratio end $\psi > 10^{-1.5}$. }
 \label{fig:SHMF_cumulative_snap_2}
\end{figure}

In Figure~\ref{fig:SHMF_cumulative_snap_99} and Figure~\ref{fig:SHMF_cumulative_snap_2}, we present the cumulative SHMF with its scatter (shown by the error bars) in TNG50 for $z = 0$ and $z = 12$, respectively. The scatter is comparable to a Poissonian fluctuation as indicated by the dashed lines in the figures, though there are also non-Poissonian features. We refer to Appendix~\ref{appendix:non_Poissonian_SHMF} for a discussion on the non-Poissonian fluctuations. For the purpose of estimating the impact of scatter in the SHMF  on the dynamical-friction heating rate, we assume Poisson fluctuations for simplicity.

To investigate the variance of the DF heating rate caused by the scatter in the SHMF, we perform a Monte Carlo sampling of subhalo masses. Following \citet{hiroshima_semi-analytical_2022}, the SHMF with Poisson fluctuations can be generated using the \textit{inverse transform sampling} method, a standard technique for drawing random samples from any given cumulative distribution function (CDF).  

For a theoretical SHMF, we first compute the mean cumulative number of subhalos, $N(>\psi)$, with a cutoff at $\psi = 10^{-4}$. We then incorporate Poisson fluctuations by drawing a random number from a Poisson distribution with expectation value $N(>10^{-4})$; this gives the total number of subhalos, $N_{\mathrm{tot},P}$. Using the inverse transform method, we generate $N_{\mathrm{tot},P}$ random numbers uniformly distributed in $[0, 1]$ and map them back to subhalo mass ratios according to the cumulative SHMF. These mass ratios define one realization of the SHMF.  

Repeating this procedure 500 times—equivalent to generating 500 host halos—yields a set of SHMFs whose scatter is purely Poissonian.

\subsection{Corrections in the DF heating formula}
\label{section:DF_correction}

The Ostriker model of gaseous dynamical friction in Equation~\ref{eq:Ostriker_DF} is based on linear perturbation, so it's only accurate when the DF wake is in the linear or quasi-linear regime. In this section, we explore the impact of nonlinear corrections. We also use TNG50 data to explore the typical values of $\left\langle I/\mathcal{M}\right\rangle$ to use in Equation~\ref{eq:Kim_DFheating}.

\subsubsection{Nonlinearity of the perturbation}

For an extended perturber with an effective softening length $r_{\rm soft}$, the strength of its gravitational perturbation relative to the gas thermal pressure can be measured by the parameter $\mathcal{A} = Gm/(c_s^2 r_{\rm soft})$. Here $m$ and $r_{\rm soft}$ refer to the mass and effective size of the subhalo perturber, while $c_s$ is the sound speed of the ambient gas in the host halo.
The perturbation is linear when $\mathcal{A} \ll 1$, and nonlinear when $\mathcal{A} \gg 1$. Numerical simulations by \citet{kim_nonlinear_2009} and \citet{bernal_gravitational_2013} show that for subsonic cases, the drag force in nonlinear perturbations reaches an asymptotic value similar to the linear case. But in the supersonic case, the nonlinear drag force is smaller than the linear result. For $\mathcal{A} \gg 1$, gravitational focusing bends incoming streamlines toward the symmetry axis; these streams collide with the upstream flow, producing a detached shock and a more front-back symmetric near wake. This reduces the net DF force compared with the linear prediction \citep{kim_nonlinear_2009, bernal_gravitational_2013}.

\citet{Morton2025} estimate the $\mathcal{A}$ numbers of subhalo perturbers in the TNG300 halo catalog at $z = 0$. They find that the conditional probability distribution of $\mathcal{A}$ values given a subhalo mass has a scatter of about one order of magnitude, but in general $\mathcal{A}$ increases with subhalo mass from $10^9 \mathrm{M}_{\odot}$ to $10^{12} \mathrm{M}_{\odot}$, changing from linear to the nonlinear regime. In this work, we check the halo catalog in TNG50 for a wide range of redshifts. $\mathcal{A}$ can be estimated by $\mathcal{A} \approx \frac{G m_{\mathrm{sub}}}{2 r_{1/2}c_s^2(T_{\mathrm{vir}})}$, where $r_{1/2}$ is the subhalo half mass radius, and $T_{\mathrm{vir}}$ is the virial temperature of the host halo.

\begin{figure*}
 \includegraphics[width=\textwidth]{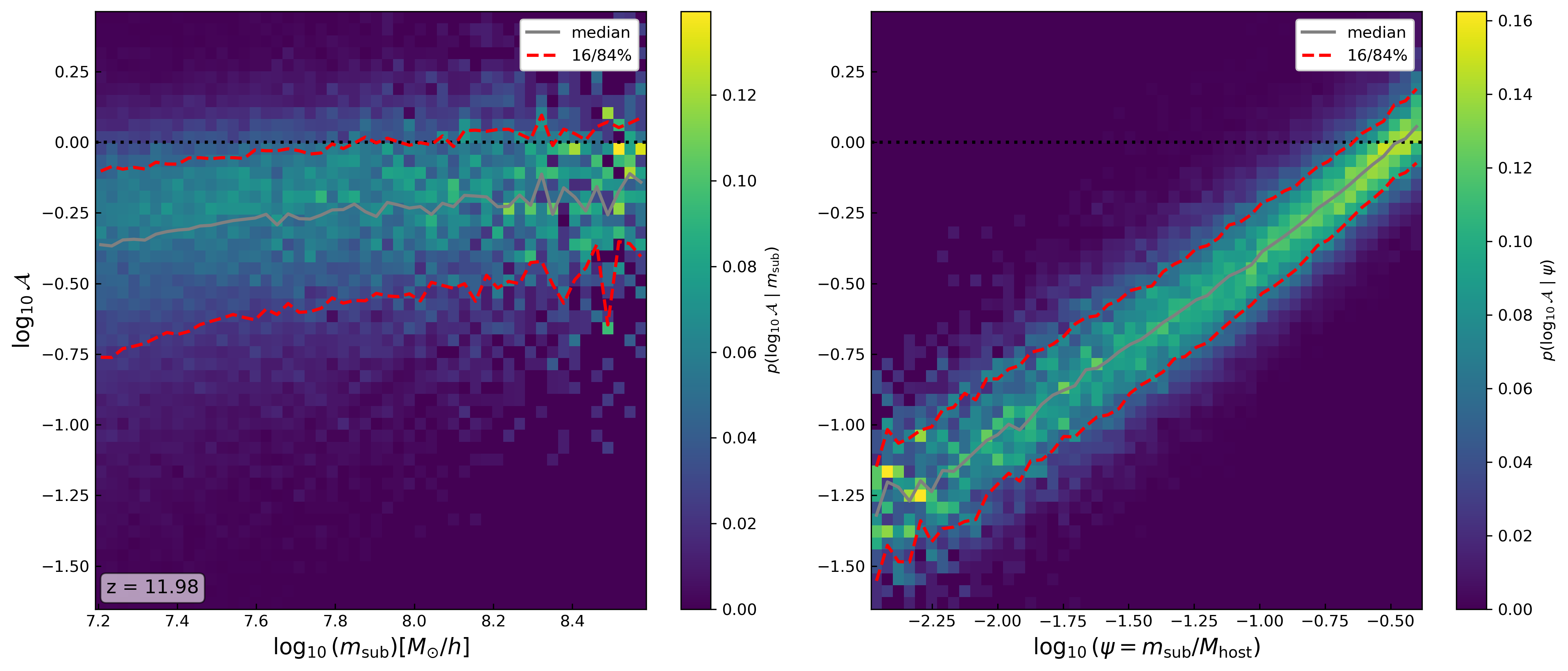}
 \caption{Conditional probability distribution of $\log \mathcal{A}$ of the subhalos at $z=12$. 
 Left panel: PDF of $\log \mathcal{A}$ for each given subhalo mass. Right panel: PDF of $\log \mathcal{A}$ for each given mass ratio. The red dashed lines bracket the 16-84 percentile, and the grey lines show the median $\log \mathcal{A}$. The black dotted lines indicate $\mathcal{A} = 1$, the boundary between linear and nonlinear perturbation. There is a strong correlation between $\mathcal{A}$ and the mass ratio $\psi$ between the subhalo and the host halo.}
 \label{fig:cond_both_Anumber_snap2}
\end{figure*}

The distribution of $\mathcal{A}$ shows similar trends across redshift in the TNG50 data. As an example, we show the conditional PDF of $\mathcal{A}$ at $z = 12$ in Figure~\ref{fig:cond_both_Anumber_snap2}. The left panel plots $\mathcal{A}$ as a function of subhalo mass, and indeed $\mathcal{A}$ gradually increases with subhalo mass. We further plot $\mathcal{A}$ as a function of mass ratio in the right panel. Again, $\mathcal{A}$ shows a strong correlation with the mass ratio $\psi$. Only in the high-mass-ratio tail, roughly $\psi \gtrsim 10^{-0.5}$, do nonlinear effects become important. This is not surprising because  $\mathcal{A}$ is defined as the strength of the subhalo's gravity relative to the sound speed of gas characterized by the host halo's mass.

\citet{kim_nonlinear_2009} find that the ratio of the nonlinear DF drag force to the linear force for supersonic models can be described by a single parameter $\eta = \frac{\mathcal{A}}{\mathcal{M}^2 -1}$. For $\eta > 2$, the drag force decreases as 
\begin{equation}
F_{\rm{DF}} = F_{\mathrm{lin}} \left(\frac{\eta}{2}\right)^{-0.45} 
\label{eq:nonlinear_correction}
\end{equation}
where $F_{\mathrm{lin}}$ is the linear force in Equation \ref{eq:Ostriker_DF}\footnote{ Note, that \citet{bernal_gravitational_2013} fit the time-dependent nonlinear force and find that it can be larger than Equation \ref{eq:nonlinear_correction}.}. In the following subsection, we apply the correction proposed by \citet{kim_nonlinear_2009} as a lower boundary to see how far the DF force can decrease from the linear case. 

\subsubsection{Mach number dependence in collisional DF}
\label{section:Mach_number_dependence}

\citet{kim_dynamical_2005} assumes that the motion of galaxies in a galaxy cluster follows a Maxwell-Boltzmann distribution. The correction factor $\langle I/\mathcal{M}\rangle$ varies with $m = \sigma_r/c_s$, where $\sigma_r$ is the one-dimensional velocity dispersion, and $\langle I/\mathcal{M}\rangle \approx 2$ for $m \approx 0.8 - 3$, the typical values for X-ray clusters. We further check the probability distribution of the Mach numbers of the subhalos in the TNG50 data for all host halo mass bins and redshifts. The Mach number is estimated as the ratio of the relative velocity between the subhalo and its host halo to the gas sound speed characterized by the host halo virial temperature, i.e., $|\mathbf{v_{\mathrm{m}}} - \mathbf{v_{\mathrm{M}}}|/c_s(T_{\mathrm{vir},\ M})$. 

\begin{figure}
  \centering
  \includegraphics[width=\columnwidth]{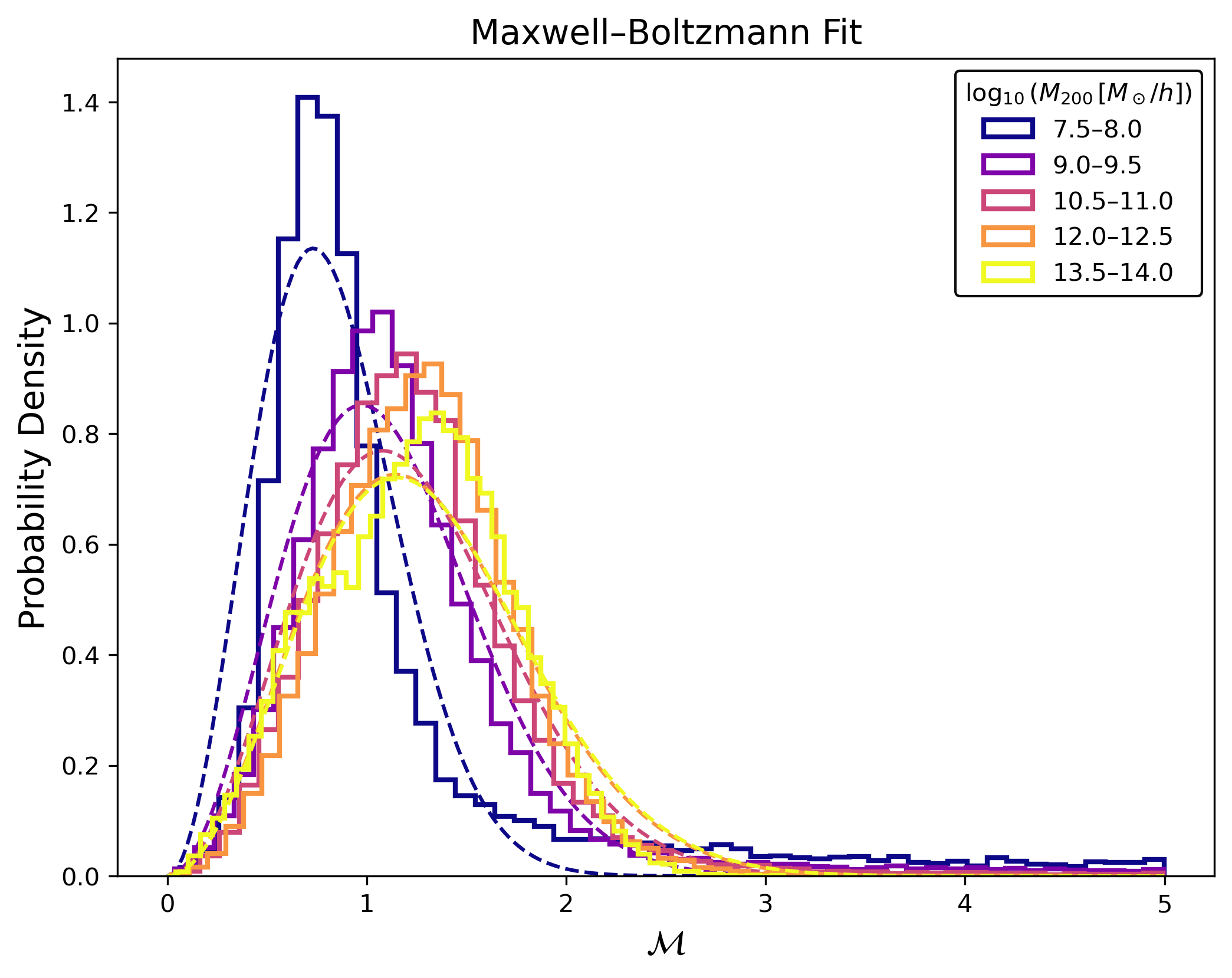}
  \caption{Mach number histogram of subhalos in TNG50-1 at $z=0$. The dashed lines show the fit results assuming a Maxwell-Boltzmann distribution. Different colors correspond to host halo mass bins from $10^{7.5} - 10^{14} M_{\odot}/h$. }
  \label{fig:mach_TNG_snap_99}
\end{figure}

Figure \ref{fig:mach_TNG_snap_99} shows the histogram of the Mach numbers of the subhalos in TNG50. The dashed lines are the Maxwell-Boltzmann fit to the distribution:

\begin{equation}
f(\mathcal{M})=\frac{4 \pi}{\left(2 \pi \sigma_{\mathcal{M}}^2\right)^{3 / 2}} \mathcal{M}^2 e^{-\mathcal{M}^2 /\left(2 \sigma_{\mathcal{M}}^2\right)}
\end{equation}

In general, the Mach number distribution exhibits a larger scatter for subhalos residing in more massive host halos. This trend likely arises from the larger velocity dispersion associated with the deeper gravitational potential wells of massive systems, as well as their more complex merger histories and environmental interactions. Overall, the Maxwell–Boltzmann distribution provides a reasonable approximation for fitting the data; however, it tends to underestimate the peak of the actual distribution and may either overestimate or underestimate the high- and low-velocity tails. Therefore, we also use a different fit, namely a Truncated Gaussian (see Appendix \ref{appendix:TG_Mach_fit}). We conclude that this fit does not change the values of the correction factor significantly.

\begin{figure}
  \centering
  \includegraphics[width=\columnwidth]{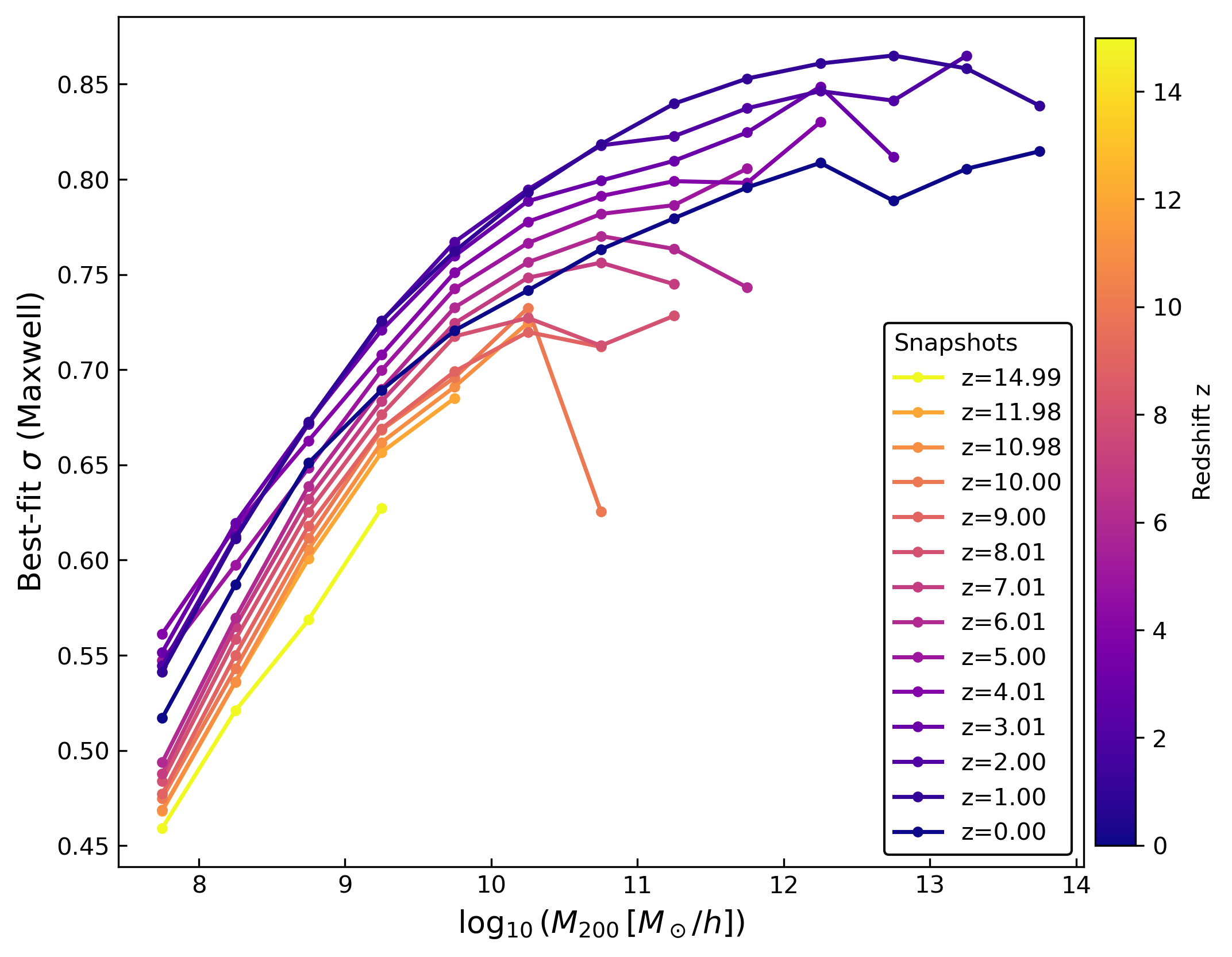}
  \caption{Maxwell-Boltzmann fit of subhalo Mach numbers for all host halo mass bins across the redshifts. $\sigma_{\mathcal{M}}$ shows an increasing trend with host halo mass. For a given host halo mass bin, $\sigma_{\mathcal{M}}$ slightly increases at low redshifts.}
  \label{fig:sigma_MB_all_hostmass_z}
\end{figure}

In Figure \ref{fig:sigma_MB_all_hostmass_z}, we plot the scatter of Mach numbers in the Maxwell-Boltzmann fits for all host halo mass bins from $z = 15$ to $z = 0$. $\sigma_{\mathcal{M}}$ increases from 0.45 to nearly 0.9 with increasing host halo mass. For a given host halo mass, $\sigma_{\mathcal{M}}$ also seems to increase at lower redshifts, but a rigorous study on this will require a more accurate measurement of Mach numbers, which is beyond the scope of this paper.



\begin{figure}
 \includegraphics[width=\columnwidth]{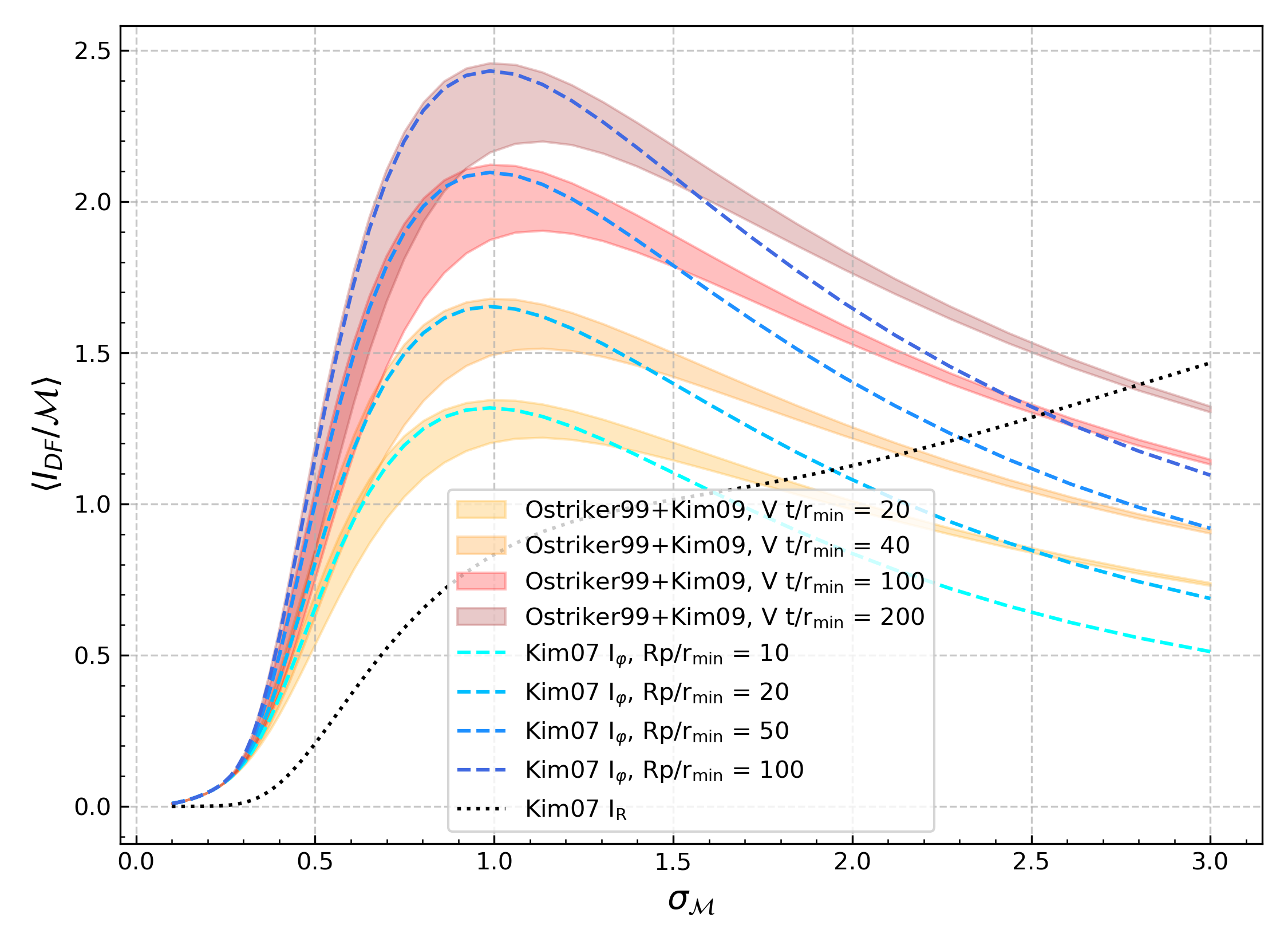}
 \caption{Average I/$\mathcal{M}$ as a function of $\sigma_{\mathcal{M}}$ assuming a Maxwell-Boltzmann distribution of Mach numbers. The orange to brown colors represent the results using the \citet{ostriker_dynamical_1999} correction for $V t/r_{\mathrm{min}}$ = 20 - 200. The shaded areas correspond to the variation of DF force due to nonlinear correction in \citet{kim_nonlinear_2009}. The cyan to blue dashed lines correspond to the azimuthal DF force in \citet{kim_dynamical_2007}'s model for a circular orbit, and the black dotted line is the radial force.}
 \label{fig:avg_I_over_M}
\end{figure}

As a result, the average correction factor is shown as a function of $\sigma_{\mathcal{M}}$ in Figure~\ref{fig:avg_I_over_M}. From bottom to top, the brown to orange colors correspond to $V t /r_{\min}$ = 20, 40, 100, and 200 in Equation \ref{eq:I_DF_Ostriker99}. Considering the nonlinear effects, the maximum $\log_{10} \mathcal{A} \approx 0.3$ across the redshifts, with Figure \ref{fig:cond_both_Anumber_snap2} giving an example. Therefore, we modify the linear correction to the nonlinear version with a maximum $\mathcal{A} = 2$, and the changes in $\left\langle I/\mathcal{M}\right\rangle$ are shown by the shaded areas, which have negligible effects. The cyan to blue dashed lines plot the correction factor from the \citet{kim_dynamical_2007} DF model, which assumes a circular orbit. The $Vt/r_{\min}$ parameters in the straight-line motion scenario can be loosely compared to $2R_p/r_{\min}$, where $R_p$ is the orbital radius in the circular-orbit model of \citet{kim_dynamical_2007}. This follows from their result that the azimuthal drag for a circular orbit is well approximated by the straight-line Ostriker formula when $V_p t \simeq 2R_p$. In our halo model, $Vt/r_{\min}$ is therefore interpreted as an effective outer-to-inner wake scale, analogous to the Coulomb logarithm, and the adopted range $Vt/ r_{\min}=20$--$200$ is used only to bracket this uncertainty.

We set $\left\langle I/\mathcal{M}\right\rangle = 1$ as a baseline model in the following calculations. This value is half of that used in \citet{kim_dynamical_2005}, primarily because we focus on a wider range of redshifts and halo masses, and are therefore fitted with a smaller $\sigma_{\mathcal{M}}$ than X-ray clusters.

In summary, the nonlinearity of the perturbation has a minor impact on the statistics of the DF heating rate, since it is only relevant for the rare subhalos with $\psi \gtrsim 10^{-0.5}$. The orbital parameters of the subhalos and the Mach number distribution exert a stronger influence, although the overall correction factor remains of order unity.

\subsection{The cosmic DF heating rates}

\label{section:total_DF_heating}
\begin{figure}
 \includegraphics[width=\columnwidth]{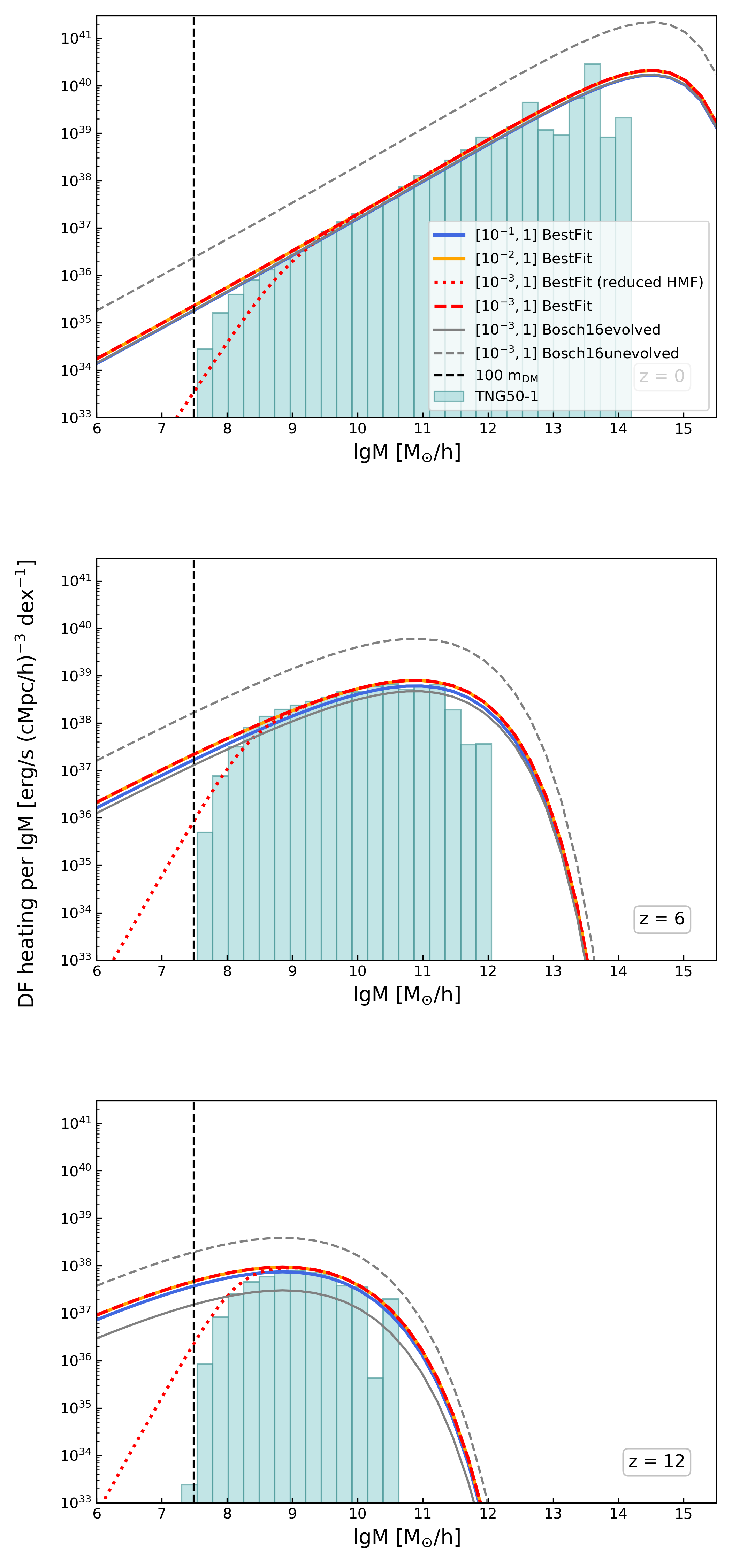}
 \caption{The HMF-weighted cosmic DF heating-rate density per $\log_{10}M_{\rm host}$ at $z = 0$ (upper panel), 6 (middle panel), and 12 (lower panel). The red, orange, and blue lines show the results using the best-fit  SHMF in Section~\ref{section:SHMF}. The red lines integrate the mass ratio range $\psi \in [10^{-3}, 1]$, while the orange and blue lines correspond to $\psi \in [10^{-2}, 1]$ and $[10^{-1}, 1]$, respectively. The dotted red line uses a reduced HMF including the selection effect near the mass resolution of TNG50. As a comparison, the results using the unevolved and evolved SHMF at $z = 0$ in \citetalias{jiang_statistics_2016} are given by the dashed and solid grey lines, respectively. The cyan histogram shows the DF heating rate in the TNG50 halo catalog, assuming $Vt/r_{\min} = 40$ for supersonic cases.}
 \label{fig:cosmic_DFheating_z0_6_12}
\end{figure}


  Based on the \textit{global DF heating model} in Section~\ref{section:ostriker}, we apply the SHMF obtained in Section~\ref{section:SHMF} and the corrections in Section~\ref{section:DF_correction} to get the SHMF-averaged DF heating rate of a single host halo, $\dot{E}_{\rm DF}(M_{\rm host},z)$. We then combine this single-host heating rate with the Sheth-Tormen \citep{Sheth1999} halo mass function (HMF)\footnote{We use the \textsc{COLOSSUS} python package \citep{Diemer2018}: \url{https://bdiemer.bitbucket.io/colossus/}.} to obtain the cosmic DF heating-rate density per logarithmic host-mass interval,
  \begin{equation}
  \frac{{\rm d}\dot{e}_{\rm DF}}{{\rm d}\log_{10} M_{\rm host}}
  =\frac{{\rm d}n}{{\rm d}\log_{10} M_{\rm host}}
  \dot{E}_{\rm DF}(M_{\rm host}, z),
\label{eq:cosmic_DFheating_perlogM}
\end{equation}
where ${\rm d}n/{\rm d}\log_{10}M_{\rm host}$ is the HMF. Figure~\ref{fig:cosmic_DFheating_z0_6_12} shows this quantity at $z = 0$ (upper panel), 6 (middle panel), and 12 (lower panel), respectively.
We also examine the impact of the subhalo mass ratio range. The red dashed line ($\psi \in [10^{-3},1]$) is identical to the orange line ($\psi \in [10^{-2},1]$), while the blue line ($\psi \in [10^{-1},1]$) is slightly lower than the others. This agrees with \citet{Dekel2007}, who found that DF heating is primarily driven by major mergers with $\psi \ge 0.05$. The cyan histogram uses the TNG50 halo catalog to calculate the total DF heating rate, and the vertical line corresponds to a mass resolution of 100 $m_{\rm{DM}}$. 
Our best-fit SHMF, together with the average correction factor, produces results similar to those obtained by directly collecting subhalos in TNG50. The DF heating rate in TNG50 at the massive end is absent due to the limited box size. It also deviates from the analytic model near the mass resolution at the lower mass end, because the selected HMF is just a fraction of the Sheth-Tormen HMF (see the dotted red line, which uses this reduced HMF). This is a resolution effect, because we require a subhalo to contain at least 50 particles to be considered as resolved, so even if the host halo reaches 100 particles, it's highly likely that it doesn't contain resolved subhalos, considering the mass ratio distribution.

At lower redshifts, the most massive halos $\sim 10^{14} M_{\odot}/h$ have the highest DF heating rate density, which is around $10^{39-40} \mathrm{erg\ s^{-1}} \mathrm{(cMpc/h)^{-3}} \mathrm{dex^{-1}}$. This is because the most massive halos experience the strongest DF heating. The peak shifts towards lower-mass halos at higher redshifts because such massive halos are much rarer at that time. Compared to AGN feedback, gaseous dynamical friction heating in the massive halos represents a secondary but physically distinct process \citep{Dekel2006, Khochfar2008}. We discuss the impact of DF in clusters in Section \ref{Section:massive_halo} and on the formation of Pop III stars and DCBH in high-z minihalos in  Section \ref{Section:minihalo}.

\section{DF heating versus cooling in low-z massive halos}

At lower redshifts, gravitational heating and DF heating have been shown to be significant in massive halos, providing an alternative mechanism for galaxy quenching beyond AGN or supernova feedback \citep{Birnboim2007, Dekel2007, Khochfar2008, Johansson2009, Birnboim2011}, especially in X-ray clusters of $10^{14-15} M_{\odot}$ \citep{ElZant2004, kim_dynamical_2005}. In this section, we present predictions for DF heating in the context of cooling within halos of $M_{\mathrm{halo}} = 10^{10-15} M_{\odot}$.

\label{Section:massive_halo}

\subsection{Halo cooling profile}
\label{section:halo_cooling_profile}

To integrate the cooling rates in a halo, we consider the gas profile in \citetalias{Dekel2007}, which assumes spherical symmetry and hydrostatic equilibrium. The virial radius $R_{\mathrm{vir}}$ is defined for an overdensity of $\Delta = 200$ relative to the critical density. The gas mass fraction is $f_{\rm g}$, and both the gas profile and the total mass profile take a generalized NFW form:

\begin{equation}
\rho(r)=\frac{\rho_{\mathrm{s}}}{x^\alpha(1+x)^{3-\alpha}}, \quad x \equiv \frac{r}{r_{\mathrm{s}}}
\label{eq:generalized_NFW_profile}
\end{equation}

where $\alpha$ is the inner slope factor; $\alpha = 0$ corresponds to a core profile, and $\alpha=1$ returns to the cusp NFW profile \citep{Navarro1997}. $r_s$ is the scale radius, and is related to the virial radius via the concentration parameter $R_{\mathrm{vir}} = c \ r_s$. The density $\rho_{\mathrm{s}}=\rho_{\mathrm{vir}} \frac{c^3}{3 A_\alpha(c)}$. For an $\alpha = 1$ cusp $A_1(x)=\ln (x+1)-\frac{x}{x+1}$, and $\alpha = 0$ core $A_0(x)=\ln (x+1)-\frac{x(3 x+2)}{2(x+1)^2}$. The halo concentration $c$ is often modeled as a function of halo mass and redshift \citep[e.g.,][]{Bullock2001}. We consider two commonly used $c(M,z)$ models \citep{Ludlow2016, Diemer2019} provided in \textsc{COLOSSUS} \citep{Diemer2018}, and the results don't change significantly. The hydrostatic equilibrium assumption results in a temperature profile that remains almost constant. Following \citetalias{Dekel2007}, we set $T(r) = T_{\mathrm{vir}}$ for simplicity.  

Therefore, the total cooling rate in a halo is 
\begin{equation}
\dot{E}_{\mathrm{cool}} = \int_0^{R_{200}} 4 \pi r^2 \Lambda(T, Z)n_{\mathrm{H}}^2(r) \ dr
\end{equation}
where $\Lambda(T, Z)$ is the volumetric cooling rate normalized by $n_{\mathrm{H}}^2$, and $Z$ is the metallicity. The hydrogen number density is $n_{\mathrm{H}} = \frac{0.76 \rho_{\rm g} (r)}{m_{\mathrm{p}}}$. Assuming $\Lambda(T, Z)$ is not time dependent and taking the density profile in Equation~\ref{eq:generalized_NFW_profile}, the total cooling rate can be written as
\begin{equation}
\dot{E}_{\mathrm{cool}} = \Lambda(T_{\mathrm{vir}}, Z) \bar{n_{\mathrm{H}}}^2 \left( \frac{4\pi}{3} R_{200}^3 \right) I_{\text{profile}} 
\end{equation}
where $\bar{n_{\mathrm{H}}}$ is the average number density corresponding to $f_{\rm g} \rho_{\mathrm{vir}}$. The profile factor

\begin{equation}
I_{\text{profile}} =
\begin{cases}
\int_0^c \frac{x^2}{(1+x)^6} dx \ \frac{c^3}{3 A_0^2(c)} = \frac{c^3(c^2+5c+10)}{30(c+1)^5} \frac{c^3}{3 A_0^2(c)} & \text{for }\alpha = 0,\\
\int_0^c\frac{1}{(1+x)^4}dx \frac{c^3}{3 A_1^2(c)} = \frac{1}{3}(1-\frac{1}{(c+1)^3}) \frac{c^3}{3 A_1^2(c)} & \text{for }\alpha = 1.
\end{cases}
\end{equation}

We use the \textsc{Grackle}\footnote{\url{https://grackle.readthedocs.io/en/latest/}} chemistry and cooling library \citep{Smith2017} to obtain $\Lambda(T,Z)$. The cooling table is provided by the \textsc{CloudyData\_UVB=HM2012.h5} file \citep{Haardt2012} and we evolve the 12 species chemistry network to reach species equilibrium for the given temperature and metallicity.

Simulations \citep[e.g.][]{Schaye2014, Rahmati2018, Muratov2017, Hafen2019, Hafen2020, Torrey2019, Proux2020} and observations \citep[e.g.][]{DeGrandi2001, Renzini2014, Molendi2016, Lehner2016, Prochaska2017} show that the gas metallicity in a halo spans a wide range from the metal-rich interstellar medium (ISM) of galaxies to the metal-poor inflows from the outer circumgalactic medium (CGM) or the intergalactic medium (IGM). Star-forming gas in galaxy centers, or in X-ray clusters with a deep potential well, typically reaches near-solar or super-solar metallicities, while the CGM can exhibit a wider range down to $10^{-3} Z_{\odot}$ due to low-metallicity accretion and incomplete recycling of galactic outflows. There is also a clear trend of increasing metallicities from $z = 5$ to $z = 0$ due to the accumulation of star formation and feedback. Therefore, we adopt $Z \in [10^{-3}, 1] Z_{\odot}$ in our calculations, with $Z_{\odot} = 0.01295$, the default value in the \textsc{Grackle} and \textsc{Cloudy} code.

\subsection{Comparison between heating and cooling}
\label{comparison_heating_cooling}

\begin{figure}
 \includegraphics[width=0.98\columnwidth]{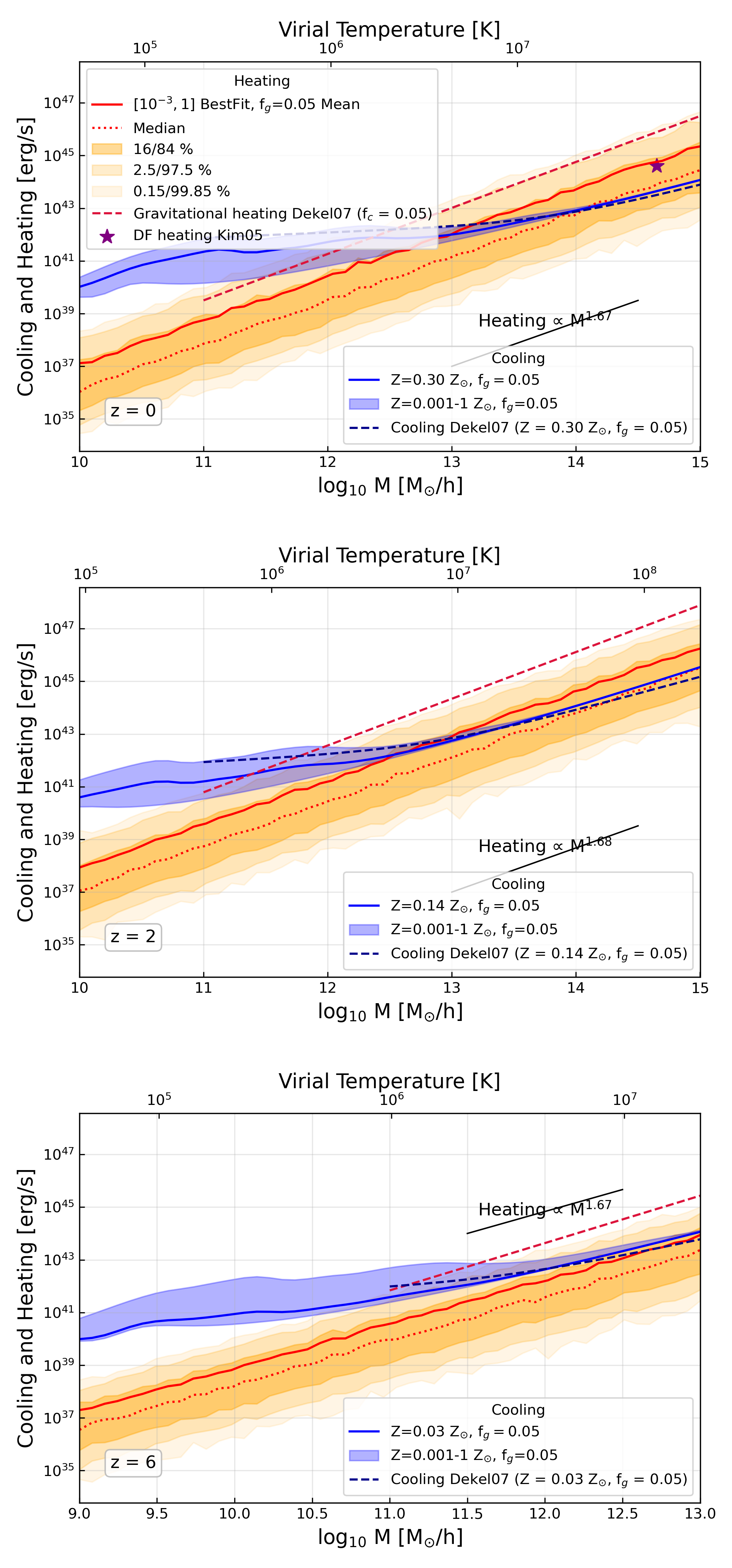}
 \caption{Dynamical friction heating and cooling rates in a single host halo at $z = 0$ (upper panel), 2 (middle panel), and 6 (lower panel). Note that we show a different x-axis range for $z= 6$ due to the lack of very massive halos. The solid red line is the DF heating rate using the best-fit SHMF in Section~\ref{section:SHMF}. The orange shaded regions show the ±1, 2, 3-sigma equivalent variation of heating rates due to the variation of SHMF (see Section~\ref{section:SHMF_scatter}), and the dotted red line corresponds to the median heating rate. The SHMF is integrated for $\psi \in [10^{-3}, 1]$, and we assume the gas fraction $f_{\rm g}$ is 0.05. The blue shaded region is the total cooling rate in the halo for metallicity $Z \in [0.001, 1] Z_{\odot}$. The solid blue lines correspond to the redshift-dependent metallicities adopted in \citetalias{Dekel2007}: $Z=0.30Z_{\odot}$, $0.14Z_{\odot}$, and $0.03Z_{\odot}$ at $z=0$, 2, and 6, respectively. As a comparison, we plot the gravitational heating rate (dashed dark-red line) and cooling rate (dashed dark-blue line) in \citetalias{Dekel2007}. The star shows the estimated DF heating rate in an X-ray cluster in \citet{kim_dynamical_2005}.}
 \label{fig:singlehost_z0_2_6}
\end{figure}


Figure \ref{fig:singlehost_z0_2_6}
shows the comparison between DF heating rate and the total cooling rate in a halo at $z =$ 0 (upper panel), 2 (middle panel), and 6 (lower panel), respectively. Here the DF heating curves are calculated using the \textit{global DF heating model} defined in Section~\ref{section:ostriker}, while the cooling rates are integrated over the gas density profile described in Section~\ref{section:halo_cooling_profile}. The effect of the radial DF heating profile is discussed separately in Section~\ref{Section:heating_profile}. We set the default gas fraction $f_{\rm{g}} =0.05$. The DF heating rate $\dot{E}_{\rm{DF}} \propto \rho_{\rm{g}} \propto f_{\rm{g}}$, and the cooling rate $\dot{E}_{\mathrm{cool}} \propto \bar{n}_{\mathrm{H}}^2 \propto f_{{\rm g}}^2$. The differences between the \citet{Ludlow2016} and \citet{Diemer2019} concentration models are negligible at $z = 0$. At higher redshifts $z \ge 2$, \citet{Diemer2019} result in a larger concentration for halos more massive than $10^{13} M_{\odot}$ and therefore a higher cooling rate by a factor of a few. For illustration purposes, we show only the results for the \citet{Ludlow2016} concentration. All the plots in this section use the core profile ($\alpha = 0$). The cooling rates using the NFW profile ($\alpha = 1$) are shown in Appendix~\ref{appendix:NFW_cooling_profile}.

The primary source of scatter in the DF heating rates is the intrinsic scatter in the SHMF. Using the Monte Carlo sampling of the SHMF described in Section~\ref{section:SHMF_scatter}, we generate 500 realizations that incorporate Poissonian fluctuations. The orange shaded regions in Figure~\ref{fig:singlehost_z0_2_6} illustrate the resulting variation in the heating rates. These span roughly two orders of magnitude for the 16–84\% (1$\sigma$ equivalent) percentile range, and up to four orders of magnitude for the 0.15–99.85\% (3$\sigma$) percentile range. The mean and median heating rates are shown by the solid and dotted red lines, respectively. The mean heating, derived from our best-fit SHMF, agrees well with the estimation of the DF heating rate in a $6.6\times 10^{14} M_{\odot}$ cluster in \citet{kim_dynamical_2005}. We also plot the gravitational heating rate in \citetalias{Dekel2007} (dashed dark-red line), which is derived from the approximate accretion rate of halos near $10^{12-13} M_{\odot}$. \citetalias{Dekel2007} estimates that the fraction of the total accreting mass in the form of halos with $\psi \ge 0.05$ is $f_{\rm{DF}} = 0.5$. The maximum DF heating is the gravitational energy from such accreted halos, and thus it can contribute a significant fraction, up to $1/3$ of the total gravitational heating. Our results, however, show that the DF heating is typically below the total gravitational heating: the mean is about 1--2 orders of magnitude smaller, while the median can be even lower because the mean is weighted toward halos with richer subhalo populations. This is most likely because they ignore the tidal stripping of subhalos, so the DF heating based on the accreted mass is treated as an upper limit, while we fit the evolved SHMF (see Figure~\ref{fig:SHMF_snap_99} for the comparison between the evolved and unevolved SHMF). Moreover, in \citetalias{Dekel2007}'s estimation of the DF heating of the subhalos, part of their gravitational energy actually `heats' the dark matter instead of being deposited to the ambient gas.

The blue shaded region shows the cooling rates with $Z\in[10^{-3},1] Z_{\odot}$. The metal lines mainly affect halos with $T_{\mathrm{vir}} \sim 10^{5-6}\,\mathrm{K}$. The cooling rates obtained by \textsc{Grackle} are broadly consistent with the cooling rate in \citetalias{Dekel2007}, which uses a redshift-dependent metallicity \citep{Dekel2006}. Nonetheless, their fitting formula seems to overestimate $\Lambda(T)$ by a factor of a few at lower temperatures.

At $z = 0$, DF heating becomes important for halos $\gtrsim 10^{13} M_{\odot}$, where the mean DF heating is more efficient than cooling. This is consistent with the threshold of gravitational heating quenching in previous papers \citep{Khochfar2008, Birnboim2011}. However, due to the large scatter of DF heating, in halos with fewer accreted subhalos, DF heating is not able to quench the central galaxy, or at least the gravitational energy from subhalos alone cannot. At higher redshifts, the heating and cooling rates for a fixed halo mass tend to increase compared to the $z = 0$ case, as seen from the comparison of the same-styled lines in different panels of Figure~\ref{fig:singlehost_z0_2_6}, which is consistent with the trend in \citetalias{Dekel2007}.
The intersection of the solid red and blue line still remains about $10^{13} M_{\odot}$. At $z \ge 4$, DF heating is generally a negligible effect. However, in the rare high-mass dark matter halos ($10^{12-13} M_{\odot}$) hosting luminous quasars, DF heating may still be important, and the total gravitational heating can dominate over cooling. 

\section{Impact of DF heating in high-z minihalos}
\label{Section:minihalo}

\subsection{Molecular hydrogen cooling versus dynamical friction heating}
\label{section:molecular_cooling}

Section \ref{Section:massive_halo} shows that DF heating is generally negligible for halos below $10^{12} M_{\odot}$ compared to the cooling rate. However, atomic cooling sharply decreases at $T \sim 10^4$ K, and it is therefore interesting to compare the DF heating with the cooling in minihalos before the epoch of reionization. We assume no metal enrichment in the minihalos, and H$_2$ is the dominant coolant. For convenience, we show the comparison at $z = 15$, the typical redshift for Pop III star formation. The DF heating rate is calculated using the same \textit{global DF heating model} as in Section~\ref{Section:massive_halo}.

\begin{figure*}
\includegraphics[width=\textwidth]{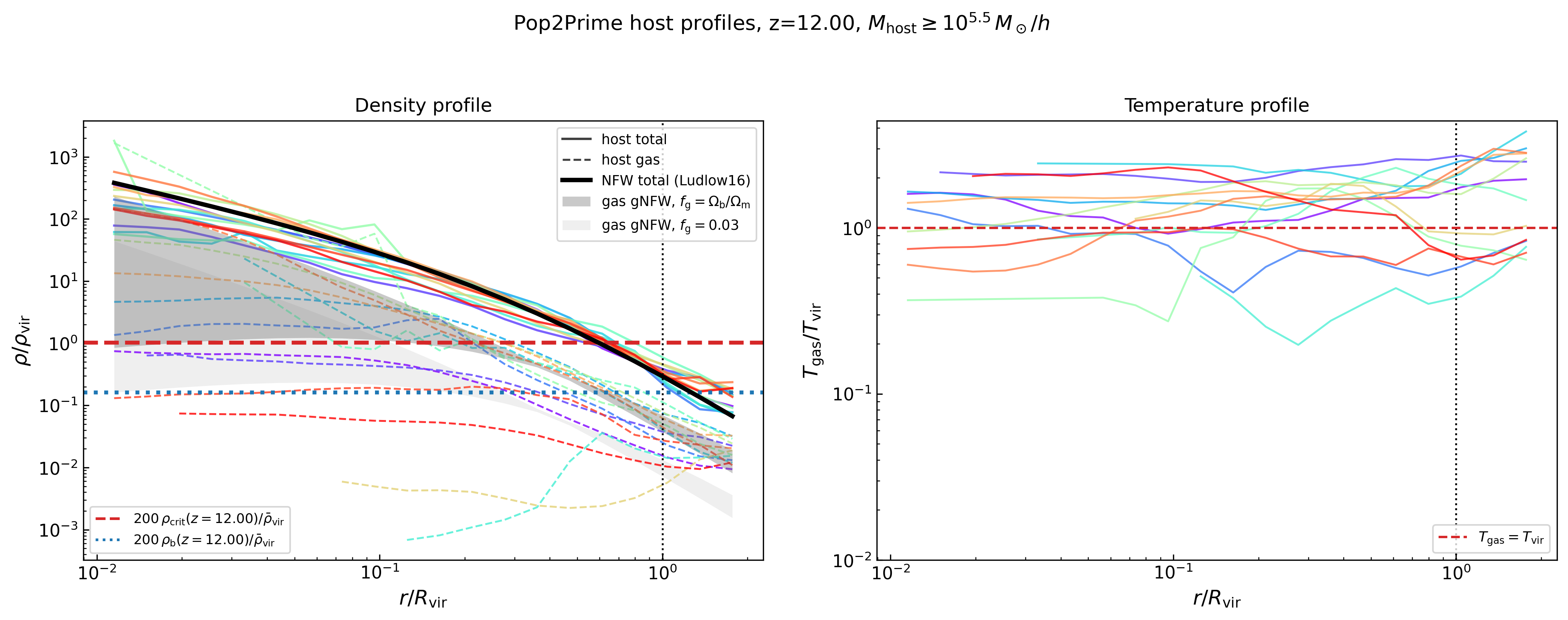}

\caption{Radial density and temperature profiles of Pop2Prime host halos at $z = 12$, selected with $M_{\rm host} > 10^{5.5}\,M_\odot/h$. Left: For each halo, the total density profile is shown as a solid colored curve and the gas density profile as a dashed curve, with colors distinguishing individual host halos. The thick black solid line shows the analytic NFW profile evaluated at the median host mass of the selected sample. The shaded bands indicate generalized NFW gas-density models spanning $-0.5 \leq \alpha \leq 1.5$: the darker band assumes the cosmic baryon fraction $f_{\rm g} = \Omega_{\rm b}/\Omega_{\rm m}$, while the lighter band shows a reduced gas fraction $f_{\rm g} = 0.03$. The horizontal red and blue lines mark $200\,\rho_{\rm crit}/\bar{\rho}_{\rm vir}$ and $200\,\rho_{\rm b}/\bar{\rho}_{\rm vir}$, respectively, where $\bar{\rho}_{\rm vir}$ is the mean virial density of the selected host sample. Right: mass-weighted gas temperature profiles normalized by the virial temperature of each host. In both panels, radial bins with shell gas mass below $1\,M_{\odot}$ are considered as unresolved, and we omit the profiles to the left of the unresolved regions.}

\label{fig:Pop2Prime_density_temperature_profile_z12}

\end{figure*}

For the high-redshift minihalos, we use the same parameterized gas-density profile as in Section~\ref{section:halo_cooling_profile} to compute the cooling profile correction. To examine the parameterized profiles, we extract the density and temperature profiles from the \textsc{Pop2Prime} data (see also Appendix~\ref{appendix:SHMF_Pop2Prime} for more details of the simulation). Figure~\ref{fig:Pop2Prime_density_temperature_profile_z12} shows the radial profiles of all the host halos with $M_{\rm host} > 10^{5.5} \rm{M}_{\odot}/h$ at $z = 12$, with each color representing an individual host halo. The total density profiles, including dark matter, generally agree with the analytic NFW total-density profile (the thick black solid line). The gas density profiles of the halos, i.e., the dashed lines, show a greater variety. The gas density can be either steeper or flatter, so in the shaded regions, we adopt $-0.5 \le \alpha \le 1.5$ for the generalized NFW profile, the same range as the parameter space in \citet{Dekel2007}. We also note that the gas fraction can be much lower than the cosmic baryon fraction. Thus, in the dark shaded region, we set $f_{\rm g} = \Omega_{\rm b}/\Omega_{\rm m}$, while in the light shaded region we set it to $0.03$. The parameter range roughly brackets the simulated halo profiles, except for a few halos with very little gas. In the right panel, we show the ratio between the temperature profiles and $T_{\rm vir}$ for each halo. In the resolved gas shells, the temperature profiles do not deviate from $T_{\rm vir}$ by more than one order of magnitude. Therefore, it is reasonable to use $T_{\rm vir}$ as a zeroth-order estimate.

\begin{figure}
 \includegraphics[width=\columnwidth]{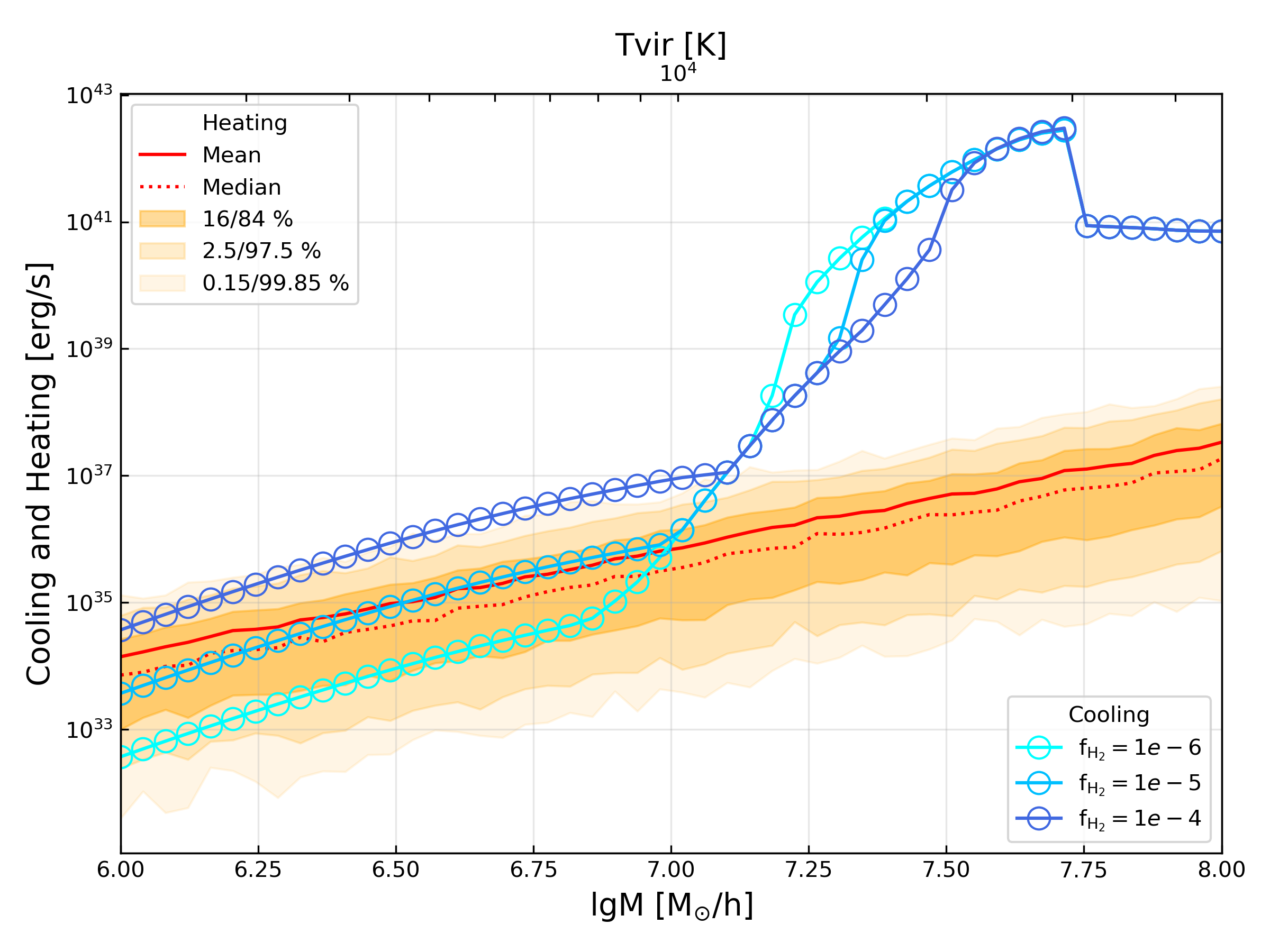}
 \caption{Comparison between the cooling rates and DF heating rates for halos with $M_{\textrm{halo}}$ between $10^6$ and $10^8 M_{\odot}/h$ at $z = 15$. The baseline cooling rates are calculated in \textsc{Grackle} at the mean halo gas density and virial temperature, and are then multiplied by the gas-density profile correction described in Section~\ref{section:halo_cooling_profile}. In the molecular-cooling regime, the H$_2$ fraction is kept fixed at the labelled values $10^{-4}$, $10^{-5}$, and $10^{-6}$.}
 \label{fig:singlehost_minihalo}
\end{figure}

Figure~\ref{fig:singlehost_minihalo} shows the comparison between the cooling rate and DF heating rates in $10^{6}-10^8 {\rm M}_{\odot}/h$ halos.
The DF heating rate integrates the SHMF for minihalos with $\psi \in [10^{-3},1]$. At this high redshift ($z = 15$), we set the default gas fraction to $f_{{\rm g}} = \Omega_{\rm {b}}/\Omega_{\rm {m}}$. As the figures in the last section show, we sample the SHMF from 500 realizations with a Poisson distribution. The solid and dotted red lines correspond to the mean and median heating rates, and the orange-shaded regions mark the 1, 2, and 3-$\sigma$ equivalent variations. The blue to cyan circles represent the cooling rates for the given halo mass. We use \textsc{Grackle} at the mean halo gas density $f_{\rm g} \rho_{\mathrm{vir}}$ and virial temperature $T_{\mathrm{vir}}$ to compute the baseline cooling rate, and then multiply it by the gas-density profile correction described in Section~\ref{section:halo_cooling_profile}. The different colors correspond to an initial H$_2$ fraction of $10^{-4}, 10^{-5}, 10^{-6}$. In the atomic cooling regime, we evolve the gas to species equilibrium to obtain the correct atomic cooling rates. While in the molecular cooling regime, we keep the H$_2$ fractions fixed at their initial values to show the cooling rates across the H$_2$ fractions. This is because, at the fixed density of $f_{\rm g} \rho_{\rm vir}$ prior to the onset of the gas collapse, the chemical timescale for reaching species equilibrium is much longer than the characteristic Pop III star formation timescale ($\sim 10$–$100,\mathrm{Myr}$). In minihalos, whether the DF heating dominates the cooling rate largely depends on the H$_2$ fraction and the variation in the SHMF.
For $f_{\mathrm{H_2}} = 10^{-5}$, the mean DF heating rate is in approximate balance with the cooling rate. Since the typical cosmological $\mathrm{H}_2$ fraction is below $10^{-5}$ \citep{Galli1998}, this indicates that DF heating by itself can suppress and regulate the formation of Pop III stars. For larger H$_2$ fractions minihalos will go through a phase during which they will be dominated by DF heating, before cooling will start dominating again. Whether Pop III stars form in such halos depends sensitively on the number of subhalos, the formation history and other environmental factors like the LW radiation background (see Section~\ref{section:PopIII_DCBH_formation} for a discussion on this). 
Above $10^4$ K, atomic cooling rates are several orders of magnitude above the DF heating rates. Therefore, in the presence of strong DF heating, the gas can still be cooled to $\sim$ 7000 K.

\subsection{The impact on Pop III star and DCBH formation}
\label{section:PopIII_DCBH_formation}

\begin{table*}
\centering
\caption{List of simulation data used in $f_{\mathrm{H}_2}-T$ diagrams.}
\label{tab:halo_list}
\begin{tabular}{lcccc}
\toprule
& $T_{\mathrm{rad}}$ (K) & \makecell{selected halos \\(Pop III star)}  & \makecell{selected halos \\(DCBH candidate)} & ref \\
\midrule
Shang2010  & $10^4$ & A & A & \citet{shang_supermassive_2010} \\
Shang2010  & $10^5$ & A & A & \citet{shang_supermassive_2010} \\
Latif2014  & $10^4$ & A, B, C, D, E & A, B, C, D, E & \citet{latif_uv_2014} \\
Latif2015  & $2 \times 10^4$ & A, B, C  & A, B, C & \citet{latif_how_2015}  \\
Latif2019  & $2 \times 10^4$ & 1, 6 & 6 & \citet{latif_uv_2019} \\
Latif2021  & $10^5$ & 1, 2, 3, 4, 5, 6 & & \citet{latif_radiation_2021} \\
Wise2019  & stellar population SED &  & MMH, LWH & \citet{wise_formation_2019} \\
Correa-Magnus2024 & stellar population SED & 1, 2, 4, 7/8, 9, 10/11 & & \citet{CorreaMagnus2024} \\

\bottomrule
\end{tabular}
\end{table*}

According to the analytic model of \citet{tegmark_how_1997}, the H$_2$ fraction in a minihalo scales as $f_{\mathrm{H}_2} \propto T^{1.52}$. Simulations in \citet{yoshida_simulations_2003} confirm that the mass-weighted average H$_2$ fraction of the minihalos scatter around this scaling relation in the absence of LW radiation. Only gas in minihalos with $f_{\mathrm{H}_2}$ above a critical value is able to collapse and form stars. 
The critical H$_2$ fraction ($f_{\mathrm{crit}}(T)$) is obtained by equating the age of the Universe at the redshift $t_{\mathrm{H}} (z)$ and the cooling timescale 
\begin{equation}
\label{eq:cooling_timescale}
t_{\mathrm{cool}} = \frac{\rho \epsilon}{(\rho/m_{\mathrm{p}}) ^2\Lambda(\rho, \ T,\ f_{\mathrm{H}_2})} 
\end{equation}
where $\epsilon = \frac{k_{\mathrm{B}} T}{(\gamma-1)\mu m_{\mathrm{p}}}$ is the specific internal energy, and $\gamma = 5/3$ is the adiabatic index. In the $f_{\mathrm{H}_2} - T$ plane, the intersection between the H$_2$ formation scaling relation and the critical H$_2$ line marks the boundary between minihalos that can or cannot form Pop III stars. 

\begin{figure}
 \includegraphics[width=\columnwidth]{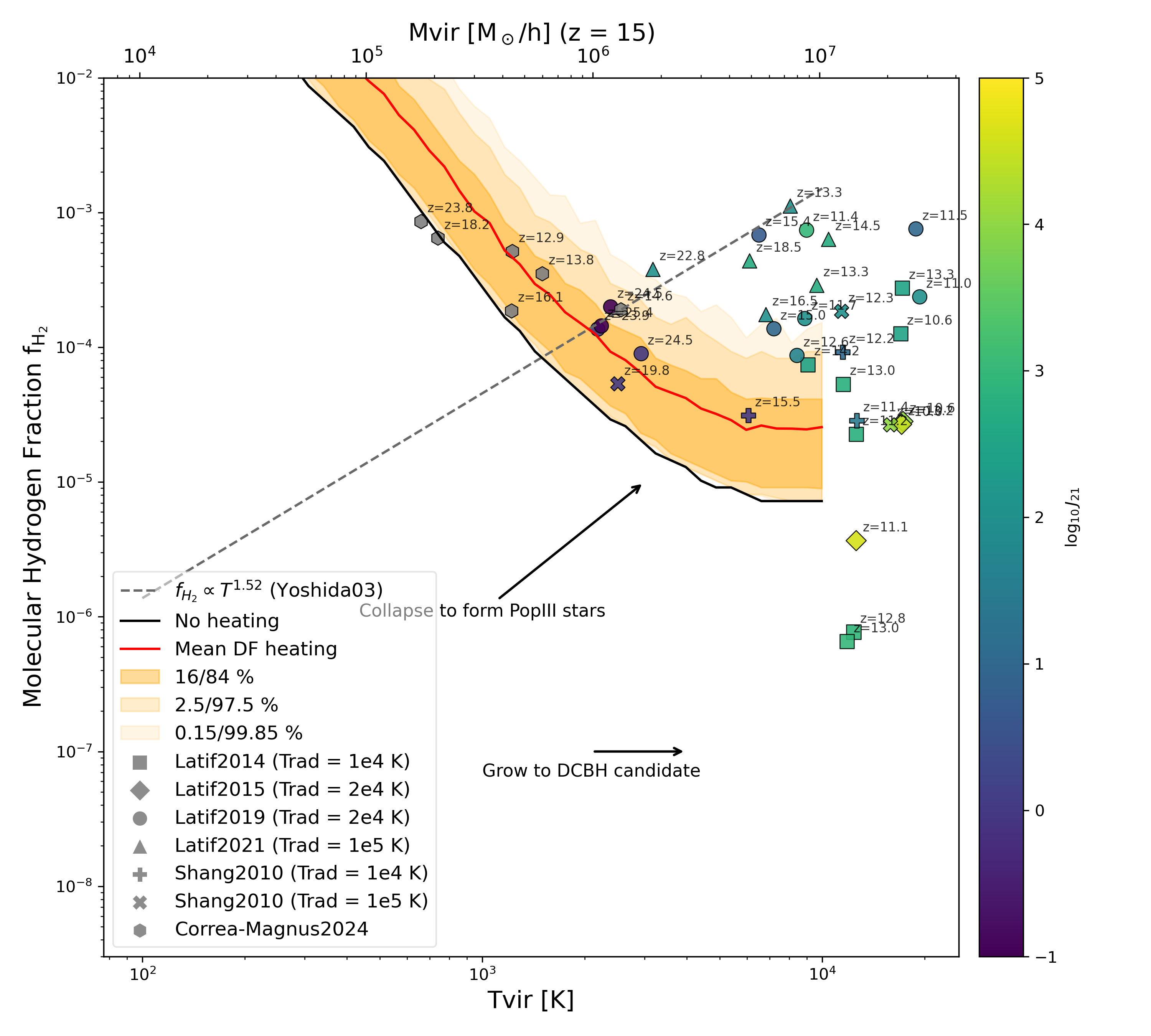}
 \caption{The critical H$_2$ fraction for star formation as a function of T$_{\mathrm{vir}}$ at $z = 15$. The solid black line is the baseline critical fraction without DF heating. The shaded regions and the solid red line correspond to the critical H$_2$ fraction when DF heating is included. Considering the variation of SHMF, the shaded regions and the solid red line show the 1, 2, 3-$\sigma$ variation of the critical $f_{\mathrm{H}_2}$ and its mean value, respectively. As a comparison, the dashed grey line shows the mass-weighted average $f_{\mathrm{H}_2}$ formed in halos in \citet{yoshida_simulations_2003} at $z = 17$. The markers are the $f_{\mathrm{H}_2}$ for halos that can form Pop III stars in previous literature listed in Table~\ref{tab:halo_list}. The marker shapes represent different papers or background UV radiation spectra types. The LW intensity $J_{21}$ is color-coded as shown in the colorbar.} 
 
\label{fig:fH2_vs_T_SHMF_sampling_z15_Mgas1e5_PopIII}
\end{figure}

\begin{figure}
 \includegraphics[width=\columnwidth]{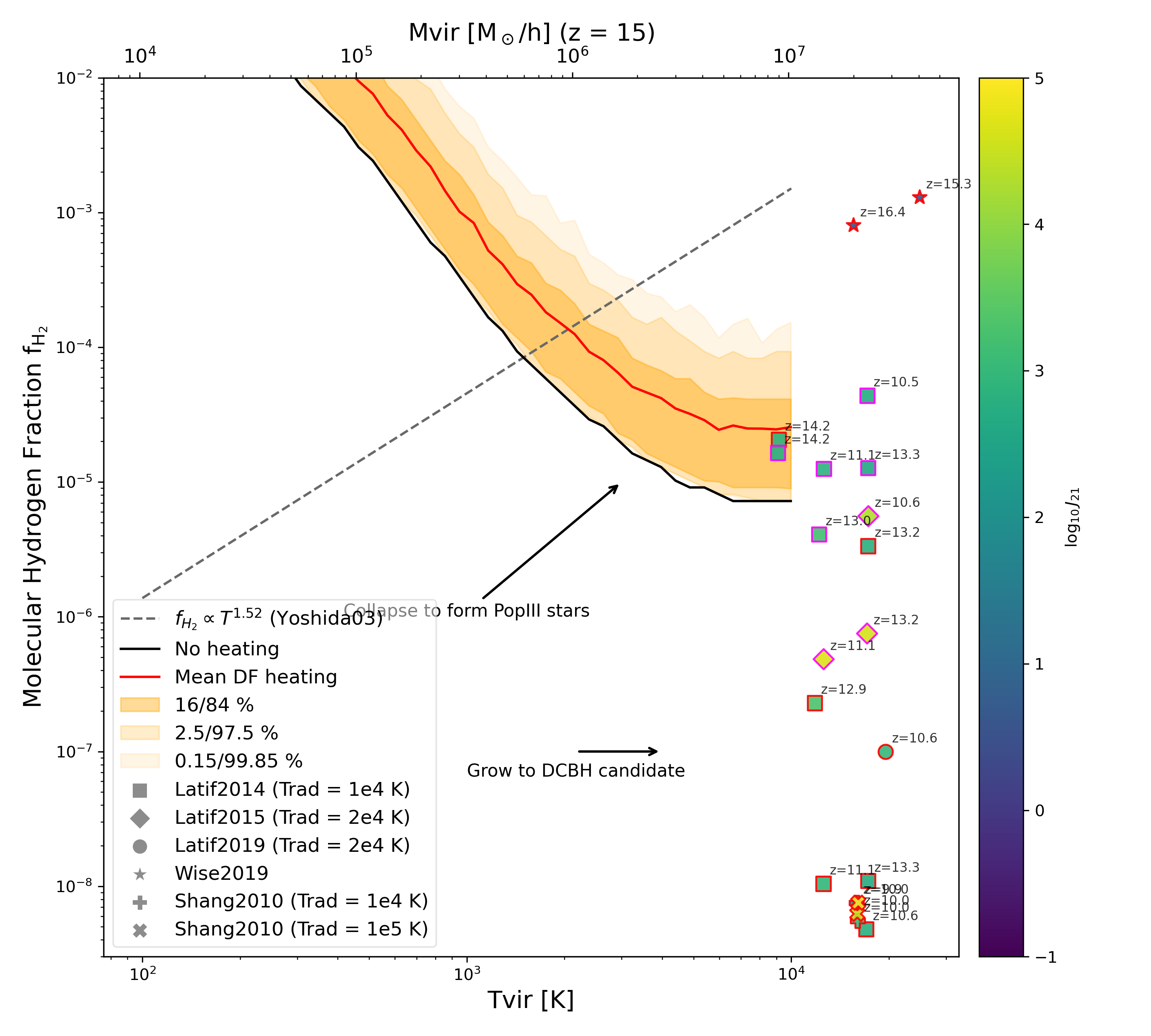}
 \caption{The lines are the same as Figure~\ref{fig:fH2_vs_T_SHMF_sampling_z15_Mgas1e5_PopIII}, but the markers are for the halos considered as DCBH candidates in the previous literature. The red-outlined markers denote a robust DCBH candidate, with the gas temperature remaining above 7000 K at the center. The magenta-outlined cases are boundary cases in which the LW intensity equals $J_{\mathrm{crit}}$ while molecular cooling occurs in the halo center.}
 
\label{fig:fH2_vs_T_SHMF_sampling_z15_Mgas1e5_DCBH}
\end{figure}

In Figure~\ref{fig:fH2_vs_T_SHMF_sampling_z15_Mgas1e5_PopIII} (Figure\ref{fig:fH2_vs_T_SHMF_sampling_z15_Mgas1e5_DCBH}), we investigate Pop III star (DCBH) formation in the $f_{\mathrm{H}_2} - T$ plane.
Based on the cooling rates calculated in \textsc{Grackle} with a fixed density ($\rho_{\mathrm{vir}} \Omega_{\rm b}/\Omega_{\rm m}$), temperature ($T_{\mathrm{vir}}$), and H$_2$ fraction, the downward trend represents the critical $f_{\mathrm{H}_2}$ needed for gravitational collapse. The solid black line is the baseline critical $f_{\mathrm{H}_2}$ without the impact of DF heating, and it agrees well with the critical line in \citet{yoshida_simulations_2003}.
In the presence of DF heating, even if the H$_2$ fraction is above the critical line, its cooling of the gas can be suppressed. We define the new cooling timescale as 
\begin{equation}
\label{eq:cooling_timescale_with_DF}
t_{\mathrm{cool}} = \frac{\rho \epsilon}{(\rho/m_{\mathrm{p}}) ^2\Lambda(\rho, \ T,\ f_{\mathrm{H}_2}) - \dot{E}_{DF}/(\frac{4}{3}\pi R_{\mathrm{vir}}^3)} 
\end{equation}
To include the effect of SHMF scatter, we sample 500 SHMF with Poisson fluctuations as before, and the orange regions correspond to the variation of the new $f_{\mathrm{crit}}$, and the solid red line is its mean. As a result of DF heating, the value of $f_{\mathrm{crit}}$ can increase by up to one order of magnitude, implying the importance of DF heating in regulating the formation of Pop III stars.

We further compare the new $f_{\mathrm{crit}}$ value with those inferred from simulated halos to see if DF heating affects the Pop III star or DCBH formation. 
Table~\ref{tab:halo_list} lists the data we collect for this comparison. The mass resolution of halos in Shang et al. and Latif et al. ranges from below $100 M_{\odot}$ to $600 M_{\odot}$. The \textsc{Pop2Prime} halo in \citet{CorreaMagnus2024} has an even higher mass resolution of $\sim 1 M_{\odot}$. 
Therefore, for a $\sim 5 \times 10^6 M_{\odot}$ halo, the DF heating from substructures above $\psi \approx 10^{-2}$ is naturally included.
At $z = 15$, the average LW background intensity in the unit of $10^{-21} \mathrm{erg\ s^{-1}\ sr^{-1}\ Hz^{-1}\ cm^{-2}}$ is $J_{21} \approx 0.1 - 1$ \citep{incatasciato_modelling_2023}. But $J_{21}$ can be orders of magnitude higher than this mean value due to radiation from nearby galaxies \citep{agarwal_optimal_2019}. In the presence of a strong LW background, H$_2$ photo-dissociation and H$^{-}$ photo-detachment lead to a $f_{\mathrm{H}_2}$ much lower than the $f_{\mathrm{H}_2} \propto T^{1.52}$ scaling relation. The simulations have experimented with a wide range of $J_{21}$ between $10^{-1} - 10^{5}$, as shown by the colorbars in Figure~\ref{fig:fH2_vs_T_SHMF_sampling_z15_Mgas1e5_PopIII} and Figure~\ref{fig:fH2_vs_T_SHMF_sampling_z15_Mgas1e5_DCBH}. Moreover, the spectral energy distribution (SED) of the first generations of galaxies largely affects the chemical reaction rates and $J_{\mathrm{crit}}$ for DCBH formation \citep{agarwal_new_2016}. The simulations by Shang et al. and Latif et al. summarised in Table~\ref{tab:halo_list} approximate the irradiating external source as a blackbody with a surface temperature of $T_{\mathrm{rad}} = 10^4 - 10^5$ K, corresponding to Pop II and Pop III stars. On the other hand, \citet{wise_formation_2019} and \citet{CorreaMagnus2024} model radiation from the local stellar population. 
The value of $J_{21}$ near the selected halos in \cite{wise_formation_2019} increases to $\sim 3$ at $z = 15$. In \citet{CorreaMagnus2024}, there is no continuous LW background; instead, the halos experience LW radiation from nearby halos in short bursts of roughly $4\,\mathrm{Myr}$ (Pop III star main-sequence lifetime) with an average $J_{21} \sim 1$.
We exclude the metal-enriched halos in \citet{CorreaMagnus2024}. For other simulations, we select only halos with published profiles in their respective papers, as listed in Table~\ref{tab:halo_list}.
In Figure~\ref{fig:fH2_vs_T_SHMF_sampling_z15_Mgas1e5_PopIII} and Figure~\ref{fig:fH2_vs_T_SHMF_sampling_z15_Mgas1e5_DCBH}, the shape of the markers represents the simulation suites and $T_{\mathrm{rad}}$, and each marker corresponds to one halo with a specific $J_{21}$. The same halo can appear multiple times as a Pop III star-forming halo or as a DCBH candidate, depending on the $J_{21}$ used. The mass-weighted H$_2$ fraction is calculated from the halo profile:
\begin{equation}
f_{\mathrm{H}_2} = \frac{1}{M_{\mathrm{gas}}} \int_0^{M_{\mathrm{gas}}} f_{\mathrm{H}_2}(r) \rho_{\rm g}(r) 4\pi r^2 dr
\end{equation}
We integrate the central $M_{\mathrm{gas}} = 10^5\ \rm{M}_{\odot}$ region, which roughly corresponds to the mass of a DCBH. 
In reality, the core of the halo may fragment into several DCBH seeds of several $10^5 {\rm M}_{\odot}$. So we have also checked the results when integrating the profile to an outer region with $M_{\mathrm{gas}} = 10^6 ~ {\rm M}_{\odot}$. However, the overall results do not change qualitatively except that a few halos have a slightly lower $f_{\mathrm{H}_2}$. For better illustration, we show the cases for Pop III star formation and DCBH seed formation separately in Figure~\ref{fig:fH2_vs_T_SHMF_sampling_z15_Mgas1e5_PopIII} and Figure~\ref{fig:fH2_vs_T_SHMF_sampling_z15_Mgas1e5_DCBH} with $M_{\mathrm{gas}} = 10^5\ \rm{M}_{\odot}$. An exception in the plots is the \textsc{Pop2Prime} halos from \citet{CorreaMagnus2024}. Since the halos only have a mass of a few $\times 10^5 {\rm M}_{\odot}$ and $M_{\mathrm{gas}} = 10^5 ~{\rm M}_{\odot}$ extends far beyond the central regions, we use $f_{\mathrm{H}_2,\ T_{\mathrm{half}}}$, i.e., the $f_{\mathrm{H}_2}$ at the point where the temperature is $(T_{\max} + T_{\min})/2$, as a proxy for the $f_{\mathrm{H}_2}$ of the cooling gas.

If the halo center remains above $\sim 7000$ K, it can be considered as a solid case for DCBH formation (red-outlined marker in Figure~\ref{fig:fH2_vs_T_SHMF_sampling_z15_Mgas1e5_DCBH}). If $J_{21}$ is less than the critical value, there may be cooling in the halo center and the halo is then considered a DCBH candidate (magenta-outlined marker in Figure~\ref{fig:fH2_vs_T_SHMF_sampling_z15_Mgas1e5_DCBH}). Snapshots of the halo profiles are taken at their collapse redshifts and are indicated by markers in the figures. 

In Figure~\ref{fig:fH2_vs_T_SHMF_sampling_z15_Mgas1e5_PopIII}, the \textsc{Pop2Prime} halos are above the $f_{\mathrm{H}_2} \propto T^{1.52}$ dashed line, and a few circles taken from \citet{latif_uv_2019} and triangles from \citet{latif_radiation_2021} scatter around the dashed line due to their LW background.
The $J_{21}$ can be higher for the triangles because their $T_{\mathrm{rad}} = 10^5$ K spectrum results in a lower H$_2$ photo-dissociation rate and a much lower H$^{-}$ photo-detachment rate \citep{shang_supermassive_2010, latif_how_2015, agarwal_new_2016}.
The other markers are below the dashed line due to LW radiation, with $f_{\mathrm{H}_2}$ ranging from $10^{-6}$ to $10^{-3}$. Compared to the critical H$_2$ fraction of Pop III star formation, the halos in the molecular cooling regime appear above the fiducial black line, i.e., $f_{\mathrm{crit}}$ without DF heating. Most of the collapsing halos also have an average $f_{\mathrm{H}_2}$ above the orange belts, indicating that DF heating plays a significant role in the formation of Pop III stars. From left to right, the halos grow to the atomic cooling regime as the redshift decreases. The squares and diamonds in the bottom-right of the figure show that star formation is more complicated in these more massive halos. Atomic cooling can at least cool the gas to around 8000 K. Further cooling and gas condensation does not necessarily require an average $f_{\mathrm{H}_2}$ above the critical line. In fact, the profiles of these halos show that only in the center $f_{\mathrm{H}_2}(r)$ reaches $10^{-3}$, but it decreases sharply below $10^{-5}$ at $r = 10^4 - 10^5$ AU (see, e.g., the $f_{\mathrm{H}_2}$ profiles of Halo B in Fig.~2 of \citealt{latif_uv_2014}, and Halo C in Fig.~5 of \citealt{latif_how_2015}).

Figure~\ref{fig:fH2_vs_T_SHMF_sampling_z15_Mgas1e5_DCBH} shows the DCBH candidate cases.
As indicated by the arrows in the figures, DCBH requires pristine atomic cooling halos, but $f_{\mathrm{H}_2}$ exhibits a wide range. The bottom-right corner is the most robust scenario for DCBH seed formation. 
Due to $J_{21} \gg J_{\mathrm{crit}}$, the average $f_{\mathrm{H}_2}$ stays below $10^{-7}$, with negligible molecular cooling in the halo center.
DF heating is not important compared to the strong LW radiation. More complicated is the upper regime with $f_{\mathrm{H}_2}$ between $10^{-6} - 10^{-4}$, where the markers overlap with the Pop III formation in Figure~\ref{fig:fH2_vs_T_SHMF_sampling_z15_Mgas1e5_PopIII}. 
Whether a DCBH can form is the result of multiple factors acting in combination. For the magenta-outlined markers, $J_{21} \approx J_{\mathrm{crit}}$ and both Pop III star and DCBH formation are possible.
For the red-outlined markers taken from \citet{latif_uv_2014}, it is found that for $J_{21} > J_{\mathrm{crit}}$ case gas in the halo collapses isothermally with $T \sim 8000$ K. Consequently, the mass accretion rate is an order of magnitude higher compared to the low LW case. 
In this case, if many substructures orbit around the halo, dynamical friction will further suppress H$_2$ cooling and even halos with $J_{21} < J_{\mathrm{crit}}$ may form DCBHs.
Thus, DCBH formation is possible even if the average $f_{\mathrm{H}_2}$ reaches $10^{-5} - 10^{-4}$. The red-outlined circle is halo 6 in \citet{latif_uv_2019}, which has a quiescent growth history and has not gone through a major merger. Its lower $f_{\mathrm{H}_2}$ is due to the less self-shielding related to the density structure in the halo. The star markers, i.e., MMH and LWH in \citet{wise_formation_2019}, are considered DCBH forming halos because of their rapid accretion rate above the critical accretion rate $\sim 0.04~M_{\odot}/\mathrm{yr}$. However, their average $f_{\mathrm{H}_2}$ is well above $f_{\mathrm{crit}}$ for star formation, even considering the impact of DF heating. As outliers in the $f_{\mathrm{H}_2} - T$ plane, the mechanism for DCBH formation, if possible, would be different from the rest of the halos. \citet{sakurai_radiative_2020} has conducted a follow-up study of these halos based on their 1D profiles, they show that a stellar mass protostar forms in center of the halo which rapidly grows to $10^5 M_{\odot}$ SMS within 2 Myr. The LW radiation of the protostar helps to dissociate H$_2$ in the inner region and increases the gas temperature. This indicates that the formation of a DCBH is still possible in the $f_{\mathrm{H}_2} \sim 10^{-2}$ regime due to the rapid gas accretion onto the protostar.

We conclude that DF heating is conducive to DCBH formation in cases with $J_{21} < J_{\mathrm{crit}}$, but the outcome of the collapse also depends on other factors such as metallicity  and accretion history, etc.

\section{Discussion on the DF heating profile}
\label{Section:heating_profile}

\subsection{Heating profile models}
\label{section:heating_profile_model}

\begin{figure}
    \centering
    \includegraphics[width=\columnwidth]{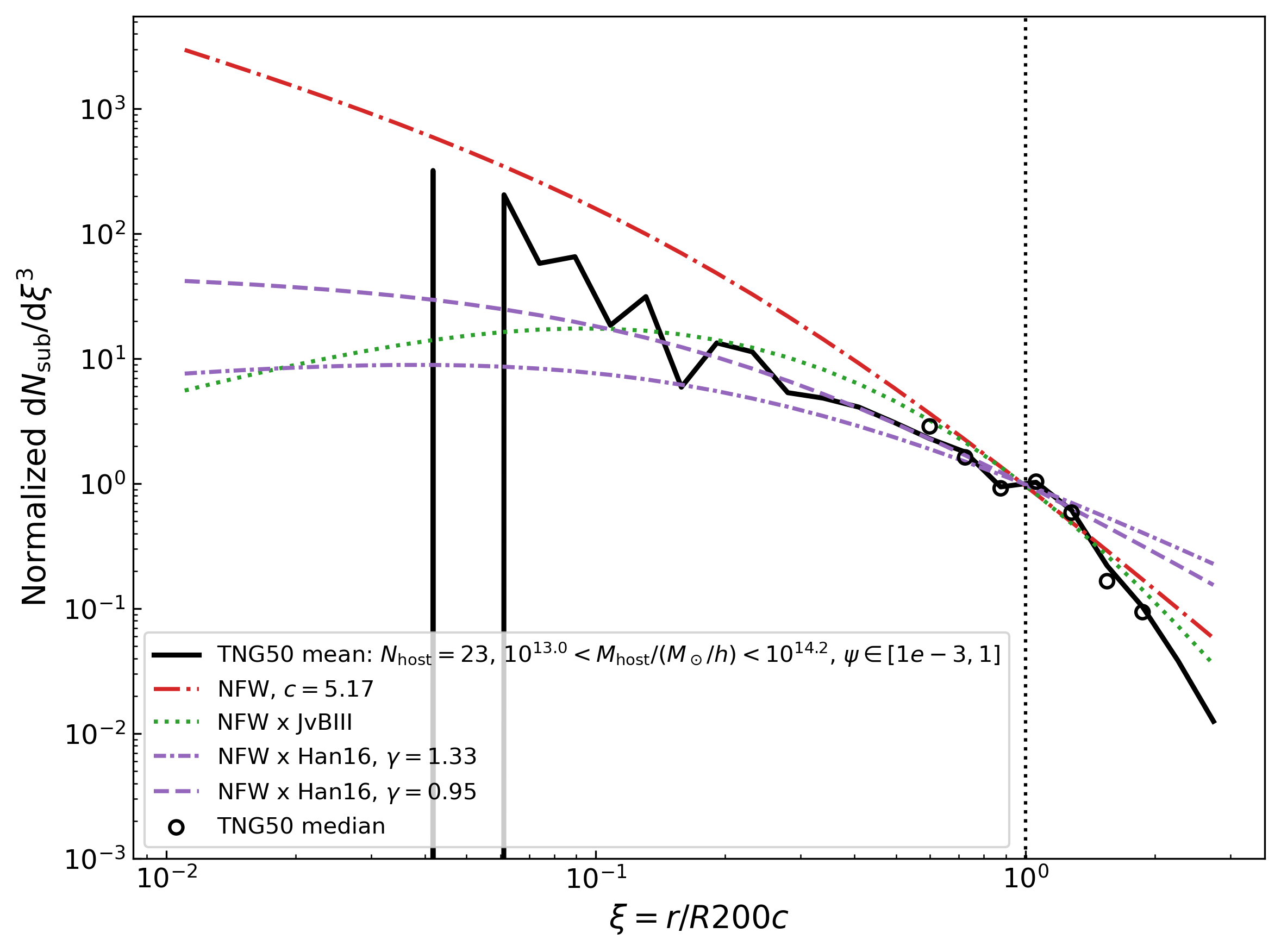}
    \caption{Radial subhalo number-distribution in TNG50 at $z = 0$. We show the normalized profile ${\rm d}N_{\rm sub}/{\rm d}\xi^3$ as a function of $\xi = r/ R_{200c}$ for host halos with $M_{\rm host} > 10^{13} M_{\odot}/h$ and subhalo mass ratio $ \psi \in [10^{-3},1]$. The black solid line and open circles show the host-averaged mean and median profiles, respectively, normalized to unity at $\xi=1$. The red dash-dotted line shows the NFW reference profile. The green dotted line shows the NFW profile multiplied by the ``radial bias'' function from \citetalias{jiang_statistics_2017}. The purple lines show the \citet{han_unified_2016} correction with their $\gamma = 1.33$ and $\gamma = 0.95$ (their best-fit for a Milky-Way-sized halo and a more massive galaxy cluster, respectively).}
    \label{fig:count_dx3_z0}
\end{figure}

\begin{figure}
    \centering
    \includegraphics[width=\columnwidth]{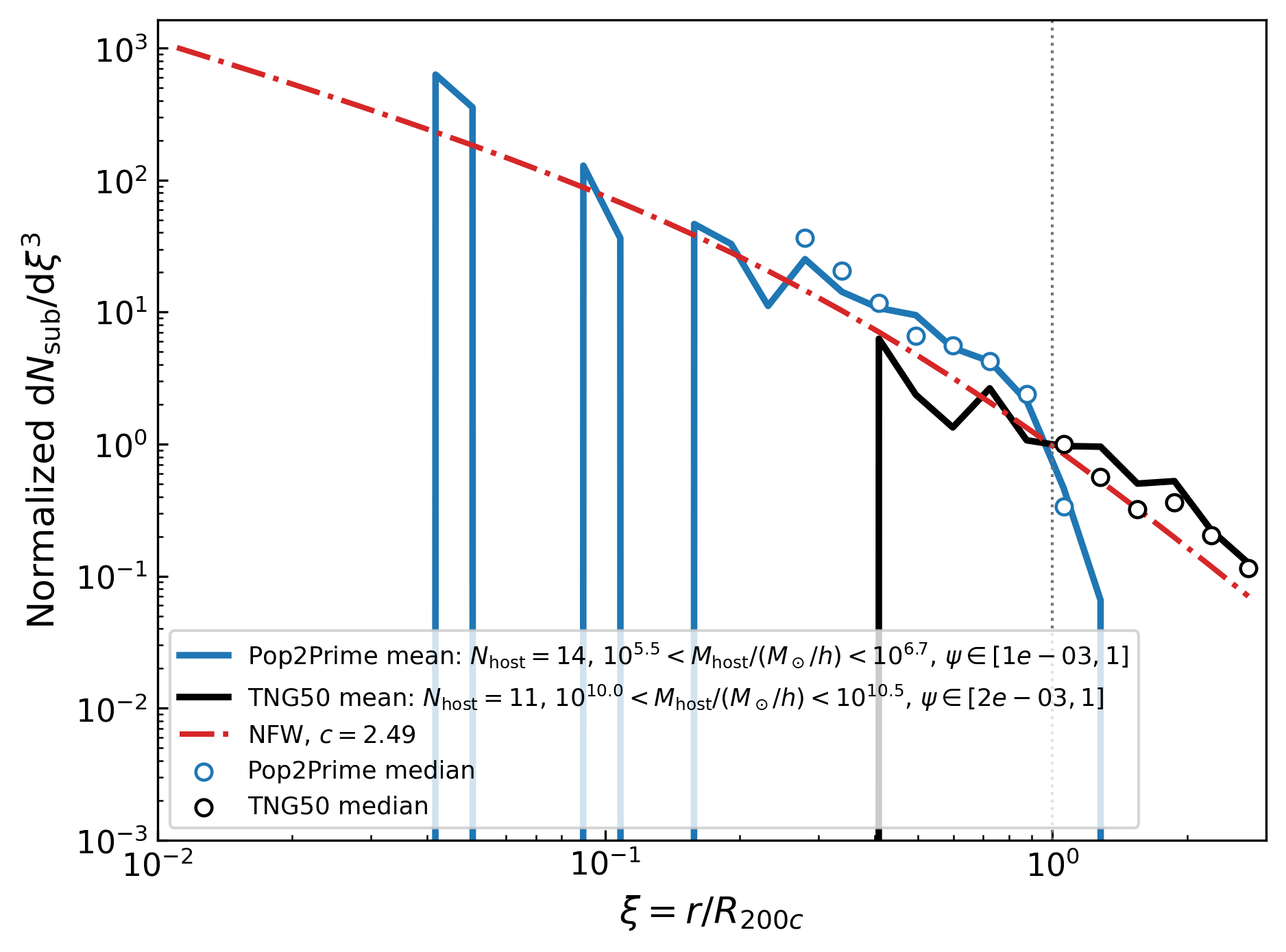}
    \caption{Same as Figure~\ref{fig:count_dx3_z0}, but at $z \simeq 12$. We also add the subhalo profile from \textsc{Pop2Prime}. The \textsc{Pop2Prime} profile is measured for host halos with $10^{5.5}<M_{\rm host}/ (M_{\odot}/h)<10^{6.7}$ and $\psi \in [10^{-3},1]$. The TNG50 profile uses host halos with $M_{\rm host}>10^{10} M_{\odot}/h$ and subhalos with $\psi\in[2\times10^{-3},1]$. The blue and black curves show the mean profiles from \textsc{Pop2Prime} and TNG50, respectively, while the open circles show their median profiles. The red dash-dotted line shows the NFW reference profile.}
    \label{fig:count_dx3_z12}
\end{figure}

In the previous sections, we have considered a global DF heating model, where the gas density is treated as a characteristic constant and the spatial distribution of subhalos is ignored. In reality, the gas density can span several orders of magnitude within a halo, and the radial distribution of subhalos can affect where the DF energy is deposited. We therefore examine the DF heating profile in this section. We particularly focus on the inner halo where radiative cooling is strongest to see if the DF heating indeed balances the central cooling region.

As discussed in Section~\ref{section:halo_cooling_profile} and Figure~\ref{fig:Pop2Prime_density_temperature_profile_z12}, the gas density profile of a halo can be described by a generalized NFW profile. The cumulative cooling rate inside a radius $r$ can then be written as

\begin{equation}
\dot{E}_{\text {cool }}(<r)=\Lambda\left(T_{\text {vir }}, Z\right){\bar{n_{\mathrm{H}}}}^2\left(\frac{4 \pi}{3} R_{200}^3\right) I_{\text {cool }}(<r)
\end{equation}

where $I_{\rm cool}(<r)$ is the same gas-density profile correction introduced in Section~\ref{section:halo_cooling_profile}, but with the integral truncated at $r$ instead of $R_{\rm vir}$. Since cooling scales as density squared, the gas fraction enters as $\dot{E}_{\rm cool}\propto f_{\rm g}^2$.

We introduce two levels of approximation for the radial DF heating profile. The first is the \textit{gas-density-only weighted model}, denoted $H_{\rm gas}(<r)$. In this model, the local DF heating rate is assumed to be proportional to the gas density, while the subhalo contribution is distributed uniformly in volume. The heating profile is normalized to the global DF heating rate $H_{\rm global}$, i.e., the $\dot{E}_{\rm DF}$ calculated in Section~\ref{section:ostriker}. This gives

\begin{equation}
H_{\mathrm{gas}}(<r)=H_{\mathrm{global}} \frac{\int_0^r \rho_{\mathrm{g}}\left(r^{\prime}\right) 4 \pi r^{\prime 2} d r^{\prime}}{\int_0^{R_{\mathrm{vir}}} \rho_{\mathrm{g}}\left(r^{\prime}\right) 4 \pi r^{\prime 2} d r^{\prime}}  \ .
\label{eq:H_gas_within_r}
\end{equation}

By construction, $H_{\rm gas}(<R_{\rm vir})=H_{\rm global}$. This model isolates the effect of the centrally enhanced gas density.

The second approximation, denoted $H_{\rm sub}$, is the \textit{subhalo spatial distribution model}, which additionally includes a radial subhalo distribution. Starting from the DF heating formula, the local heating rate per unit volume can be written schematically as
\begin{equation} 
\frac{dH}{dV} = \frac{4\pi G^2 M^2 \rho_{\rm g}(r)}{c_s} \int \psi^2 \left\langle \frac{I}{\mathcal{M}} \right\rangle \frac{d^2N}{d\psi\,dV}(\psi,r)\,d\psi , 
\label{eq:subhalo_spatial_model_original}
\end{equation}

where $\psi=m/M$ is the subhalo-to-host mass ratio, $\langle I/ \mathcal{M}\rangle$ denotes the Mach-number averaged DF factor, and $\frac{d^2N}{d\psi\,dV}$ is the joint mass ratio and spatial distribution of the subhalos. The global model effectively replaces $\rho_{\rm g}(r)$ by a characteristic halo gas density and integrates over the whole virial volume. 

To construct a smooth radial correction, we make several simplifying assumptions. First, we assume that the Mach-number averaged DF factor, $\langle I/\mathcal{M}\rangle$, can be taken out of the radial integral. Here $\langle I/\mathcal{M}\rangle$ represents the population-averaged velocity factor of the subhalo ensemble. We assume that its radial dependence is a secondary effect compared with the stronger radial variation of the gas density and the subhalo spatial distribution. \citet{Faltenbacher2005} studied galaxy motions in simulated clusters and found that galaxies are typically mildly supersonic, with a mean Mach number around 1.4 and no strong dependence on cluster-centric radius. For dark-matter subhalos, \citet{Diemand2004} compared the velocity distributions of subhalos and dark matter in both the inner and outer regions of simulated halos. They found that subhalos have a positive velocity bias relative to the dark matter, but the bias increases only mildly toward smaller radii. We therefore leave the radial velocity bias as a higher-order correction, and absorb $\langle I/\mathcal{M}\rangle$ into the overall normalization of the heating profile.

Second, we assume that the $\psi$-dependent SHMF and the subhalo spatial distribution are separable,

\begin{equation}
\frac{d^2N}{d\psi\,d\xi}
\approx
\frac{dN}{d\psi}\,u(\xi),
\label{eq:separable_subhalo_distribution}
\end{equation}
where $\xi \equiv r/R_{\rm vir}$ is the dimensionless radius, i.e., $dV = 4\pi r^2 dr = 4\pi R_{\rm vir}^3 \xi^2 d\xi$. The subhalo mass function $dN/d\psi$ is the same as Equation~\ref{eq:SHMF_form}, and $u(\xi)$ is the normalized radial probability distribution of subhalos. This assumption is adopted in the unified subhalo distribution model in \citet{han_unified_2016}, meaning that the different mass subhalos have the same shape of spatial distribution, except for the SHMF as the amplitude. The simulations in \citet{vandenbosch_segregation_2016} also show little segregation of subhalos by their present-day mass along the radius.

We assume that the subhalos are located within $\xi_{\max}$. The normalization of $u(\xi)$ is
\begin{equation}
\int_0^{\xi_{\max }} u(\xi) d \xi=1
\end{equation}
For a simulated or analytic subhalo number density profile with a chosen threshold mass ratio $\psi_{\rm th}$, 
\begin{equation}
u(\xi)=\frac{1}{N_{\mathrm{sub}}\left(>\psi_{\mathrm{th}}\right)} \frac{d N\left(>\psi_{\mathrm{th}}\right)}{d \xi}
\end{equation}
The previous studies have found that the subhalo number density profile is typically shallower than an NFW profile. In \citetalias{jiang_statistics_2017}, the radial distribution of subhalos is described by

\begin{equation}
\frac{dN_{\rm sub}}{d\xi^3} = \phi(\xi) \left.\frac{dN}{d\xi^3}\right|_{\rm NFW}
\end{equation}

where $\left.\frac{dN}{d\xi^3}\right|_{\rm NFW} \propto \frac{1}{c \xi(1+c \xi)^2} \equiv \frac{1}{r/r_s(1+r/r_s)^2}$ is the normalized NFW profile of the host halo. The ``radial bias function'' $\phi(\xi)$ is well fit by

\begin{equation}
\phi(\xi) =
2^\mu \frac{\xi^\eta}{(1+\xi)^\mu}
, \quad \eta = 2.0, \mu = 4.0
\end{equation}

\citet{han_unified_2016} uses a different function to fit, with $\phi(\xi) = \xi^{\gamma}$, where $\gamma = 1.33$ and $\gamma = 0.95$ for their Milky Way-sized and cluster-sized halos, respectively. In the \citet{han_unified_2016} model, the spatial distribution of subhalos in their accretion or infall mass bin follows the NFW profile of the host halo; it is the stronger tidal stripping at a smaller radius that leads to the radial bias function, or the shallower profile of subhalos by their present-day mass.

We further use the TNG50 and \textsc{Pop2Prime} data to examine the subhalo spatial distribution. Figure~\ref{fig:count_dx3_z0} shows the radial number-density profile of subhalos with $\psi \ge 10^{-3}$ in TNG50 at $z=0$. The profile is normalized to unity at the virial radius. The black solid line is the average of the 23 selected host halos above $10^{13} M_{\odot}/h$ at $z = 0$, and the open circles represent their median. The profile is clearly less concentrated than the NFW profile shown by the red dash-dotted line, which uses the concentration parameter of the median host halo mass. The profile is broadly consistent with the \citet{jiang_statistics_2017} and \citet{han_unified_2016} results, except for the less-resolved and noisy central region $< 10^{-1} R_{\rm vir}$. 

Figure~\ref{fig:count_dx3_z12} is the result at $z = 12$. We select halos with different mass ranges from TNG50 and the high-resolution \textsc{Pop2Prime} data. For TNG50, we require $M_{\rm host} > 10^{10} M_{\odot}/h$, which results in 11 halos. Considering the limited mass resolution, we apply a higher mass ratio threshold $\psi \ge 2\times10^{-3}$ so that the subhalo mass can be resolved. For the \textsc{Pop2Prime} data, we select 14 halos with $M_{\rm host} > 10^{5.5} M_{\odot}/h$, the mass relevant for Pop III star or DCBH formation in Section~\ref{Section:minihalo}. Due to the small number of the host halos, the results should be considered as a rough description instead of rigorous statistics. Nevertheless, the profiles suggest that the strong low-redshift bias has not yet developed in these high-redshift halos, likely because subhalos have experienced less tidal stripping and orbital evolution. We therefore assume an NFW shape, or $\phi(\xi) = 1$ for high-z minihalos in the following section.

We also note that some TNG50 subhalos are located outside $R_{200c}$, because the \textsc{SUBFIND} algorithm does not apply a spherical overdensity cut, while we additionally require the distance between the host halos and the \textsc{ROCKSTAR} subhalos $\lesssim R_{200c}$ in the \textsc{Pop2Prime} data. This is slightly inconsistent with the \textit{global DF heating model}. So in Section~\ref{section:DF_heating_profile_results}, we compare the results with $\xi_{\max} = 1$ and $\xi_{\max} = 2$ to see how sensitively the DF heating in the inner region depends on the subhalo spatial distribution. Moreover, the difference between the \textsc{Pop2Prime} and TNG50 data in Figure~\ref{fig:count_dx3_z12} is not only because of the halo mass range and simulation resolution, but also the choice of subhalo finder. It is difficult for \textsc{SUBFIND} to identify the subhalos in the central region, while \textsc{ROCKSTAR} may be better (see Figure 10 in \citet{ForouharMoreno2025} and Figure 8 in \citet{Han2017} for a comparison with \textsc{HBT+}).

Based on Equation~\ref{eq:separable_subhalo_distribution}, we can rewrite Equation~\ref{eq:subhalo_spatial_model_original} as

\begin{equation}
\frac{d H_{\rm sub}}{d \xi}=\frac{4 \pi G^2 M^2 \rho_{\rm g}(\xi)}{c_s} u(\xi) \left\langle\frac{I}{\mathcal{M}}\right\rangle\int_{\psi_{\mathrm{th}}}^1 \psi^2 \frac{d N}{d \psi} d \psi .
\label{eq:dHsub_dxi}
\end{equation}
The cumulative heating profile is 
\begin{equation}
H_{\rm sub}(<r)=\int_0^\xi \frac{d H_{\rm sub}}{d \xi^{\prime}} d \xi^{\prime}
\end{equation}

In this \textit{subhalo spatial distribution model}, the SHMF sets the overall $\psi^2$-weighted heating amplitude, while $u(\xi)$ provides a smooth radial redistribution. We note that the $H_{\rm gas}$ model is the special case of Equation~\ref{eq:dHsub_dxi} corresponding to a spatially uniform subhalo number density: setting $u(\xi)\propto\xi^2$ (i.e. $dN/d\xi^3={\rm const}$) with $\xi_{\max}=1$ leaves the gas density $\rho_{\rm g}(\xi)$ as the only radial dependence, and $H_{\rm sub}(<r)$ then reduces exactly to $H_{\rm gas}(<r)$ in Equation~\ref{eq:H_gas_within_r}. In this sense, $H_{\rm gas}$ provides a lower bound on the central heating rate, as it neglects the central concentration of the subhalos. 

As a caveat, the $H_{\rm sub}$ model assumes that the DF heating is deposited locally at the subhalos. In a more realistic scenario, the energy is spread along the subhalo wakes \citep[see the distributed-heating treatment in][]{kim_dynamical_2005}, so $H_{\rm sub}$ likely overestimates the heating in the central region. A more complete treatment is beyond the scope of this paper and should be examined with future simulations. The two models should therefore be regarded as bracketing estimates under different assumptions rather than a single refined answer.

The models above provide the radial profile factors used to evaluate the cumulative heating and cooling rates within a chosen radius. In the following section, we first examine these dimensionless factors, and then apply them to the full heating--cooling comparison within the inner halo.

\subsection{Results}

\label{section:DF_heating_profile_results}

\begin{figure}
    \centering
    \includegraphics[width=\columnwidth]{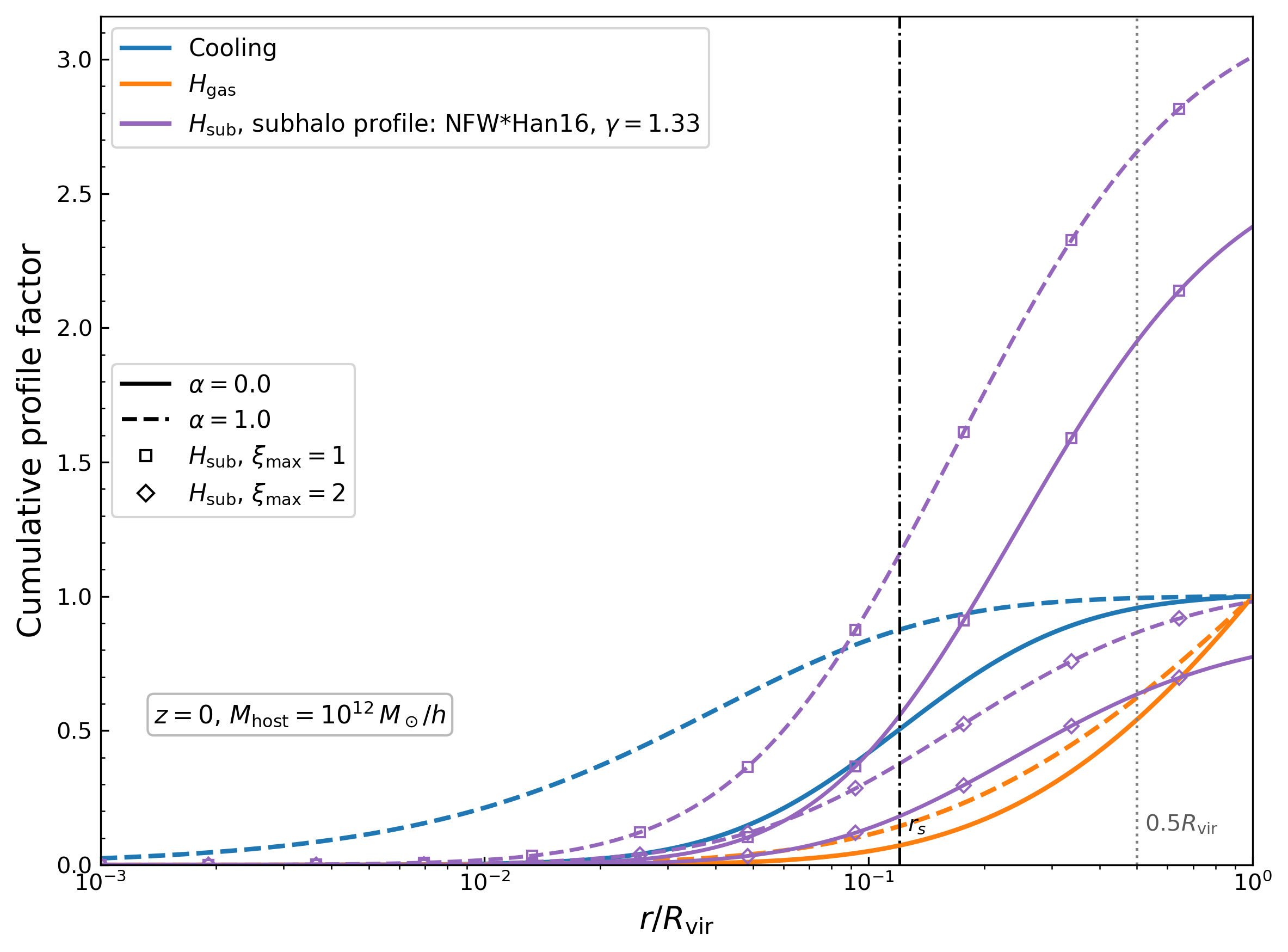}
    \caption{Cumulative cooling and heating profile factors for a $M_{\rm host}=10^{12}\,M_\odot/h$ halo at $z=0$. The blue and orange curves show the normalized cumulative cooling profile and the $H_{\rm gas}(<r)/H_{\rm global}$ cumulative heating profile correction, respectively. The purple curves show $H_{\rm sub}(<r)/H_{\rm global}$, which additionally includes the subhalo radial distribution using the \citet{han_unified_2016} profile with $\gamma=1.33$. Solid and dashed lines correspond to gas-profile slopes $\alpha=0$ and $ \alpha=1$, respectively, while square and diamond markers denote $\xi_{\max} =1$ and $\xi_{\max}=2$ for $H_{\rm sub}$. Vertical lines mark $r_s$ and $0.5R_{\rm vir}$.}
    \label{fig:cooling_heating_profile_z0}
\end{figure}

\begin{figure}
    \centering
    \includegraphics[width=\columnwidth]{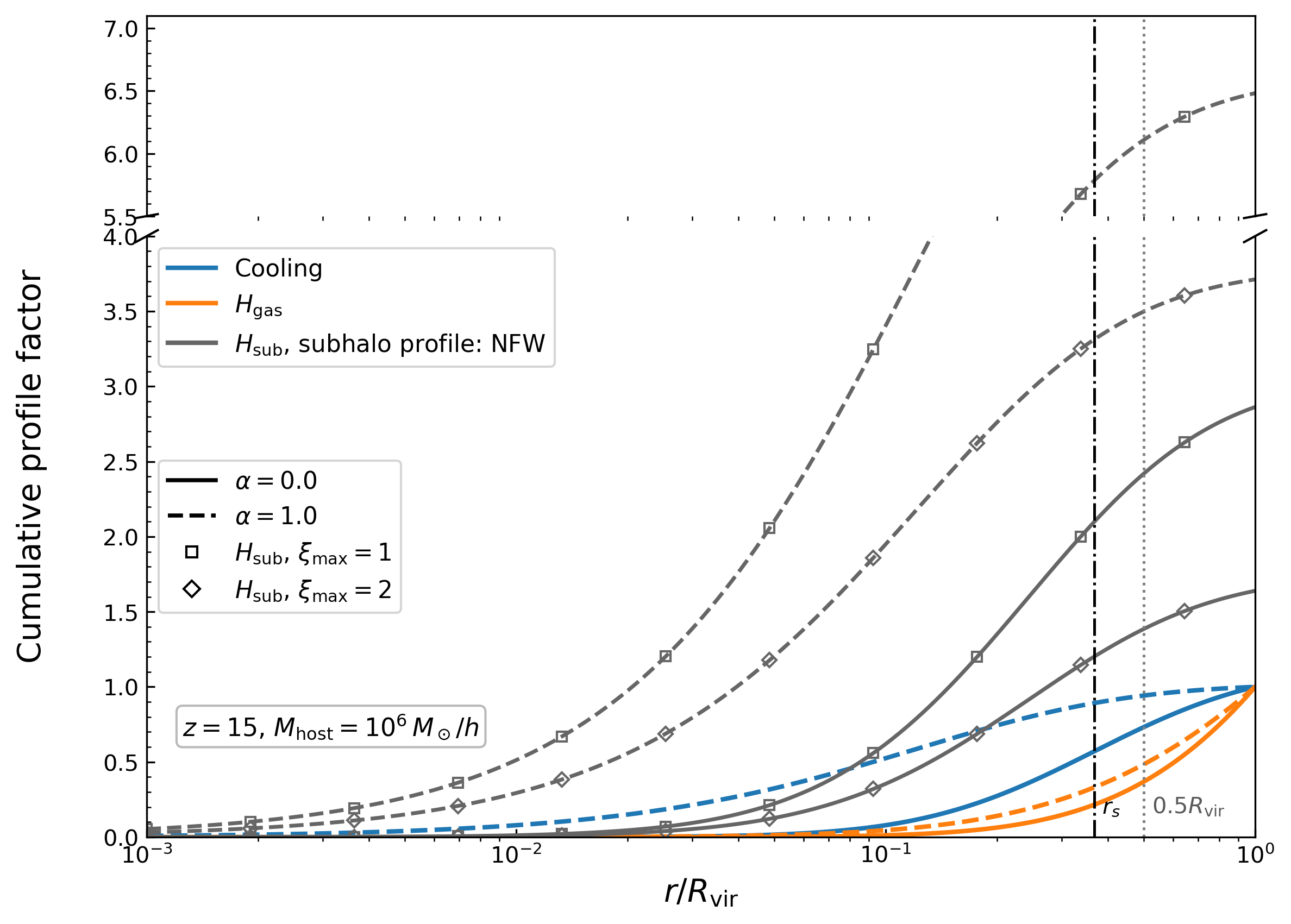}
    \caption{Same as Figure~\ref{fig:cooling_heating_profile_z0}, but for a $M_{\rm host} =10^{6}\,M_\odot/h$ halo at $z=15$. Here $H_{\rm sub}(<r)$ uses an NFW subhalo radial profile, as shown by the grey lines. The $y$-axis is broken to show both the low-amplitude profile factors and the larger $H_{\rm sub}$ values on a linear scale.}
    \label{fig:cooling_heating_profile_z15}
\end{figure}

We now apply the profile models to two representative cases. Figure~\ref{fig:cooling_heating_profile_z0} is the cumulative heating or cooling profile correction factor for a $M_{\rm host}=10^{12}\, M_\odot/h$ halo at $z=0$, while Figure~\ref{fig:cooling_heating_profile_z15} is for a $M_{\rm host} =10^{6}\, M_\odot/h$ halo at $z=15$. The cooling curves $I_{\rm cool}(<r)/I_{\rm cool}(<R_{\rm vir})$ are normalized to the global cooling rate at $R_{\rm vir}$. For the heating rate, we show $H_{\rm gas}(<r)/H_{\rm global}$ and $H_{\rm sub}(<r)/H_{\rm global}$ to compare with the global model. The \textit{gas-density-only weighted model} is normalized so that $H_{\rm gas}(<R_{\rm vir})=H_{\rm global}$. In contrast, $H_{\rm sub}$ is obtained by directly integrating the local heating rate in Equation~\ref{eq:dHsub_dxi}, including both the gas density and the subhalo radial distribution. Thus, $H_{\rm sub}(<R_{\rm vir})$ is not forced to equal $H_{\rm global}$; the ratio $H_{\rm sub}/H_{\rm global}$ simply measures the profile correction. For $H_{\rm gas}$ (orange lines), the local heating rate is proportional to the gas density $\rho_{\rm g}(r)$, whereas the cooling rate (blue lines) is proportional to $\rho_{\rm g}^2(r)$. This leads to overcooling in the central region. This effect is stronger for a cuspy NFW gas profile than for a cored profile, as shown by the comparison between the dashed and solid lines. $H_{\rm sub}$ shows a larger variety depending on the parameters. We adopt the \citet{han_unified_2016} subhalo spatial distribution with $\gamma = 1.33$ for the Milky Way-sized halo in Figure~\ref{fig:cooling_heating_profile_z0} (purple lines). For $\xi_{\max} = 1$, the heating rate can be as large as 3 times the baseline global heating estimate. When the subhalo radial distribution is normalized out to $\xi_{\max}=2$, the enclosed $H_{\rm sub}$ drops lower than the cooling profile due to a smaller fraction of the subhalo population lying within a fixed inner radius. We also use the vertical lines marking $r_{s}$ and $0.5R_{\rm vir}$ to indicate the inner region of the host halo. Except for the $H_{\rm sub}, \xi_{\max}=1$ case, the cumulative heating profile factor is typically smaller than the cooling profile factor, but the gap is within one order of magnitude. The $z = 15$ result for a smaller halo in Figure~\ref{fig:cooling_heating_profile_z15} is similar, but the $H_{\rm sub}$ is larger due to the NFW subhalo spatial distribution without any radial bias correction (grey lines).

\begin{figure*}
    \centering
    \includegraphics[width=\textwidth]{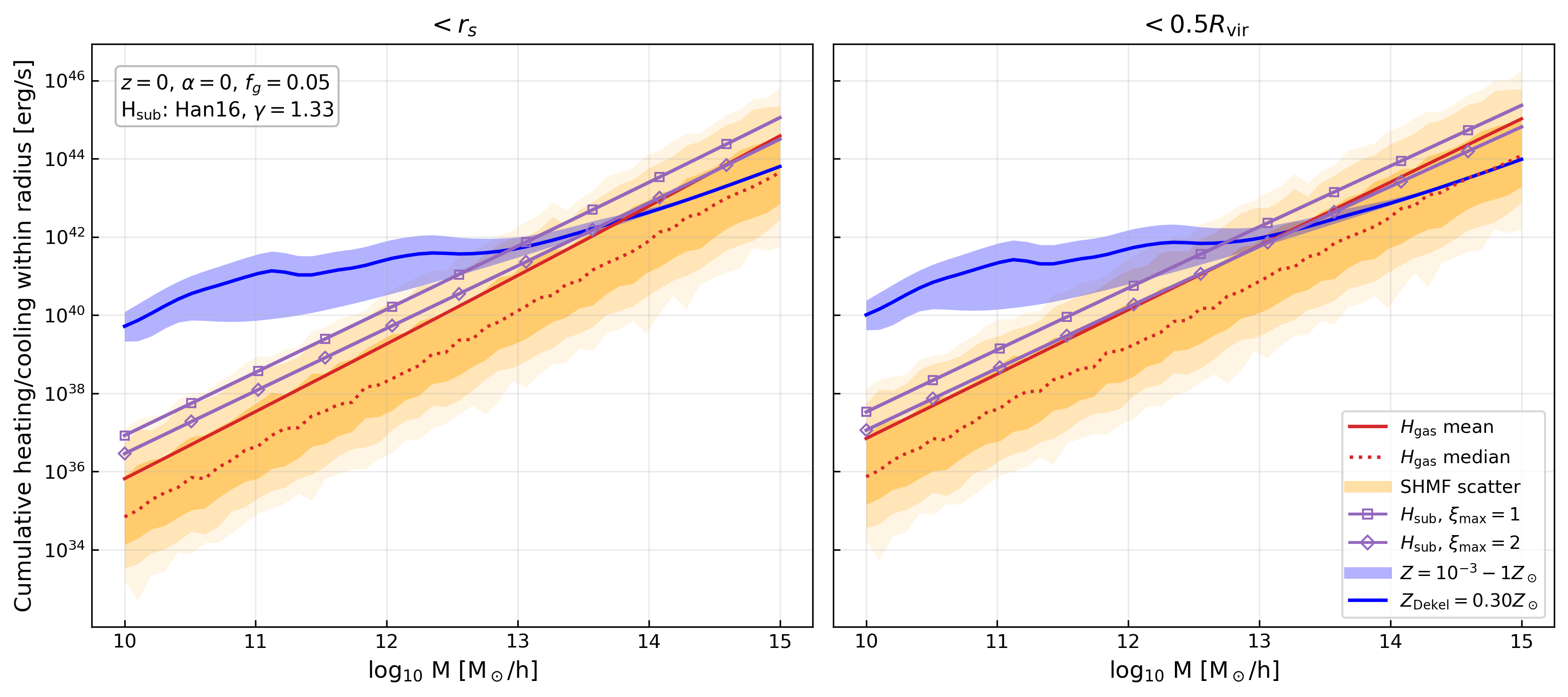}
    \caption{Cumulative DF heating and radiative cooling rates within selected radii for low-redshift massive halos. We show the rates enclosed within $r_s$ and $0.5R_{\rm vir}$ at $z=0$, assuming $\alpha=0$ for the gas-density profile and $f_{\rm g}=0.05$. The red solid and dotted lines show the mean and median $H_{\rm gas}$ heating rates, respectively, while the orange band shows the scatter from stochastic sampling of the SHMF. The purple curves show the mean $H_{\rm sub}$ heating rates using the \citet{han_unified_2016} subhalo radial distribution with $\gamma=1.33$; square and diamond markers correspond to $\xi_{\max}=1$ and $\xi_{\max}=2$, respectively. The blue shaded region brackets the cooling rates for $10^{-3}Z_\odot \leq Z \leq Z_\odot$, and the blue solid line shows the cooling rate for the Dekel metallicity model.}
\label{fig:cooling_heating_within_radius_z0}
\end{figure*}

\begin{figure*}
    \centering
    \includegraphics[width=\textwidth]{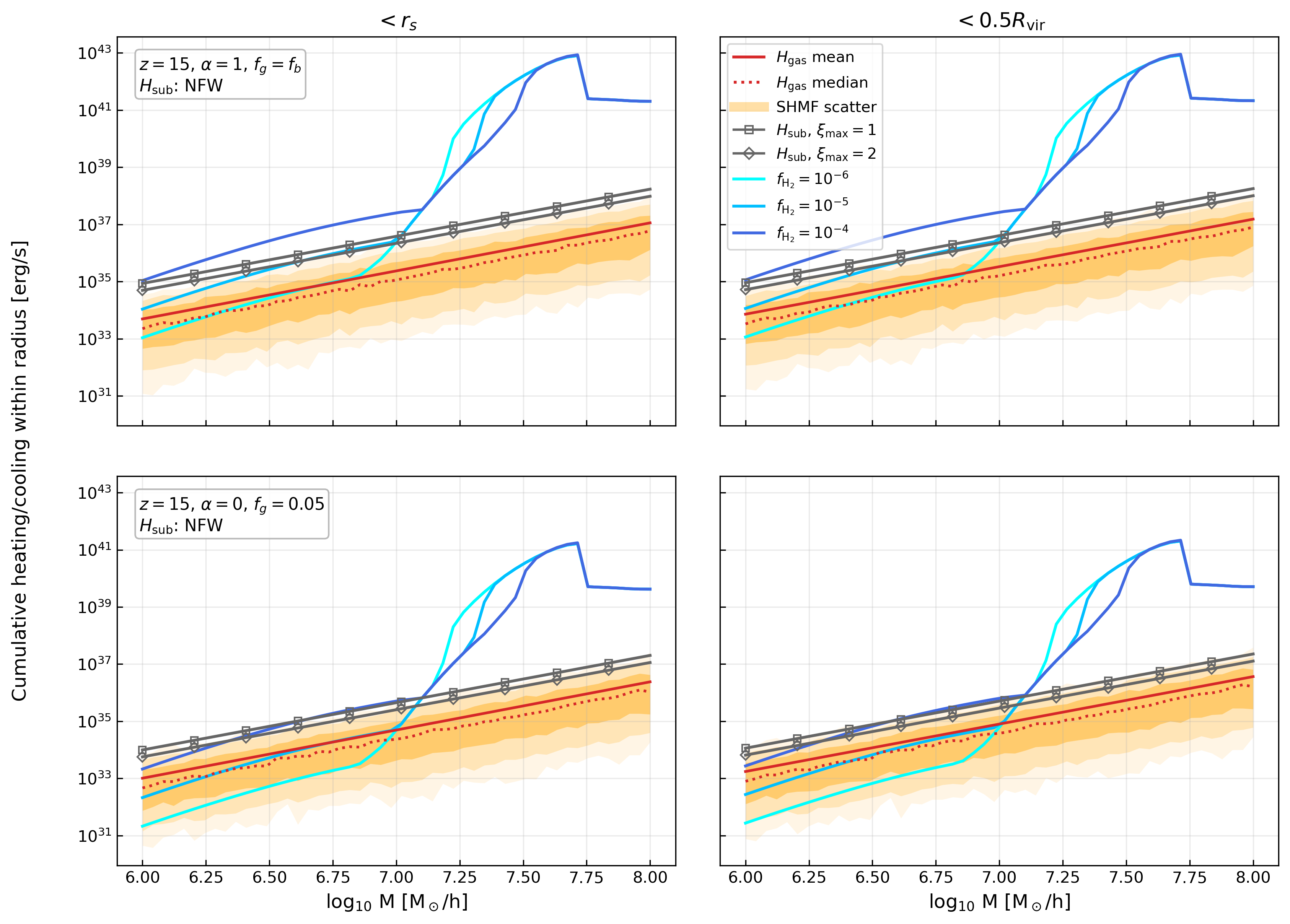}
    \caption{Cumulative DF heating and radiative cooling rates within selected radii for high-redshift minihalos at $z=15$. The left and right columns show the rates enclosed within $r_s$ and $0.5R_{\rm vir}$, respectively. The top row shows the gas-rich, centrally concentrated case with $ \alpha=1$ and $f_{\rm g}=\Omega_{\rm b}/\Omega_{\rm m}$, while the bottom row shows the gas-poor, cored case with $\alpha=0$ and $f_{\rm g}=0.05$. The red solid and dotted lines show the mean and median $H_{\rm gas}$ heating rates, respectively, and the orange band shows the SHMF sampling scatter. The grey curves show the mean $H_{\rm sub}$ heating rates assuming an NFW subhalo radial distribution; square and diamond markers correspond to $\xi_{\max}=1$ and $\xi_{\max}=2$, respectively. The blue curves show the cooling rates for $f_{\rm H_2} =10^{-6}$, $10^{-5}$, and $10^{-4}$.}
    \label{fig:cooling_heating_within_radius_z15}
\end{figure*}

We then combine the global heating rates from Section~\ref{Section:DF_method} with the profile corrections from Section~\ref{section:heating_profile_model}, treating the gas fraction $f_{\rm g}$ as a free parameter. The DF heating rate scales linearly with $f_{\rm g}$, whereas the radiative cooling rate scales as $f_{\rm g}^2$. This combined model allows us to compare DF heating and cooling in the inner halo.

Figure~\ref{fig:cooling_heating_within_radius_z0} shows the cumulative DF heating and cooling rates within $r_s$ and $0.5 R_{\rm vir}$ for $z = 0 $ halos. If the rates are integrated to $R_{\rm vir}$, this figure will return to the upper panel of Figure~\ref{fig:singlehost_z0_2_6}, as the $H_{\rm gas}$ model is normalized to the global heating rate. We additionally show the $H_{\rm sub}$ results with the mean SHMF in the purple lines, using the same parameters as Figure~\ref{fig:cooling_heating_profile_z0}. For massive halos, the $H_{\rm sub}, \xi_{\max}=2$ results (purple diamond line) are close to the $H_{\rm gas}$ results, while $H_{\rm sub}, \xi_{\max}=1$ (purple square line) is higher by a factor of a few. Compared with Figure~\ref{fig:singlehost_z0_2_6}, in the inner region $<r_s$, the mean heating line may cross the cooling line at a higher halo mass. However, the results do not change qualitatively. 

Figure~\ref{fig:cooling_heating_within_radius_z15} is the inner region version of Figure~\ref{fig:singlehost_minihalo}.
The top row shows the gas-rich, centrally concentrated gas profile with $\alpha = 1, f_{\rm g} = \Omega_{\rm b}/\Omega_{\rm m}$, where overcooling is more significant in the central region due to the $\rho_{\rm g}^2(r)$ dependence. The bottom row shows a gas-poor, cored profile with $\alpha = 0, f_{\rm g} = 0.05$. The additional grey line shows the $H_{\rm sub}$ results using the same parameters as Figure~\ref{fig:cooling_heating_profile_z15}. The comparison between the heating lines and the cooling lines shows that the $f_{\rm H_2}$ that balances the heating can increase to $\sim10^{-4}$ or decrease to $\sim 10^{-6}$, depending on the DF heating profile. In a more realistic scenario, the $f_{\rm H_2}(r)$ profile should also be taken into account, which is beyond the analytic model.

We conclude that the radial profile correction changes the enclosed DF heating rate by factors of a few, and up to roughly one order of magnitude in the most concentrated cases. This is important for the central heating--cooling comparison, but remains subdominant to the halo-to-halo scatter from the SHMF. The global model is therefore a useful first-order estimate, while the profile correction refines the normalization within the inner halo.

\section{Discussion and Conclusions}
\label{Section:conclusion}

In this work, we studied the heating of gas in a halo by dynamical friction from subhalos and studied its impact on star formation across different astrophysical contexts.

To this aim, we first derived the evolved SHMF from the TNG50 halo catalog. We validated our fitting by comparing the $z = 0$ SHMF with the previous results from \citetalias{vandenbosch_statistics_2016}. We also found that the SHMF at high redshifts tends to be closer to the \citetalias{vandenbosch_statistics_2016}'s unevolved SHMF, and we fit the shape of the SHMF as a function of redshift. 
The main scatter in the DF heating rate is a reflection of the scatter in the  SHMF, which is mostly  Poissonian for mass ratios relevant to DF heating.
We also examined the influence of the perturbation's nonlinearity and the Mach number correction, though the nonlinear perturbations have a negligible impact, and the Mach number correction results in a factor $\sim$ 1 - 2 in the heating rate.

We then show the total DF heating in host halo mass bins. Mainly subhalos above mass ratio $\psi \sim 0.05$ contribute to dynamical friction.
It scales with the host halo mass, but at high redshifts it contributes little to the heating of the Universe because of the rarity of massive halos. 

As a first application, we studied DF heating and cooling in halos at low redshifts. The cooling rates are calculated in \textsc{Grackle} using halo profiles from \citetalias{Dekel2007}. Our results are in agreement with those for massive clusters by \citet{kim_dynamical_2005} and those by \citetalias{Dekel2007} on gravitational heating. However, since we include the actual scatter in the SHMF, we are able to quantify the scatter in the heating rates in contrast to the earlier studies.

Consistent with previous research, DF heating is only significant in massive halos $\gtrsim 10^{13} M_{\odot}/h$, possibly leading to galaxy quenching \citep[][e.g.]{Khochfar2008}. 
At higher redshifts $z \gtrsim 4$, DF heating can be negligible in most halos, except for very rare high-mass dark matter halos ($10^{12-13} M_{\odot}$) hosting luminous quasars. 

We then explored the impact of DF heating on Pop III star and DCBH formation in the high-redshift Universe and compared our results with numerical simulations. DF heating is comparable to molecular cooling in pristine minihalos at $z \approx 15$, although it is not as strong as atomic cooling above $10^4$ K. 
DF heating increases the critical H$_2$ fraction $f_{\mathrm{crit}}$ for Pop III star formation in minihalos by up to one order of magnitude, and the average H$_2$ fractions from simulated halos are mostly above $f_{\mathrm{crit}}$ assuming a mean DF heating rate.
DF heating can suppress Pop III star formation in minihalos with $f_{\mathrm{H}_2} \approx 10^{-6} - 10^{-4}$, which allows for the formation of DCBHs in atomic cooling halos, also see \cite{L25}.

We also tested the effect of radial profile corrections on the heating--cooling comparison. Including the gas-density profile and the subhalo spatial distribution changes the enclosed DF heating rate by factors of a few, and up to roughly one order of magnitude in the most concentrated cases. Comparing the rates within $r_s$ and $0.5R_{\rm vir}$ shows that the global model remains a useful first-order estimate, although profile corrections are needed for a more accurate assessment of the central heating--cooling balance.

By construction, our DF heating model is an approximation of the complex physical processes at play. For example, in our analysis of DF heating at high redshift, we assume that the SHMF has a universal analytic shape across the halo mass range. This is broadly consistent with the TNG50 and \textsc{Pop2Prime} data. However, in forthcoming work, we will use high-resolution N-body simulations to better quantify the SHMF at high redshifts. 
We will also analyze DF heating in self-consistent cosmological hydrodynamics simulations to study the distribution of DF heating in halos and other baryonic processes involved, such as turbulence.

Our results show that the formation of DCBHs can be aided by DF heating. 
In contrast to other DCBH scenarios that require a burst of LW radiation \citep[][e.g]{agarwal_revised_2015}, heating in rapidly growing halos \citep{wise_formation_2019} or turbulence in halos \citep{latif_turbulent_2022}, DF heating is naturally occurring continuously in $\Lambda$CDM as subhalos orbit within their host. 
Our results suggest that this is an important factor to include in semi-analytic BH seeding models \citep{agarwal_ubiquitous_2012, Dijkstra2014, habouzit_number_2016, Valiante2017, lupi_forming_2021, OBrennan2025} and that numerical simulations require sufficient mass resolution to capture the relevant subhalo population.

\section*{Acknowledgements}

ZW thanks the China Scholarship Council (CSC) under Grant No. 202106190038 for their financial support.
SK acknowledges funding via STFC Small Grant ST/Y001133/1. MAL thanks the UAEU for funding via UPAR grants No. G00005454. For the purpose of open access, the authors have applied a Creative Commons Attribution (CC BY) license to any Author Accepted Manuscript version arising from this submission.

\section*{Data Availability}
Our analysis data will be made available upon request to the corresponding author.



\bibliographystyle{mnras}
\bibliography{bib_0, bib_1} 




\appendix

\section{SHMF from the \textsc{Pop2Prime} simulations}
\label{appendix:SHMF_Pop2Prime}

\begin{figure}
 \includegraphics[width=\columnwidth]{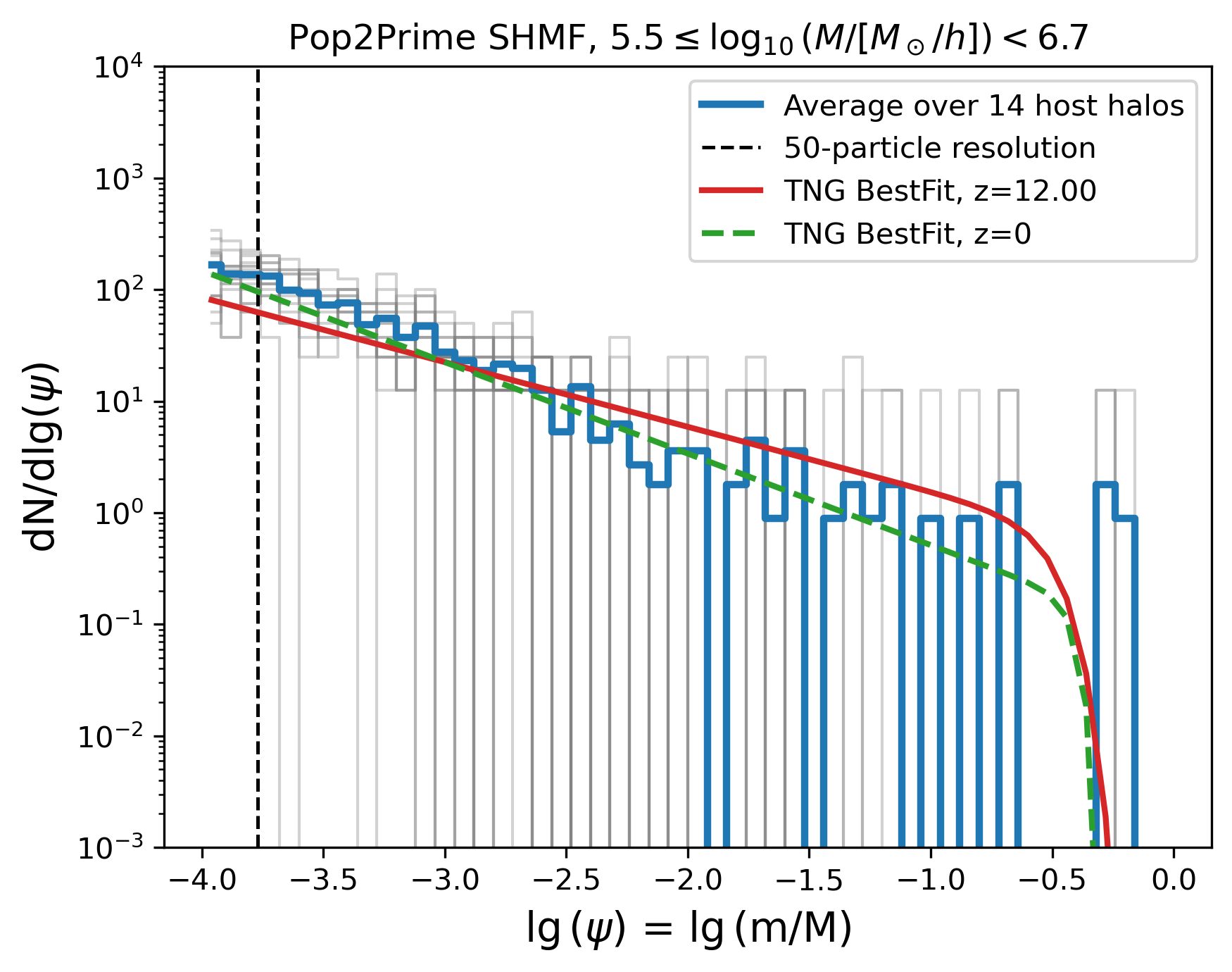}
 \caption{The SHMF for \textsc{Pop2Prime} halos $\ge 10^{5.5} M_{\odot}/h$ at $z = 12$. There are 14 host halos in this mass range, and the light grey lines show the SHMF for each individual of them. The solid blue line shows the mean SHMF, compared with the TNG $z = 0$ (green) and $z = 12$ (red) mean SHMF from Section~\ref{section:SHMF}. The vertical dashed line is the ratio between a resolved subhalo mass (50 particles) and $10^{5.5} M_{\odot}/h$.} 
 \label{fig:SHMF_Pop2Prime_z12}
\end{figure}

\begin{figure}
 \includegraphics[width=\columnwidth]{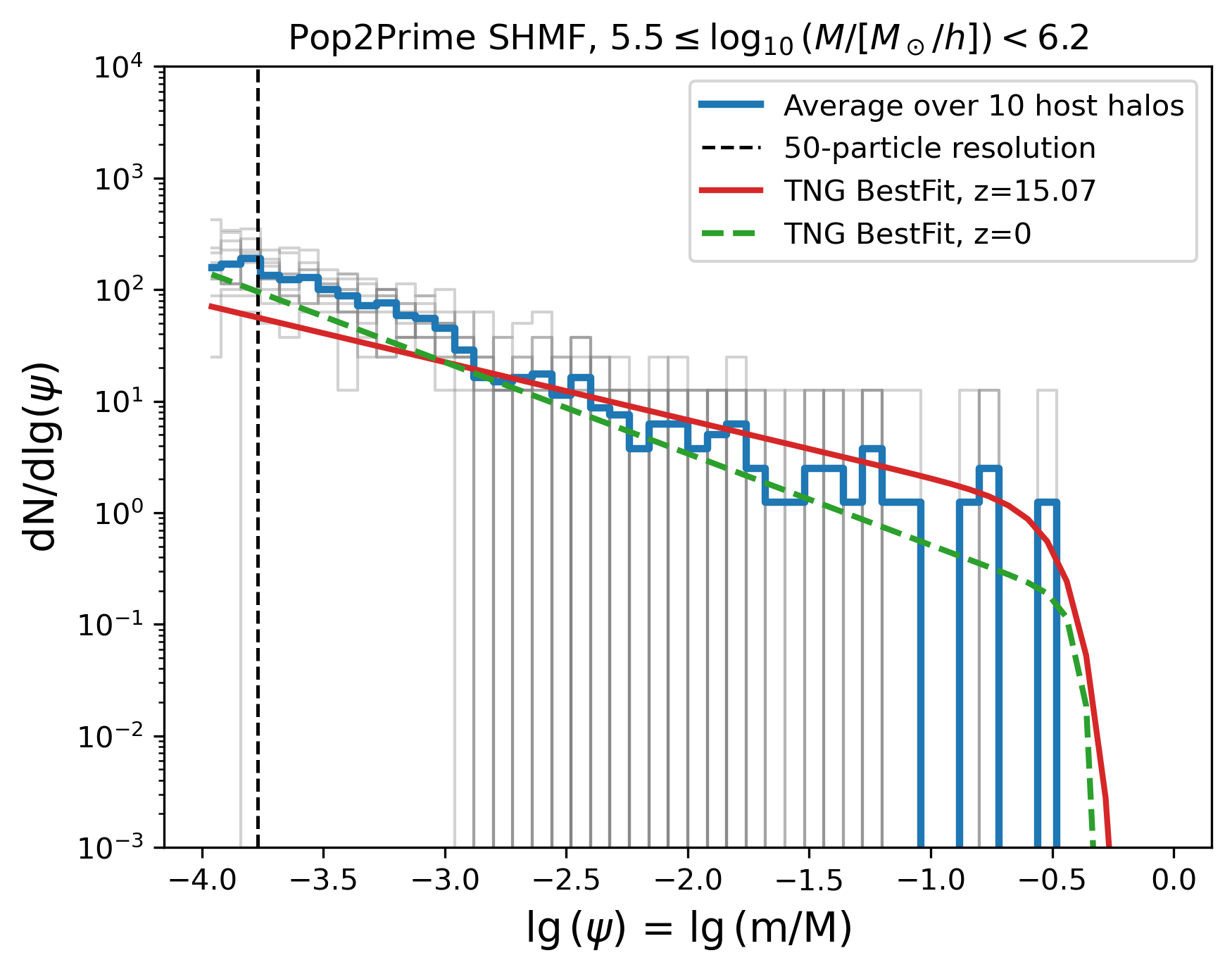}
 \caption{Same as Figure~\ref{fig:SHMF_Pop2Prime_z12}, but at $z = 15$.} 
 \label{fig:SHMF_Pop2Prime_z15}
\end{figure}

Due to the limited mass resolution of TNG50 and the absence of halos below $10^{7.5}, M_{\odot}/h$, in this appendix we present the SHMF from the $\textsc{Pop2Prime}$ simulation as a supplement. The simulation data analyzed in this work were presented in \citep{CorreaMagnus2024} and \citep{Smith2024}, which focus on Pop III star-forming halos. It is a variant of the original $\textsc{Pop2Prime}$ simulation \citep{Smith15}.

The simulation is based on the $\textsc{Enzo}$ AMR code \citep{Bryan2014, BrummelSmith2019}. It has a box size of $500$ ckpc$/h$ and zooms in on a halo reaching $1.7\times10^7, M_{\odot}$ by $z \sim 10$. It achieves a very high mass resolution of $m_{\rm DM} \sim 1, M_{\odot}$, so the SHMF is well resolved even in the low mass-ratio regime. Subhalos are identified using the $\textsc{ROCKSTAR}$ algorithm \citep{behroozi_rockstar_2013}.

Figures~\ref{fig:SHMF_Pop2Prime_z12} and~\ref{fig:SHMF_Pop2Prime_z15} show the resulting SHMF at $z = 12$ and $z = 15$, respectively. We select all host halos above $10^{5.5}, M_{\odot}/h$, roughly the threshold mass for Pop III star formation. This yields 14 host halos with $M_{\mathrm{vir}} \in [10^{5.5}, 10^{6.7}], M_{\odot}/h$ at $z = 12$, and 10 host halos with $M_{\mathrm{vir}} \in [10^{5.5}, 10^{6.2}], M_{\odot}/h$ at $z = 15$. We plot the SHMF for each individual host halo as grey histograms and show the mean SHMF with the blue histogram. In the higher mass-ratio regime, $\psi \sim 10^{-1}$, the $\textsc{Pop2Prime}$ SHMF appears to lie between the high-redshift TNG result and the $z=0$ TNG result. In the lower-$\psi$ regime, the slope of the TNG $z=0$ SHMF shows better agreement. This suggests that the SHMF for minihalos can also be fitted with Equation~\ref{eq:SHMF_form}. However, due to the limited number of host halos, the data lack the statistical power required to provide a precise fit. We leave this to future work.

\section{Non-Poissonian fluctuations in SHMF}
\label{appendix:non_Poissonian_SHMF}

\begin{figure}
 \includegraphics[width=\columnwidth]{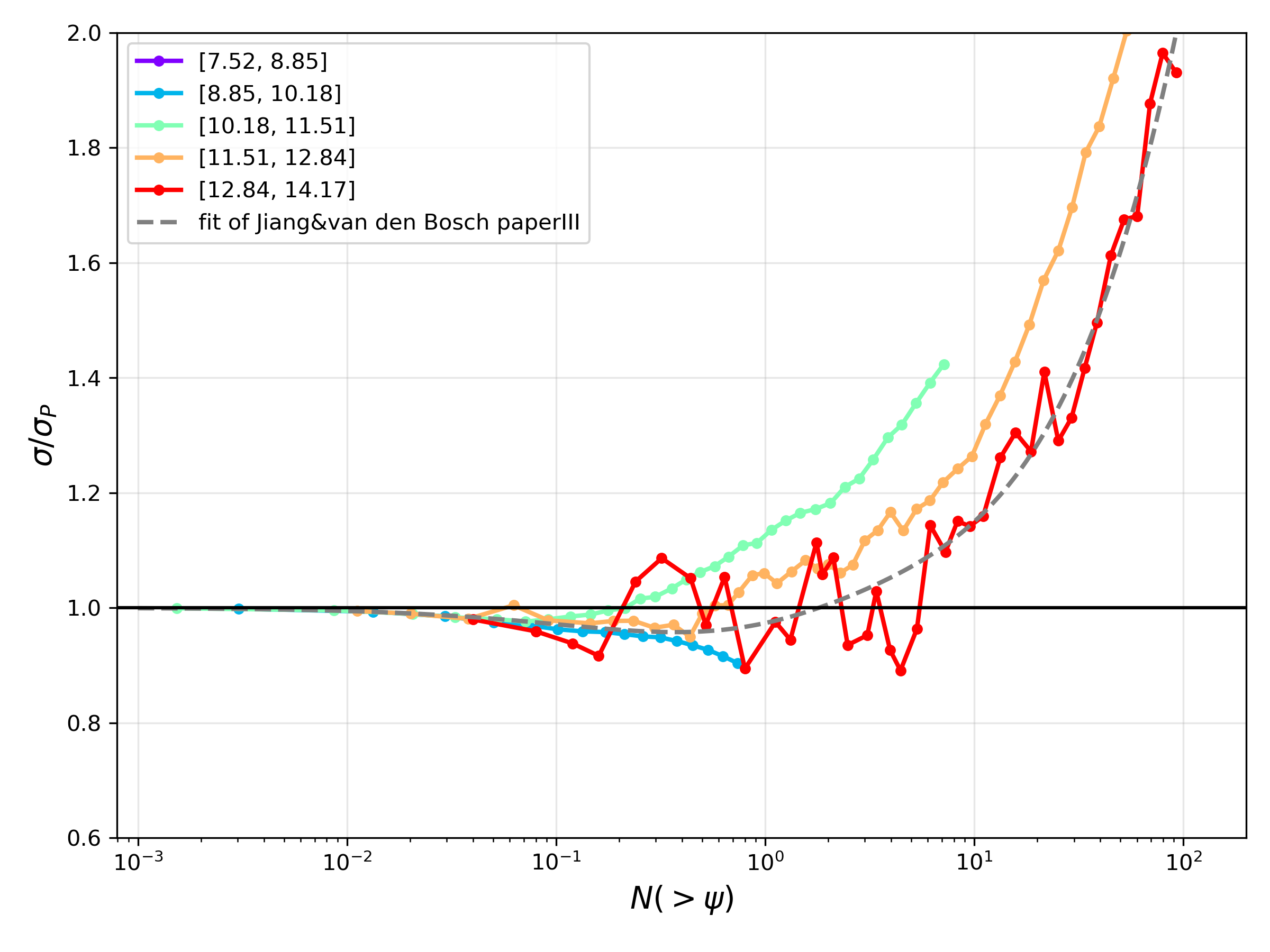}
 \caption{$\sigma /\sigma_{P}$, the scatter of the SHMF relative to a Poissonian scatter at $z = 0$. The grey dashed line is taken from the fit in \citetalias{jiang_statistics_2017} for $M_{\mathrm{halo}} \sim 10^{13.75-14.8} M_{\odot}/h$. The $\sigma /\sigma_{P}$ for the most massive bin (the red line) generally agrees with their result, with fluctuations due to low number statistics. The scatter transits from sub-Poissonian to super-Poissonian at $\langle N(\geq \psi) \rangle \approx 2$. Other bins show a similar trend, but don't quantitatively agree with \citetalias{jiang_statistics_2017}. The lowest mass bin has no data due to subhalos being unresolved.} 
 \label{fig:SHMF_sigma_ratio_snap_99}
\end{figure}

\begin{figure}
 \includegraphics[width=\columnwidth]{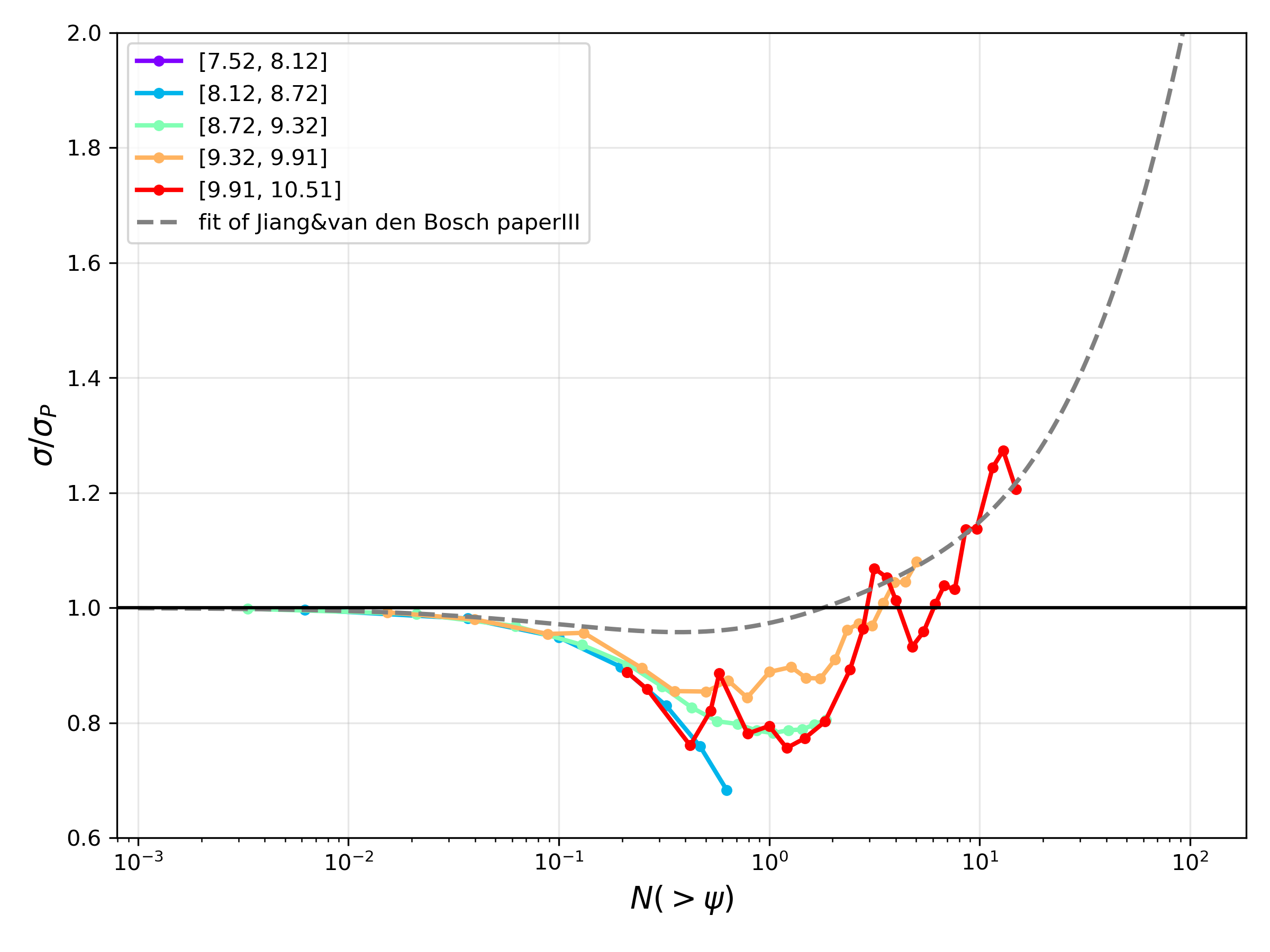}
 \caption{Same as Figure~\ref{fig:SHMF_sigma_ratio_snap_99} but at $z = 12$. The non-Poissonian scatter shows a similar trend to the most massive halos at $z = 0$. A more quantitative and statistically robust analysis will require simulations with higher mass resolution and a larger volume.} 
 \label{fig:SHMF_sigma_ratio_snap_2}
\end{figure}

Typically, the non-Poisson fluctuations are not captured by semi-analytic models. According to \citet{boylan-kolchin_theres_2010}, in N-body simulations, the scatter of SHMF becomes super-Poissonian, i.e. $\sigma > \sigma_{\rm{Poisson}} = \sqrt{N}$, when the mass ratio $\psi$ is smaller than $5 \times 10^{-3}$. \citetalias{jiang_statistics_2017} further find that at the higher mass ratio end, or when the average number of subhalos becomes small ($\langle N(\geq \psi) \rangle < 2$), the scatter of SHMF has a sub-Poissonian nature, i.e. $\sigma < \sigma_{\rm{Poisson}}$, though the ratio $\sigma /\sigma_{\rm{Poisson}}$ is still close to 1.  

By comparing the error bars with the dashed lines in  Figure~\ref{fig:SHMF_cumulative_snap_99} and Figure~\ref{fig:SHMF_cumulative_snap_2}, we observe that the scatter of the SHMF transitions from super-Poissonian to sub-Poissonian as $\psi$ increases, with $\langle N(\geq \psi) \rangle \approx 2$ approximately marking the boundary. This is shown more clearly in Figure~\ref{fig:SHMF_sigma_ratio_snap_99} and Figure~\ref{fig:SHMF_sigma_ratio_snap_2}, which plot $\sigma/\sigma_{\mathrm{Poisson}}$ as a function of $\langle N(\ge \psi)\rangle$. However, the fitting of $\sigma/\sigma_{\rm{Poisson}}$ in \citet{boylan-kolchin_theres_2010} and \citetalias{jiang_statistics_2017} accurately describes only the most massive host-halo bin at $z = 0$, and cannot be reliably extended to other redshifts or to host halos of different masses, though they do show similar trends.

Since the dynamical-friction heating is predominantly contributed by subhalos with $\psi > 10^{-2}$, and the sub-Poissonian behavior in this regime deviates only mildly from the Poisson expectation, we defer a systematic investigation of the non-Poissonian nature of the scatter to future work. We therefore assume Poisson fluctuations for estimating the impact of SHMF scatter on the dynamical-friction heating rate.

\section{Truncated Gaussian fit of subhalo Mach number distribution in TNG50}
\label{appendix:TG_Mach_fit}

\begin{figure}
  \centering
  \includegraphics[width=\columnwidth]{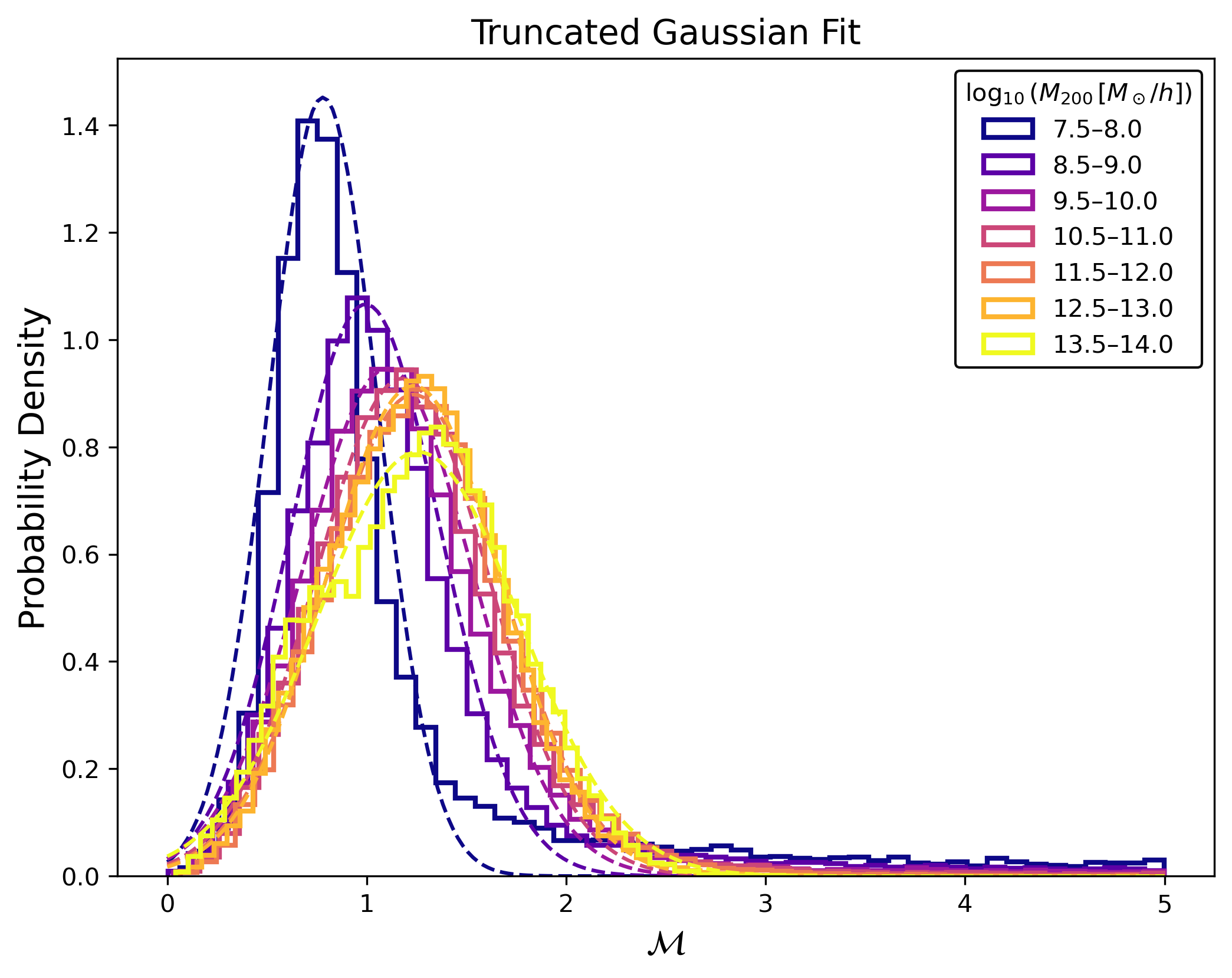}
  \caption{Same as Figure~\ref{fig:mach_TNG_snap_99}, but the dashed lines use a normalized Gaussian fit truncated at $\mathcal{M} = 0$. }
  \label{fig:mach_TGfit_TNG_snap_99}
\end{figure}

\begin{figure*}
 \includegraphics[width=\linewidth]{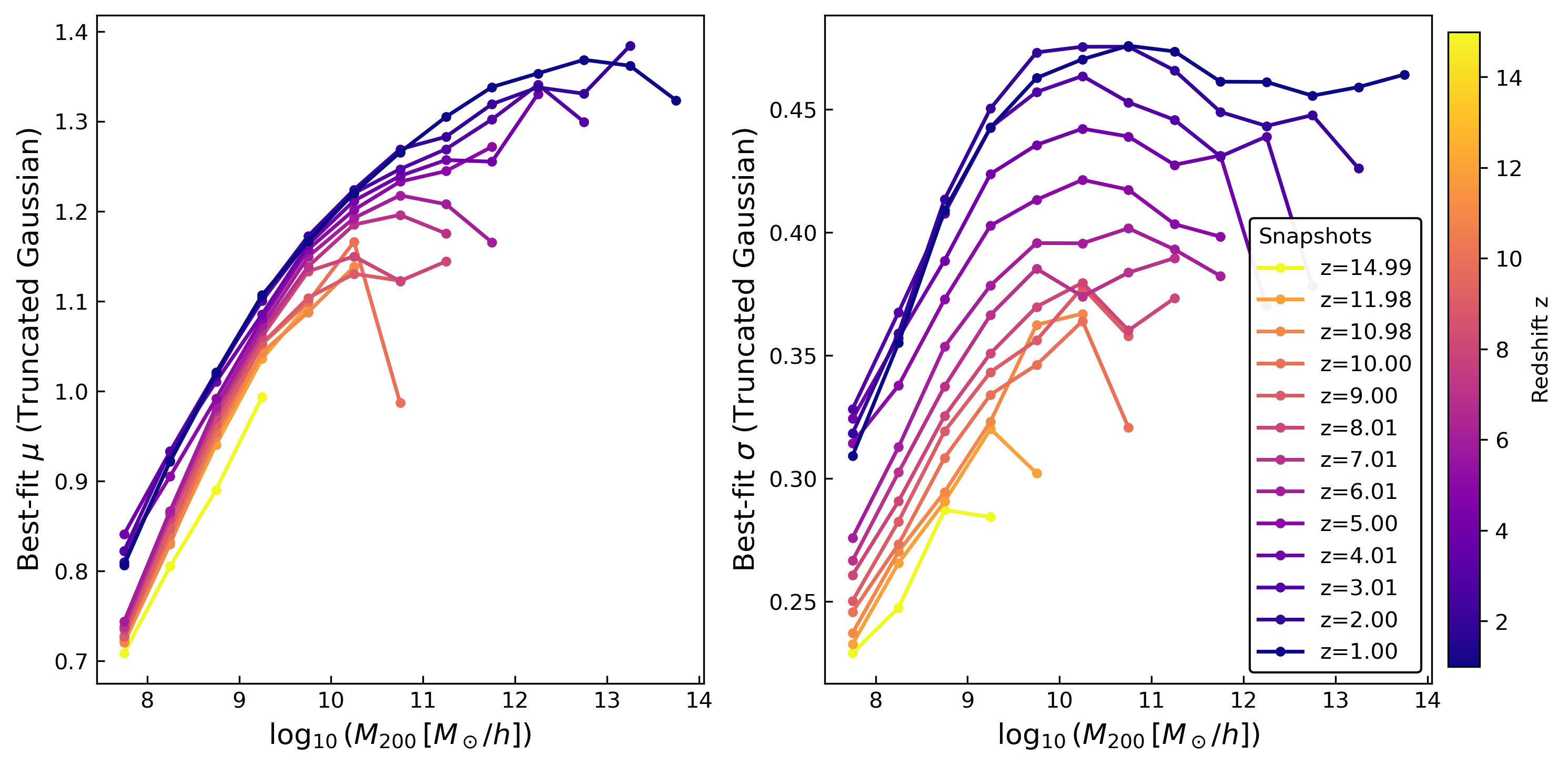}
 \caption{Truncated Gaussian fit of subhalo Mach numbers for all host halo mass bins across the redshifts. }
 \label{fig:mu_sigma_TG_fit}
\end{figure*}

\begin{figure*}
 \includegraphics[width=\linewidth]{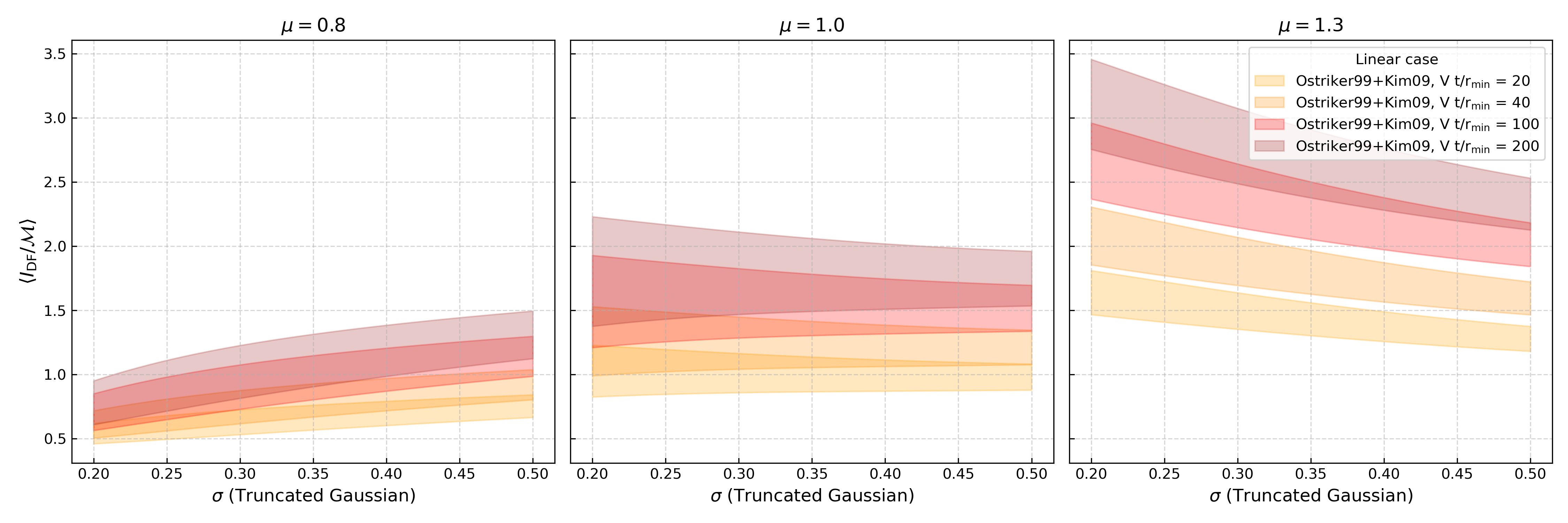}
 \caption{Average DF correction factor I/$\mathcal{M}$ for the truncated Gaussian distribution of Mach numbers. The 3 panels correspond to $\mu$ = 0.8, 1.0, and 1.3 as typical values of the fitting results in Figure~\ref{fig:mu_sigma_TG_fit}. Same as Figure~\ref{fig:avg_I_over_M}, different colors correspond to different orbital parameters $Vt/r_{\mathrm{min}}$. The shaded areas show the impact of the nonlinear perturbation.}
 \label{fig:avg_I_over_M_TGfit}
\end{figure*}

In Section~\ref{section:Mach_number_dependence}, we use a Maxwell-Boltzmann distribution to fit the Mach numbers of the subhalos to estimate the impact of the Mach number correction in gaseous DF. However, the Maxwell-Boltzmann distribution doesn't capture the peaks of the higher mass groups accurately, as shown by the comparison between the solid and dashed yellow lines in Figure~\ref{fig:mach_TNG_snap_99}. Therefore, in this appendix, we present the fitting results for a Gaussian distribution of the Mach numbers. The Gaussian distribution is truncated at $\mathcal{M} = 0$ and then normalized, i.e.,

\begin{equation}
f(\mathcal{M}; \mu, \sigma)=
\begin{cases}
\dfrac{1}{\sigma}\,
\dfrac{\varphi\!\left(\dfrac{\mathcal{M}-\mu}{\sigma}\right)}
{1-\Phi\!\left(\dfrac{-\mu}{\sigma}\right)}, 
& \mathcal{M} \ge 0, \\[10pt]
0, 
& \mathcal{M} < 0 .
\end{cases}
\end{equation}

where $\varphi(\xi)=\frac{1}{\sqrt{2 \pi}} \exp \left(-\frac{1}{2} \xi^2\right)$, and $\Phi(\xi)=\frac{1}{2}(1+\operatorname{erf}(\xi/ \sqrt{2}))$.

Figure~\ref{fig:mach_TGfit_TNG_snap_99} shows the new fit at $z = 0$. With 2 free parameters, the dashed lines agree better with the peaks of the Mach number distribution taken from TNG50. Still, for the lower mass halos like those in the $10^{7.5} - 10^8 M_{\odot}/h$ bin, the truncated Gaussian fit tends to underestimate the contribution from supersonic cases, but the original Maxwell-Boltzmann fit captures the long tail better.

In Figure~\ref{fig:mu_sigma_TG_fit}, we show the evolution of the best-fit $\mu$ and $\sigma$ across redshifts and host halo masses. Both $\mu$ and $\sigma$ increase with host halo mass. According to the right panel, for a fixed host halo mass, the lower redshift ones exhibit a larger scatter of subhalo Mach numbers. However, in the best-fit $\sigma$ range $\sim 0.2 - 0.5$, the average Mach number correction $\langle  I/\mathcal{M}\rangle$ doesn't have a significant change, as illustrated in Figure~\ref{fig:avg_I_over_M_TGfit}. The 3 panels in Figure~\ref{fig:avg_I_over_M_TGfit} correspond to 3 representative values of $\mu$. For $\mu = 0.8$, the average correction factor varies between 0.5 and 1.5, depending on orbital parameters and nonlinear corrections. As $\mu$ increases to 1.0, the $\langle  I/\mathcal{M}\rangle$ can increase to 2, and it further increases to 3 for $\mu = 1.3$. While $\langle  I/\mathcal{M}\rangle$ changes with host halo masses, it's still a factor of a few. We conclude that different fitting procedures will not qualitatively change our results in the paper.

\section{Comparison between cooling and heating in massive halos: NFW gas profile}
\label{appendix:NFW_cooling_profile}
\begin{figure}
 \includegraphics[width=\columnwidth]{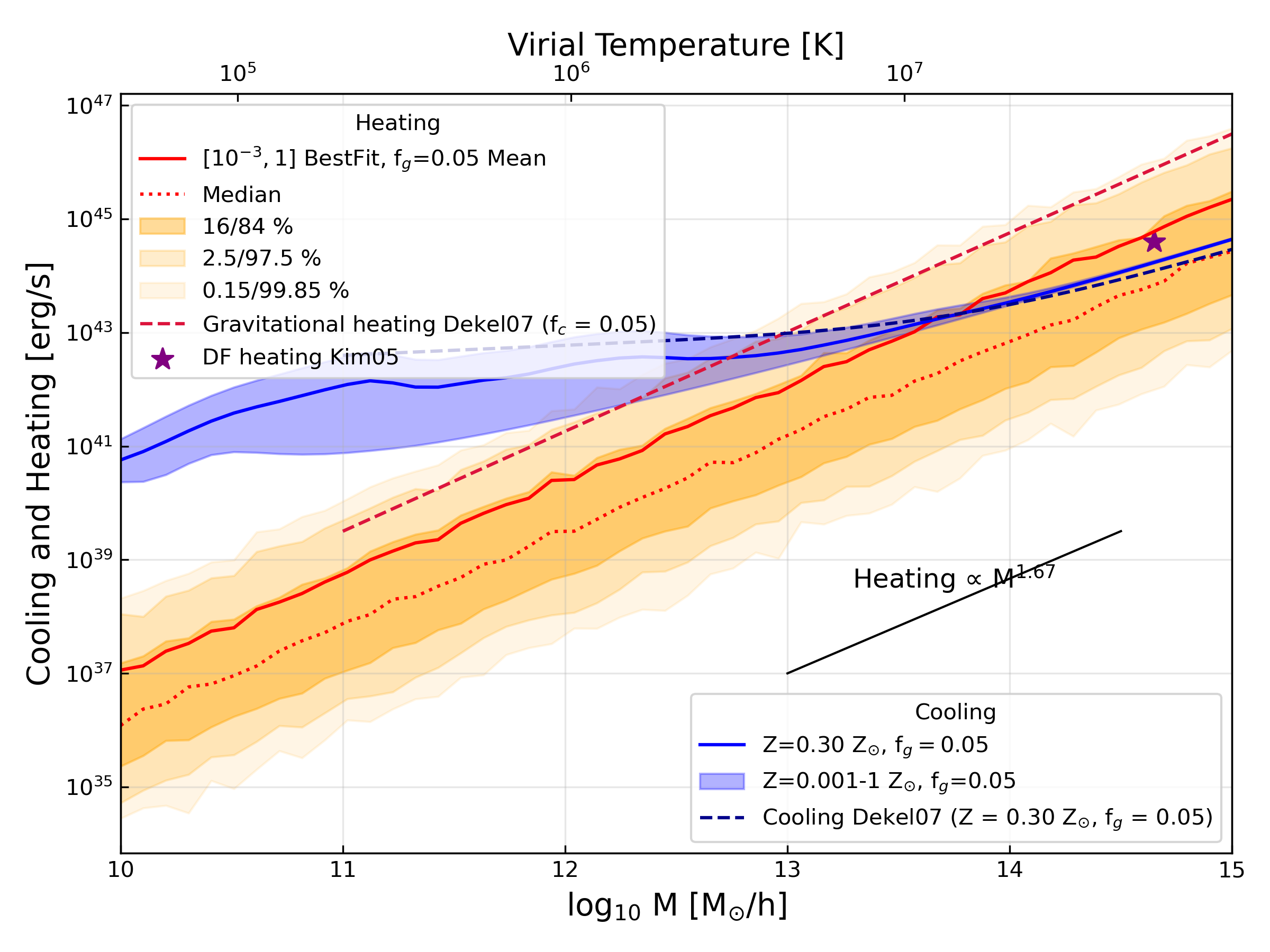}
 \caption{Same as the upper panel of Figure~\ref{fig:singlehost_z0_2_6}, but the halo gas density profile is the cusp NFW profile ($\alpha = 1$ in Equation~\ref{eq:generalized_NFW_profile}).}
 \label{fig:singlehost_z0_NFW}
\end{figure}

Section~\ref{Section:massive_halo} parameterizes the halo profiles like \citetalias{Dekel2007} and the default gas density profile is the core profile, i.e., $\alpha = 0$ in Equation~\ref{eq:generalized_NFW_profile}. \citetalias{Dekel2007} has conducted a parameter study of the influence of gas density profile and total mass profile on the minimum halo mass for galaxy quenching. 

Our experiments with the gas density profiles lead to similar results. In Figure~\ref{fig:singlehost_z0_NFW}, we show the comparison between cooling and heating in halos at $z= 0$ as an example. The only difference from the upper panel of Figure~\ref{fig:singlehost_z0_2_6} is that the cooling rate is based on the $\alpha = 1$ NFW profile. The cuspy profile leads to a higher cooling rate by a factor of a few, but doesn't qualitatively change the importance of DF heating in the most massive halos.

\bsp	
\label{lastpage}
\end{document}